\documentclass[hidelinks,12pt]{article}
\usepackage[utf8]{inputenc}
\usepackage{amsmath}
\usepackage{graphicx}
\usepackage{booktabs}
\usepackage{float}
\usepackage{setspace}
\usepackage[table]{xcolor}
\usepackage[round,authoryear]{natbib}
\usepackage{longtable}  % for multipage tables
\usepackage{hyperref}  % Load hyperref last
\usepackage{dsfont}
\usepackage{eurosym}
\usepackage[labelfont=bf]{caption}
\usepackage{setspace}
\usepackage{arydshln}  % For dashed lines in tables
\usepackage{mathtools}   % For \mathclap
\usepackage{tabularx} % for better behaved tables

\usepackage{xltabular}

\usepackage{tikz}
\usetikzlibrary{arrows}

\usepackage{rotating}    % to rotate things
\usepackage{multirow}    % For merged rows
\usepackage{array}       % For advanced column formatting
\usepackage{pdflscape}   % For landscape pages

\usepackage{amsthm}
\usepackage{amsfonts}
\usepackage{makecell}

\usepackage{enumitem} % To indent lists manually

\usepackage{algorithm} % for pseudocode
\usepackage{algpseudocode} % for pseudoccode

\setcitestyle{aysep={}}  % Removes comma between author and year
\bibpunct{(}{)}{;}{a}{}{,}  % Economic Journal citation style
\newtheorem{proposition}{Proposition}
\newtheorem{appendixproposition}{Proposition}

\renewcommand*{\thefootnote}{\fnsymbol{footnote}}

\usepackage[margin=1in]{geometry} 

\usepackage{subcaption}

\begin{document}

\setcounter{page}{0}
\title{Measuring the Unmeasurable? \\ Systematic Evidence on Scale Transformations in Subjective Survey Data} 

\author{Caspar Kaiser\footnote{Warwick Business School, University of Warwick; Wellbeing Research Centre, University of Oxford.} \\
Anthony Lepinteur\footnote{Department of Behavioural and Cognitive Science, University of Luxembourg; IZA Institute of Labour Economics, Bonn. \\ 
We thank 
Dan Benjamin,
Anita Braga,
Matthew Cashman,
Andrew E. Clark, 
Ada Ferrer-i-Carbonell,
Elena Fumagalli,
Leonard Goff,
Ori Heffetz,
Martijn Hendricks,
Richard Heys,
Christos Makridis,
Giorgia Menta, 
Andrew Oswald,
Kelsey O'Connor,
Alberto Prati,
Marco Ranaldi,
Carsten Schröder,
Claudia Senik,
Leonie Steckermeier,
Robert Stüber, 
Mattie Toma,
as well as seminar and conference participants at the London School of Economics, University of Leeds, Brno, Groningen and Alcala, Warwick Business School, DIW, Freie Universität Berlin, General Conference of ISQOLS Rotterdam, Regional Conference of ISQOLS Johannesburg, General Conference IARIW 2024, LISER, Nanyang Technological University, Measuring Progress Workshop (STATEC), for their helpful comments and suggestions on earlier drafts of the paper. We are also grateful to the authors of the papers included in \textit{WellBase} for sharing replication materials, clarifying methodological details, and engaging constructively with our replication effort. All errors remain our own. This project received ethical approval from the University of Warwick's Humanities and Social Sciences Research Ethics Committee (Ref.: 228/23-24).}}

\renewcommand*{\thefootnote}{\arabic{footnote}}

\date{\today}

\maketitle
\label{sec:titlepage}

\vspace{-18pt}

\begin{abstract} 
\setstretch{1}

%Ordered response scales are ubiquitous in economics, but their interpretation rests on an untested assumption: that numerical labels reflect equal psychological intervals. The contribution of this paper is to provide a systematic assessment of this linearity assumption, developing a general framework to quantify how easily empirical results can be overturned when it is relaxed. Using original experimental data, we show that respondents use survey scales in ways that deviate from linearity, but only mildly so. Focusing on wellbeing research, we then replicate 40,000+ coefficient estimates across more than 80 papers published in top economics journals. Coefficient signs are remarkably robust to the mild departures from linear scale-use we document experimentally. However, estimates of relative effect sizes, which are crucial for policy applications, are unreliable even under these modest non-linearities. \\

%AER restriction (100 words)
Ordered response scales are ubiquitous in economics, but their interpretation rests on an untested assumption: that numerical labels reflect equal psychological intervals. We develop a framework to quantify how relaxing this assumption affects empirical results. Using new experimental evidence, we show that scale use is only mildly non-linear. Replicating over 40,000 estimates from more than 80 papers, we find that coefficient signs and significance are largely robust, but relative magnitudes are not. Even modest non-linearities generate substantial variation in implied trade-offs.

\vspace{0.5cm}


\end{abstract}

\newpage

\section{Introduction}
\vspace{-.5em}

Ordered response scales, or `Likert scales', are a standard instrument for measuring latent constructs like political preferences,  risk attitudes, wellbeing, trust, etc. These scales are easy to administer and, for many disciplines, have proved pivotal for answering questions that cannot otherwise be answered with behavioural data. 

Yet scepticism over the validity and use of such Likert scale measures remains. Three concerns underlie such scepticism. The first concern focuses on whether commonly used survey items really do capture the underlying constructs of interest — such as attributes of utility functions (e.g. risk aversion) or utility itself (e.g., `subjective wellbeing'). See, for example, \cite{Bertrand2001} or \cite{benjamin2023happiness}. The second concern asks whether responses are comparable across people and time: does a reported ``6 out of 10'' mean the same for you as for me, or for me today as for me a year ago? See e.g. \cite{angelini2014danes}, \cite{fabian2022scale}, \cite{kaiser2022using}, \cite{benjamin2023adjusting} or \cite{prati2026possible}.  The third concern involves the relationship between the numerical labels that researchers attach to ordered response categories (i.e., ``1'', ``2'', ``3'', etc.) and how these map onto the unobserved latent variable that researchers are trying to measure. 

We focus on this third concern. The core issue is this: \emph{we do not know the functional form of the relationship between reported scale values and the underlying latent variable}. Even if all respondents use the scale in approximately the same way, does a one-unit difference on the response scale represent the same magnitude of change in the latent variable across all parts of the scale? Or is this relationship non-linear, with differences between certain response categories representing larger gaps in the underlying construct than others? 

Although this issue applies to any construct measured with Likert scales, much of the methodological work focused on wellbeing. This is unsurprising: Economists have studied life satisfaction and happiness scales for over fifty years (e.g. \cite{easterlin1974does} or \cite{vanpraag1971welfare}). The modern study of wellbeing began in the 1990s, linking it to income, unemployment and macroeconomic conditions \citep{clark1994unhappiness, oswald1997happiness, blanchflower2004well}. Today, wellbeing scales inform government policy, as seen in the UK Treasury's 2021 \textit{Green Book} \citep{treasury2021}. 
 
Within that literature, \cite{ferrer2004important} were among the first to address the linearity concern. They showed that coefficients estimated from an ordered logit or probit models are similar to those based on OLS regressions. Nevertheless, \cite{oswald2008curvature} highlighted how a potentially non-linear ``reporting function'' (i.e. the mapping from underlying states to survey responses) could distort estimates of non-linear effects, such as estimates of the curvature of the income-to-wellbeing relationship. That paper also provided some empirical evidence to suggest that the reporting function is close to linear. 

Focusing on coefficient signs, \cite{schroder2017revisiting} provided conditions under which single-covariate regression results can be sign-reversed when allowing for a non-linear reporting function. They also showed that such sign reversals can indeed occur in practice; as did \cite{bloem2022how} who broadened the analysis to a wider class of non-linear functions. 

\cite{bond2019sad} generalised these ideas. They demonstrated that virtually all empirical findings based on Likert scales can be reversed via some monotonic transformations of the response scale. They argued that without strong assumptions about the distribution of the latent concept within response categories and about the functional form of the reporting function, it is impossible to draw definitive conclusions about the sign of differences between groups.\footnote{\cite{liu2023happy} propose using survey response times to overcome the identification problem raised by \cite{bond2019sad}. Their approach exploits \textit{chronometric effects}: decisions tend to be faster when the latent state is further from the reporting threshold. They show that response times thereby contain information about the distribution of the latent variable within categories, helping to relax the assumptions needed in standard ordered response models.} In turn, \cite{kaiser2023how} identified effect heterogeneities across the distribution of wellbeing as the underlying mechanism that drives potential sign reversals. They derived a condition under which coefficients in OLS regressions with multiple covariates are reversible and applied this condition to a selected set of covariates.\footnote{These papers all focus on how a potentially non-linear reporting function may affect estimates of the conditional mean of underlying wellbeing. Rankings of the conditional median, in contrast, are invariant to such non-linear transformations \citep{chen2022robust,bloem2022analysis}.}

However, we currently lack systematic evidence on how serious these concerns are.  Existing studies have only analysed a small number of \textit{selected} datasets and variables. Even if results can be reversed in principle, we have no measure of how `easy' it is to obtain such reversals, and thus how concerned we should be in practice. We also have surprisingly little direct evidence on how respondents actually interpret survey scales. This makes it difficult to assess which transformations are empirically plausible. Finally, while much attention has focused on coefficient signs, we know little about how non-linear transformations affect statistical significance or the relative magnitudes of estimates.

We address these gaps. To do so, we first propose a set of desiderata that a measure of departure from linear scale use should satisfy, and characterise the class of functions that meets them. We then show how, for any such `cost' function $C$, finding the least non-linear transformation capable of reversing a coefficient's sign, altering its significance, or shifting a coefficient ratio can be cast as a constrained optimisation problem. For special cases of these problems, we give simple closed-form solutions. We provide Stata routines on \href{https://github.com/casparwarwick/reversals}{\color{blue}{GitHub}}.

To make full use of this framework, it helps to know which values for this cost $C$ are empirically plausible. Therefore, using new experimental data from multiple, complementary elicitation strategies, we provide direct evidence on how respondents use response scales. We find that deviations from linearity are present but small and tightly concentrated. Many respondents use the scale linearly or near-linearly, and extreme non-linear scale use is rare. Individual heterogeneity in scale use is largely idiosyncratic and unrelated to observables. 

We then ask how much any plausible departures from linear scale use matter in practice. Re-running a regression on a single dataset, or illustrating the issue with a limited number of selected examples, would be unlikely to provide a meaningful answer. We therefore reproduce the quasi-universe of wellbeing literature published in top-tier economics journals over the past fifteen years, creating an extensive database we call \textit{WellBase}. This means reproducing 73 papers, with more than 1,600 regressions and 28,000 coefficients. Using this dataset, we systematically assess the vulnerability of the published literature.\footnote{Throughout, we primarily focus on the behaviour of OLS estimators under monotonic transformations of the response scale. It might instead be natural to analyse ordered response scales through threshold-crossing models \citep{klein2002shift}. However, as shown by \cite{bond2019sad}, very similar concerns apply to ordered probit/logit models. Moreover, our systematic replication exercise in Section \ref{sec:wellbase_evidence} shows that the published economics literature overwhelmingly applies OLS to such scales. The central question, thus, is how robust this practice is.}

Approximately 25\% of results published in leading economic journals are reversed with some transformation that has a \textit{plausible} cost. Restricting ourselves to interpreting wellbeing data as merely ordinal (i.e. allowing for any departure from linear scale use), increases this share to about 60\%. The relationship between the `cost' of deviating from linearity and the risk of sign reversal is concave. Crucially, the probability that a given estimate can be sign-reversed is systematically related to identifiable features of research design. Certain design choices, like leveraging natural experiments, are associated with notably lower risks. Likewise, more precisely estimated coefficients are much less prone to reversals under \textit{plausible} transformations.

We also examine risks of `significance reversals'. Estimates originally significant at the 0.1\% level prove highly robust: roughly 95\% remain significant at the 5\% level even under a purely ordinal interpretation. However, estimates with p-values between 0.01 and 0.05 are highly vulnerable even under \textit{plausible} transformations. The potential for non-linear scale use therefore makes reliable statistical inference considerably more challenging. Turning to relative magnitudes, we focus on unemployment and income. While coefficient signs for these determinants are fairly robust, their relative magnitudes are highly sensitive to scale use assumptions: Marginal rates of substitution between unemployment and income can vary substantially even under \textit{plausible} deviations from linearity. 

Our findings generalise beyond wellbeing scales. To show this, we reproduce 16 papers (23,104 coefficients) published in top-five economics journals. Each of these use Likert-scales to measure e.g. risk aversion, social trust, or political preferences. The prevalence and predictors of reversal for these measures closely mirror our wellbeing results.

The next section develops our cost-function approach. Section \ref{sec:scale_use_evidence} empirically assesses respondents' scale interpretations. Section \ref{sec:wellbase_evidence} describes \textit{WellBase} and presents our results based on it. Section \ref{sec:guidelines} shows how to implement our approach in practice and provides some guidance for assessing robustness to non-linear scale transformations via an illustrative example. Section \ref{sec:discussion} concludes. The appendices provide proofs, additional discussion, and further results. Replication codes are available \href{https://osf.io/u2kh5}{\color{blue}{here}}.

\section{Evaluating robustness to scale transformations}
\label{sec:background}
\vspace{-.5em}

Standard OLS regressions implicitly treat the numerical labels on the response scale as reflecting equal intervals in the underlying latent variable. This section develops a framework for assessing how much depends on that assumption.
 
Previous work has established conditions under which coefficient signs can be reversed by monotonic transformations of the response scale \citep{schroder2017revisiting, bond2019sad, bloem2022how, kaiser2023how}. We build on these contributions in two ways. First, we broaden what it means for a transformation to `overturn' a result, extending the analysis from coefficient signs to statistical significance and to ratios of coefficients. Second, we introduce a cost function approach that makes it possible to quantify the \textit{minimal} departure from linearity needed to overturn a result. 

In principle, the approach applies to any bounded ordered scale and to any linear estimator. For example, as shown in Supplementary Appendix \ref{sec:analytical}, it extends to fixed-effects and two-stage least squares estimators, and to continuous outcomes.
 
\subsection{Set-up}
\label{sec:setup}
 \vspace{-.5em}

Consider a dataset containing responses to a survey question. For each individual $i$, responses are recorded using ordered categories: $r_i \in \{1, 2, \ldots, k, \ldots, K\}$. These responses measure an underlying but unobservable state $s_i$.
 
However, the functional relationship between $r_i$ and $s_i$ is unknown. This uncertainty motivates our analysis.\footnote{\cite{bond2019sad} also raise a second concern, namely that typically we only observe a small number of response categories, and that individuals sharing the same response category may differ systematically in their underlying state in ways that correlate with covariates. Under sufficiently adverse within-category distributions, this discreteness alone can flip the sign of estimated regression coefficients, even when scale use is perfectly linear. In Supplementary Appendix \ref{sec:discreteness} we provide empirical evidence to suggest that this second concern is minor. Specifically, when eliciting both discrete and continuous measures of life satisfaction, the two yield near-identical regression coefficients.} We could transform $r_i$ using any positive monotonic function $f$ to obtain $\tilde{r}_i = f(r_i)$. Different transformations correspond to different interpretations of the response scale. The identity function $f(r) = r$ treats the scale as cardinal. Non-linear transformations alter the assumed `distances' between response categories. Following \citet{oswald2008curvature}, we can interpret $f$ as the inverse of a `reporting function' that maps underlying states to survey responses.

Write an OLS regression of $\tilde{r}_i$ on $\mathbf{X}_i$ as
\begin{equation}
\label{eq:ols_reg}
    \tilde{r}_i = \mathbf{X}_i \hat{\boldsymbol{\beta}}^{(\tilde{r})} + e_i,
\end{equation}

where $e_i$ denotes the residuals and $\hat{\beta}^{(\tilde{r})}_m$ is the coefficient on covariate $X_{im}$. Different choices of $f$ yield different coefficient estimates $\hat{\beta}^{(\tilde{r})}_m$. In particular, the sign, significance, and relative magnitude of any estimated coefficient may all depend on the transformation used. \citet{kaiser2023how} derived a condition based on a decomposition of $\hat{\beta}^{(\tilde{r})}_m$ into coefficients from a series of binary regressions (cf. Section \ref{sec:optimisation}), under which the sign of $\hat{\beta}^{(\tilde{r})}_m$ is invariant to the choice of $f$. When this condition fails, which empirically it often does,  there exist transformations that reverse the sign of the estimated coefficient. 

A purely ordinal interpretation of the response scale treats all positive monotonic transformations as equally admissible. However, it is not clear that we really should treat all positive monotonic transformations as equally admissible. Instead, it seems that more non-linear transformations are less plausible (and further away from common practice) than less non-linear transformations. We thus introduce a way to measure how \textit{far} a given transformation departs from linearity, and then show how to make practical use of such a measure. Thereafter, we provide evidence on how respondents' scale use actually departs from linearity.

\subsection{Desiderata for a measure of non-linear scale use}
\label{sec:theorem_p}
\vspace{-.5em}

We here postulate and argue for five desiderata that a measure $\Pi$ of departure from linear scale use should usefully satisfy. We will use the term `cost function' for any $\Pi$ satisfying these desiderata (reflecting our intuition that increasingly non-linear transformations are increasingly `costly' departures from the standard assumption). 

We need some additional notation for this purpose. Let $l_k$ denote the value assigned to response category $k$ in the original coding of $r_i$, with $l_k = k$ under linear scale use. Let $\tilde{l}_k \equiv f(l_k)$ denote the value assigned to category $k$ under some transformation $f$, and let $\Delta_k \equiv \tilde{l}_{k+1} - \tilde{l}_k$ denote the gap between adjacent transformed labels. Without loss of generality, we here rescale so that $\tilde{l}_K - \tilde{l}_1 = 1$, or equivalently $\sum_{k=1}^{K-1} \Delta_k = 1$. Let $\mathcal{D} = \{\Delta \in \mathbb{R}^{K-1} : \Delta_k \geq 0 \text{ for all } k, \text{ and } \sum_{k=1}^{K-1} \Delta_k = 1\}$ denote the set of all valid gap vectors.

\vspace{8pt}

\noindent \textbf{Desideratum 1} (Extremal values). \textit{$\Pi(\Delta) = 0$ when $\Delta_k = 1/(K-1)$ for all $k$, and $\Pi(\Delta) = 1$ when $\Delta_j = 1$ for some $j$ and $\Delta_k = 0$ for all $k \neq j$.}

\vspace{4pt}

This desideratum sets any cost function's minimum to occur at uniform gaps (which is equivalent to linear scale use) and the maximum at single-jump scale use, where a single pair of adjacent categories covers the entire range of the scale (while all other categories are collapsed to the same value). By fixing the minimum at zero and the maximum at one, it forces all cost functions to a common range. 

\vspace{8pt}

\noindent \textbf{Desideratum 2} (Permutation invariance). \textit{$\Pi$ is invariant under permutations of the gap indices: $\Pi(\Delta_1, \ldots, \Delta_{K-1}) = \Pi(\Delta_{\sigma(1)}, \ldots, \Delta_{\sigma(K-1)})$ for any permutation $\sigma$.}

\vspace{4pt}

Without specific evidence on scale use, we have no reason to treat gaps at different positions on the scale differently. A compression between categories 2 and 3 should carry the same cost as an identical compression between categories 8 and 9. Permutation invariance ensures this. It means that the cost function measures the \textit{extent} of departure from linearity, but remains agnostic about its \textit{shape}.\footnote{It may be that researchers eventually have evidence about both the extent \textit{and} shape of non-linear scale use. For example one might find that respondents place disproportionately larger distances near one end of the scale. In the absence of such evidence, we should not encode a preference into our cost function.}

\vspace{8pt}

\noindent \textbf{Desideratum 3} (Strict spread sensitivity). \textit{If $\Delta$ is a strict mean-preserving spread of $\Delta'$, then $\Pi(\Delta) > \Pi(\Delta')$.}

\vspace{4pt}

A gap vector in which some gaps are very large and others very small represents a greater departure from linearity than one in which all gaps are similar. Strict spread sensitivity captures that redistributing `mass' from smaller gaps to larger ones (while keeping the mean fixed) must increase the cost. This desideratum is analogous to the Pigou-Dalton transfer principle in the measurement of inequality.

\vspace{8pt}

\noindent \textbf{Desideratum 4} (Continuity). \textit{$\Pi$ is continuous on $\mathcal{D}$.}

\vspace{4pt}

Two transformations that differ only slightly in their gap vectors should receive similar costs. Without continuity, arbitrarily small changes in scale interpretation could produce large jumps in measured non-linearity. 

\vspace{8pt}

\noindent \textbf{Desideratum 5} (Monotonic additive separability). \textit{$\Pi$ has the form $\Pi(\Delta) = g\bigl(\sum_{k=1}^{K-1} \varphi_k(\Delta_k)\bigr)$ for some monotonic function $g$ and some functions $\varphi_k$.}

\vspace{4pt}

Motivated by the same concerns as noted for Desideratum 2 (Permutation invariance), this imposes that one gap's deviation from uniformity does not interact with another's.

\vspace{10pt}

Jointly, these five desiderata characterise the following class of functions.

\begin{proposition}
\label{prop:plausibility}
Any cost function $\Pi: \mathcal{D} \tiny{\rightarrow} [0,1]$ satisfying Desiderata 1-5 can be written as:
$$\Pi(\Delta) = h\left(\frac{\sum_{k=1}^{K-1} \psi(\Delta_k)}{\max_{\Delta' \in \mathcal{D}} \sum_{k=1}^{K-1} \psi(\Delta'_k)}\right)$$
where $\psi: \mathbb{R}_+ \rightarrow \mathbb{R}$ is continuous and strictly convex with a unique minimum at $\Delta_k = 1/(K-1)$, and $h: [0,1] \rightarrow [0,1]$ is continuous and increasing with $h(0) = 0$ and $h(1) = 1$.
\end{proposition}

\noindent The proof is in Appendix \ref{sec:proofs_theorem}.   

\subsection{The SD-based cost function}
\label{sec:variance}
 \vspace{-.5em}
 
Proposition \ref{prop:plausibility} does not uniquely identify any particular cost function. Yet, in what follows we will primarily focus on the following cost function:

\begin{equation}
\label{eq:cost}
C(\Delta) = \frac{\text{SD}(\Delta)}{\text{maxSD}(\Delta)}.
\end{equation}

Here, $\text{SD}(\Delta)$ denotes the standard deviation of the gap vector $\Delta$, and $\text{maxSD}(\Delta)$ denotes its maximum feasible value under the constraints $\Delta_k \geq 0$ and $\sum_k \Delta_k = 1$. Thus, this cost function is simply the ratio of the standard deviation of gaps between adjacent categories to the largest such achievable standard deviation.\footnote{In Appendix \ref{sec:sd_properties}, we first show the SD-based cost function of Equation \ref{eq:cost} indeed satisfies Proposition \ref{prop:plausibility}, and then that $\text{maxSD}(\Delta) = \sqrt{K-2}/(K-1)$.} 

\begin{figure}[!tb]
    \caption{Examples of scale transformations with different costs $C$.}
    \centering
    \label{fig:c_examples_11}
    \includegraphics[width=0.8\textwidth]{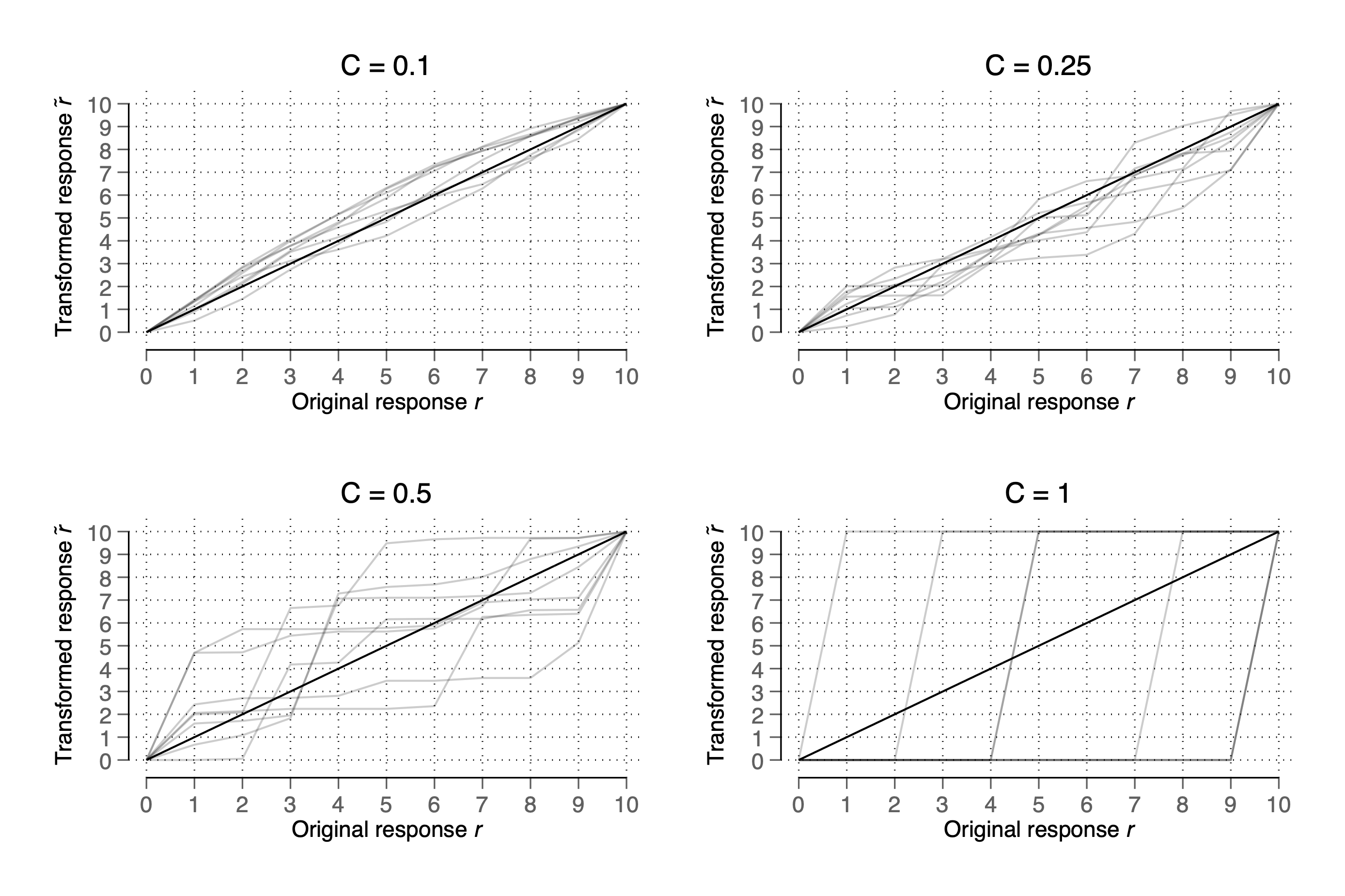}
    \begin{minipage}{0.975\textwidth}
    \footnotesize
    \noindent \textbf{Notes:} Each panel shows several randomly selected ways to transform an 11-point response scale, all sharing the same SD-based cost $C$ (displayed at the top). The horizontal axis represents the original scale $r$. The vertical axis shows the transformed scale $f(r) = \tilde{r}$. The straight 45-degree line represents linear scale use. As $C$ increases from 0 to 1, transformations increasingly depart from this linear benchmark. At $C = 1$, the scale collapses to a single jump. Both axes are displayed on the conventional $0$-$10$ range. $C$ depends only on the relative spacing of adjacent categories and is invariant to this choice.
    \end{minipage}
\end{figure}

We focus on this particular function for three reasons. First, this cost function conforms to our visual intuitions about what different degrees of non-linearity should look like. Figure \ref{fig:c_examples_11} displays several randomly selected transformations of an 11-point scale with several given cost levels. At $C = 0.1$, all transformations remain fairly close to the 45-degree line. In contrast, at $C = 0.5$, deviations from linearity are substantial, but not completely erratic. At $C = 1$, the scale collapses to a single jump. Second, as we show in Appendix \ref{sec:homogen}, this cost is \textit{linearly homogeneous}. Consider mixing any gap vector $\Delta$ with the uniform (linear) gap vector $\mathbf{u} = (1/(K{-}1), \ldots, 1/(K{-}1))$ using weight $\lambda$ to obtain $\Delta^{(\lambda)} = \lambda \Delta + (1-\lambda) \mathbf{u}$. Then, linear homogeneity implies that $C(\Delta^{(\lambda)}) = \lambda  C(\Delta)$. This gives the SD-based cost the intuitive interpretation that a transformation with (say) $C = 0.1$ is `10\% non-linear', in that its gap vector has the same cost as a hypothetical 10-to-90 mixture of maximally non-linear and uniform spacing. Third, using this cost-function often makes it feasible to arrive at a closed-form solution for the optimisation problems discussed in the next section.

%====================================
% FIGURE: Examples for C using 11 categories
%====================================
 
That said, in practice little hinges on our particular choice of cost function: All of our empirical analyses and conclusions are robust to alternatively using the variance-based function $C(\Delta)=\text{Var}(\Delta)/\text{maxVar}(\Delta)$ or the normalised Theil index $C(\Delta)=\frac{\sum_{k=1}^{K-1}\Delta_k\ln((K-1)\Delta_k)}{\ln(K-1)}$. %, with the convention $0\ln 0 = 0$

\subsection{Finding least non-linear reversals}
\label{sec:optimisation}
\vspace{-.5em}

Suppose a researcher has estimated a coefficient $\hat{\beta}_m^{(r)}$ using the standard linear coding of the response scale. They may now want to know whether (I) there exist some alternative codings under which the sign of this coefficient reverses, or (II) under which it loses or gains statistical significance, or (III) under which its ratio to another coefficient changes substantially. If such alternative codings exist, they will want to know how non-linear they need to be. The less non-linear the required transformations, the more fragile the original result.
 
We can view this as a constrained optimisation problem. Specifically, we search for the gap vector (i.e. the vector of differences in the transformed labels we assign to each response category)  $\Delta = (\Delta_1, \ldots, \Delta_{K-1})$ that minimises the cost $C(\Delta)$ subject to three constraints. The first two constraints are common across applications:
 
\vspace{8pt}
 
\noindent \textbf{Constraint 1} (Scale normalisation). Gaps sum to the total range of the scale: $\sum_{k=1}^{K-1} \Delta_k = 1$.
 
\vspace{8pt}

Recall that $C$ only depends on the relative spacing of the differences between labels, but we do need to some normalisation across the original and transformed coding.\footnote{Some papers study potential stretching or compression of the scale across respondents while maintaining the linearity assumption; see e.g. \cite{benjamin2023adjusting} or \cite{fabian2022scale}. Combining these approaches with ours may be a natural direction for future work.} 

\vspace{8pt}

\noindent \textbf{Constraint 2} (Monotonicity). All gaps are strictly positive: $\Delta_k > 0$ for all $k$.

\vspace{8pt}

This constraint could in principle be relaxed, for instance to accommodate focal-value rounding \citep{barrington2024econometrics}. However, currently our experimental evidence does not indicate violations of monotonicity (cf. Appendix \ref{sec:evidence_indiv_appendix}).

The third constraint depends on whether we are reversing coefficient signs, altering coefficient ratios, or changing statistical significance:
 
\vspace{8pt}
 
\noindent \textbf{Constraint 3a} (Sign reversal). The sign of the coefficient changes: $\text{sgn}(\hat{\beta}_m^{(\tilde{r})}) \neq \text{sgn}(\hat{\beta}_m^{(r)})$. Here, $\hat{\beta}_m^{(\tilde{r})}$ denotes the coefficient under the tranformed gap vector $\Delta$.
 
\vspace{8pt}
 
\noindent \textbf{Constraint 3b} (Target ratio). Coefficients $\hat{\beta}_m^{(\tilde{r})}$ and $\hat{\beta}_n^{(\tilde{r})}$ satisfy a target ratio $\rho = \hat{\beta}_m^{(\tilde{r})} / \hat{\beta}_n^{(\tilde{r})}$.
 
\vspace{8pt}
 
\noindent \textbf{Constraint 3c} (Target significance). The p-value associated with coefficient $\hat{\beta}_m^{(\tilde{r})}$ equals some target $\kappa = p(\hat{\beta}_m^{(\tilde{r})})$.
 
\vspace{8pt}
 
The general optimisation problem is then
\begin{equation}
\label{eq:optimization}
\Delta^* = \underset{\Delta}{\text{argmin}} \; C(\Delta), \quad \text{subject to Constraints 1, 2, and either 3a, 3b, or 3c.}
\end{equation}
The value $C^* = C(\Delta^*)$ here gives the minimal departure from linearity needed to overturn the corresponding feature of the original result.\footnote{More generally, for any cost threshold $\bar{C}$, the set of coefficients $\hat{\beta}_m^{(\tilde{r})}$ attainable by gap vectors with $C(\Delta) \leq \bar{C}$ defines an identified set for the coefficient under the maintained assumption that departures from linearity do not exceed $\bar{C}$. Our approach can thus be viewed as seeking numerical bounds for partial identification \citep{tamer2010partial, molinari2020microeconometrics}. The purely ordinal case corresponds to $\bar{C} = 1$.}

The next subsections discuss each version of the third constraint. Our Stata commands \texttt{coeff\_reverser} and \texttt{mrs\_reverser}, available on \href{https://github.com/casparwarwick/reversals}{\color{blue}{GitHub}}, implement these procedures.

\subsubsection{Sign reversals}
\label{sec:sign_reversal_method}
\vspace{-.5em}

Consider first sign reversals. As briefly mentioned earlier, \citet{kaiser2023how} provide a useful decomposition of $\hat{\beta}_m^{(\tilde{r})}$ in terms of coefficients from binary regressions. Specifically, define $d_{ki} \equiv \mathds{1}(r_i \leq k)$ and let $\hat{\beta}_{km}^{(d)}$ denote the coefficient on $X_{im}$ from regressing $d_{ki}$ on $\mathbf{X}_i$. Then, for any transformation with labels $\tilde{l}_1, \ldots, \tilde{l}_K$, we have:

\begin{equation}
\label{eq:decomp}
\hat{\beta}_m^{(\tilde{r})} = \sum_{k=1}^{K-1} (\tilde{l}_k - \tilde{l}_{k+1}) \hat{\beta}_{km}^{(d)}.
\end{equation}

Since $(\tilde l_k - \tilde l_{k+1})<0$ for all positive monotonic transformations, this immediately yields a non-reversibility condition: \textit{if all $\hat{\beta}_{km}^{(d)}$ share the same sign, no positive monotonic transformation can reverse the sign of $\hat{\beta}_m^{(\tilde{r})}$.} See Appendix \ref{sec:decomposition} for details. When the condition fails, the optimisation problem of Equation \eqref{eq:optimization} (with Constraint 3a) becomes feasible.

This decomposition also makes solving the problem computationally cheap, as the $K{-}1$ coefficients $\hat{\beta}_{km}^{(d)}$ need to be estimated only once, and each candidate transformation then requires only recomputing the weights $(\tilde{l}_k - \tilde{l}_{k+1})$ and forming the sum. For any cost function in the class characterised by Proposition \ref{prop:plausibility}, the resulting problem is convex and can hence be reliably found by standard numerical solvers.

For the SD-based cost function of Equation \eqref{eq:cost}, we can even obtain a simple analytical solution. Let $\mu_b \equiv \tfrac{1}{K-1}\sum_k \hat\beta_{km}^{(d)}$ and $V_b \equiv \tfrac{1}{K-1}\sum_k (\hat\beta_{km}^{(d)} - \mu_b)^2$ denote the mean and variance of the dichotomised coefficients. We can then state:

\begin{proposition}[Minimum-cost sign-reversing transformation]
\label{prop:sign_reversal_closed_form}
Suppose the sign-reversal of $\hat\beta_m^{(r)}$ is feasible, and suppose that the monotonicity constraint (Constraint 2) is non-binding at the optimum. Then the minimum cost of sign reversal under the SD-based cost function is $C^{\star} \;=\; \frac{\bigl|\hat\beta_m^{(r)}\bigr|}{(l_K - l_1)\sqrt{(K-2)\,V_b}}$, attained by the normalised gap vector $\Delta_k^{\star} = \frac{1}{K-1} + \frac{\hat\beta_m^{(r)}(\hat\beta_{km}^{(d)} - \mu_b)}{(K-1)(l_K - l_1)\,V_b}$.
\end{proposition}

The proof is in Appendix \ref{sec:proofs_sign}.\footnote{Analogous result hold for any cost function in the class of Proposition \ref{prop:plausibility} whose inner function is quadratic, i.e.\ $\psi(\Delta_k) = (\Delta_k - 1/(K-1))^2$. This includes the variance-based cost $\mathrm{Var}(\Delta)/\mathrm{maxVar}(\Delta)$. All such cost functions share the same optimal gap vector $\Delta_k^\star$ and only differ by the reported cost value.} Here $\hat\beta_m^{(r)}$ is the coefficient estimated under the original (linear) coding of the scale, and $l_K - l_1$ is the range of that coding. For the standard coding $l_k = k$ we have $l_K - l_1 = K-1$. The optimal gap vector $\Delta_k^{\star}$ is here reported in the normalised convention $\sum_{k=1}^{K-1}\Delta_k = 1$. To express it in the units of the original coding, multiply by $l_K - l_1$. The cost $C^{\star}$ is itself scale-invariant. In Supplementary Appendix \ref{sec:algorithm}, we give an analogous proposition (and associated algorithm) that covers the full case in which the monotonicity constraint is binding at the optimum. We omit that material here for brevity, and because the relevant intuitions are already well-captured by Proposition \ref{prop:sign_reversal_closed_form}. For example, the proposition shows that larger $|\hat\beta_m^{(r)}|$ makes reversal harder, while greater heterogeneity in $\hat\beta_{km}^{(d)}$ makes it easier. 

\subsubsection{Coefficient ratios}
\label{sec:ratio_method}
\vspace{-.5em}

When $\hat{\beta}_n^{(\tilde{r})}$ in the denominator is not reversible, the ratio $\hat{\beta}_m^{(\tilde{r})}/\hat{\beta}_n^{(\tilde{r})}$ is bounded by the minimum and maximum values of $\hat{\beta}_{km}^{(d)}/\hat{\beta}_{kn}^{(d)}$ across $k = 1, \ldots, K-1$.\footnote{\citet{kaiser2023how} give this result, but make an error in stating these bounds without imposing the non-reversibility of $\hat{\beta}_n^{(\tilde{r})}$. Clearly, when the denominator is itself reversible the ratio is unbounded. We give a corrected proof in our notation in Supplementary Appendix \ref{sec:proof_ratios}.} For any target ratio $\rho$ within these bounds, the minimum-cost transformation can be found by solving the problem of Equation \eqref{eq:optimization} subject to Constraint 3b. As in the sign-reversal case, the decomposition of Equation \eqref{eq:decomp} makes this computationally cheap. Likewise, the constraint $\hat\beta_m^{(\tilde r)} - \rho\hat\beta_n^{(\tilde r)} = 0$ is linear in the gap vector $\Delta$, and the resulting problem is convex for any cost function in the class of Proposition \ref{prop:plausibility}.
 
Finally, for the SD-based cost function, a simple analytical solution is again available. Specifically, defining $c_k \equiv \hat\beta_{km}^{(d)} - \rho\hat\beta_{kn}^{(d)}$ with mean $\mu_c$ and variance $V_c$, we have:

\begin{proposition}[Minimum-cost ratio-targeting transformation]
\label{prop:ratio_closed_form}
Suppose $\rho$ lies strictly within the bounds given by the minimum and maximum values of $\hat\beta_{km}^{(d)}/\hat\beta_{kn}^{(d)}$ across $k = 1, \ldots, K-1$, $\hat\beta_n^{(r)}$ is not reversible, and the monotonicity constraint is non-binding at the optimum. Then the minimum SD-based cost of attaining the target ratio $\rho$ is $C^{\star}_{\rho} = \frac{(|\hat\beta_m^{(r)} - \rho\hat\beta_n^{(r)}|)}{(l_K - l_1)\sqrt{(K-2)V_c}}$.
\end{proposition}
 
The proof is in Appendix \ref{sec:proofs_ratios_closed_form}. Supplementary Appendix \ref{sec:mrs_full_analytic} gives a generalisation for the case in which monotonicity binds at the optimum. 

\subsubsection{Statistical significance}
\label{sec:significance_method}
 \vspace{-.5em}
 
Significance reversals are less tractable because p-values depend on the variance-covariance matrix of $\hat{\boldsymbol{\beta}}^{(\tilde{r})}$. This matrix takes the standard form $(\mathbf{X}'\mathbf{X})^{-1} \mathbf{X}' \hat{\boldsymbol{\Omega}} \mathbf{X} (\mathbf{X}'\mathbf{X})^{-1}$, where $\hat{\boldsymbol{\Omega}}$ depends on the regression residuals $\tilde{\mathbf{e}}$. Since the residuals change with the transformation, $\hat{\boldsymbol{\Omega}}$ must in principle be recomputed for each candidate.
 
However, as we show in Appendix \ref{sec:residual_derivation}, the residuals under any transformation can be expressed as a weighted sum of residuals from the dichotomised regressions, yielding $\tilde{\mathbf{e}} = \sum_{k=1}^{K-1} (\tilde{l}_k - \tilde{l}_{k+1})\mathbf{e}_{dk}$. Therefore, $\hat{\boldsymbol{\Omega}}$, and hence the p-value associated with any coefficient under any transformation, can be constructed without re-estimating any underlying regressions.
 
But, unlike for sign reversals and ratios, we fail to obtain simple analytical bounds on the range of attainable p-values. Instead, we numerically maximise and minimise $p(\hat\beta_m^{(\tilde r)})$ subject to Constraints 1 and 2 to obtain bounds on the attainable p-values. Within these bounds, for any target $\kappa$, the minimum cost is then found by numerically solving Equation \eqref{eq:optimization} subject to Constraint 3c. Although this problem is not convex, Python's SciPy \texttt{minimize} routine tends to converge to the same solution from different initial label configurations. This suggests that we do tend to find the global optimum. 

%====================================
%====================================

\section{How are response options interpreted?}
\label{sec:scale_use_evidence}
\vspace{-.5em}

The previous section was based on the idea that more extreme departures from a linear interpretation of the response scale are increasingly unlikely. 

However, there is little direct evidence on the extent to which respondents interpret adjacent response categories as equally spaced. Existing work in e.g. psychology is more indirect and concerned with survey design more broadly.\footnote{See \cite{krosnick1997designing} for in-depth discussion of ordered scales in psychology.} Work from psychophysics on how people interpret numbers suggest that, for bounded intervals analogous to survey scales, subjective and objective values are roughly linear \citep{Banks1981, Banks1974, Schneider1974}. Earlier contributions in economics also point to near-linearity in scale use \citep{van1988household, van1991ordinal, van1999measurement}. More recently, \cite{kaiser2022scientific} show that reported satisfaction in domains such as jobs, housing, and health predicts subsequent quitting actions in a near-linear fashion. 

Unfortunately, however, none of these studies provide sufficient data to form a clear empirical benchmark for our cost $C$. 

\subsection{Eliciting scale use with interactive sliders}
\vspace{-.5em}

We used an interactive slider task to directly elicit how respondents interpret the spacing between response categories. This was implemented in a Prolific online survey.

\begin{figure}[!t]
    \caption{The interactive slider exercise - An example}
    \centering
    \label{fig:example_slider}
    \includegraphics[width=0.9\textwidth]{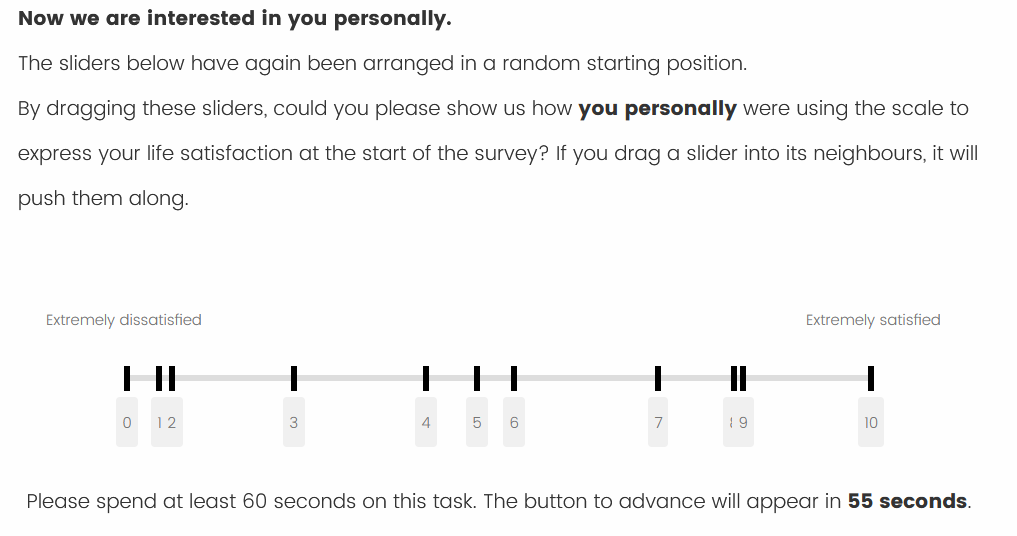}
    \begin{minipage}{0.975\textwidth}
\footnotesize
\noindent \textbf{Notes:} This figure displays a screenshot of the interactive slider exercise used to elicit respondents’ perceived distances between adjacent response categories. Respondents were asked to position sliders to indicate how large they perceived the gaps between successive values on a life satisfaction scale (0–10). The initial placement of all sliders is randomized across respondents to avoid anchoring or default effects. Participants were required to spend at least 60 seconds on the task before the validation button appeared.
\end{minipage}
\end{figure}

After answering a standard life satisfaction question on an 11-point scale (`\textit{Overall, how satisfied are you with your life nowadays?}'), respondents are shown a set of sliders corresponding to each response category (with the endpoints fixed). They are asked to position these sliders on a line so that the distance between adjacent categories reflects how they interpreted the scale when answering the question (see Figure \ref{fig:example_slider} for an example and this \href{https://wbs.qualtrics.com/jfe/form/SV_4T30LFjKnQnkfhI}{\color{blue}link} for an interactive demo).

The slider task produces, for each respondent $i$, perceived labels $(l_{1i},\ldots,l_{Ki})$ for the response categories. We use these labels to construct an individual-level cost of non-linearity, $C_i$, using the cost function defined in Section \ref{sec:variance}. Thus, $C_i=0$ corresponds to a respondent who interprets all adjacent response options as equally spaced, while larger values of $C_i$ indicate more non-linear individual scale use.

All sliders are initially placed at random positions. This is to avoid imposing any implicit default scale structure. Starting from a fixed or linear configuration could anchor responses and bias reported scale use. By contrast, random initialization should ensure that any alignment across respondents reflects their choices rather than the interface.\footnote{Random slider positions imply an average cost of about 0.3 in our SD-based cost.}

These random starting points additionally introduce  variation that is informative about adjustment behaviour. If moving sliders is effortful, respondents may only partially adjust them toward their intended positions. In that case, final responses will remain mechanically related to the random initial positions. We exploit this feature to correct for imperfect adjustment. Intuitively, we interpret respondents’ final slider positions as the outcome of a trade-off between a desire to report their true scale (which yields utility from accuracy) and a cost of effort required to move sliders. Assuming a quadratic cost of effort, this allows us to estimate the extent of incomplete adjustment. Using this estimate then yields an adjusted set of labels $(l^+_{1i},\ldots,l^+_{Ki})$, from which we compute a bias-corrected individual non-linearity cost $C_i^+$. See Supplementary Appendix \ref{sec:correction} for the formal model underlying this correction.

Finally, for respondents whose raw reported scale use is close to linear ($C_i<0.1$), we ask whether they intended to make the spacing between categories equal. For those who confirm this, we treat their responses as consistent with linear scale use by setting their $C_i=0$.

Our survey design includes three additional safeguards to ensure that the reported scale interpretations are meaningful. Respondents first complete a comprehension check in which they are asked to reproduce a pre-specified non-linear configuration (i.e., a larger gap between ``3'' and ``4'' than between ``7'' and ``8''). Second, respondents are required to spend at least 60 seconds on the task. Third, if a respondent attempts to confirm their answer without having moved any slider, a warning message is displayed asking them to confirm whether this was intentional. The full survey is available \href{https://wbs.qualtrics.com/jfe/form/SV_0P7kzJ8jZxOscaq}{\color{blue}here}.

\subsection{Main results on scale use}
\label{sec:exp_evidence}
\vspace{-.5em}

We implement this approach in an online survey collected in April 2026. Participants were recruited via Prolific and we sought for the samples to be nationally representative of the adult population of the UK ($N=549$) and the US ($N=578$). See Table \ref{tab:dataset_description} for further details on data collection and Table \ref{tab:descriptives_prolific26} for descriptive statistics. 
Results are displayed in Table \ref{tab:main_cost}, which reports average non-linear scale-use costs across samples and adjustment protocols. We use bootstrapping with 500 replications to obtain 95\% confidence intervals and focus on the SD-based cost function given in Equation \ref{eq:cost}.\footnote{We report results for alternative functions in Table \ref{tab:other_cost}. We will use these numbers in Section \ref{sec:evidence_sign_reversals_alt} and show that the extent of the risk of sign reversal does not hinge on the choice of a particular cost function.}

\begin{table}[!t]\centering
\caption{Average deviations from linearity with bootstrapped confidence intervals}
\label{tab:main_cost}
\resizebox{1.01\textwidth}{!}{
\begin{tabular}{lcccccc}
\toprule
& \multicolumn{6}{c}{Average cost measures} \\   \cmidrule(lr){2-7}
 & (1) & (2) & (3) & (4) & (5) & (6) \\
\midrule
\multicolumn{7}{l}{\textbf{Panel A. US and UK}} \\
SD-based &     0.184 &     0.173 &     0.165 &     0.149 &     0.198 &     0.174 \\
 & [0.173,     0.194] & [0.162,     0.183] & [0.155,     0.177] & [0.138,     0.161] & [0.186,     0.209] & [0.161,     0.187] \\
Observations &      1127 &      1035 &       883 &       883 &       883 &       883 \\
\midrule
\multicolumn{7}{l}{\textbf{Panel B. US}} \\
SD-based &     0.195 &     0.182 &     0.173 &     0.156 &     0.205 &     0.180 \\
 & [0.180,     0.210] & [0.166,     0.198] & [0.155,     0.191] & [0.137,     0.175] & [0.187,     0.224] & [0.160,     0.201] \\
Observations &       578 &       528 &       442 &       442 &       442 &       442 \\
\midrule
\multicolumn{7}{l}{\textbf{Panel C. UK}} \\
SD-based &     0.172 &     0.163 &     0.158 &     0.142 &     0.190 &     0.167 \\
 & [0.158,     0.186] & [0.149,     0.177] & [0.143,     0.172] & [0.126,     0.158] & [0.174,     0.206] & [0.149,     0.185] \\
Observations &       549 &       507 &       441 &       441 &       441 &       441 \\
\midrule
\multicolumn{7}{l}{\textbf{Specification:}} \\
Passed attention check & . & \checkmark & \checkmark & \checkmark & \checkmark  & \checkmark \\
Passed comprehension check & . & . & \checkmark & \checkmark &  \checkmark & \checkmark \\
Linear adjustment & . & . & . & \checkmark & . & \checkmark \\
Effort adjustment & . & . & . & . & \checkmark & \checkmark \\
\bottomrule
\end{tabular}

}

\begin{minipage}{1\textwidth}
\footnotesize
\noindent \textbf{Notes:} This table reports estimates of average SD-based costs $C$ with bootstrapped confidence intervals (500 replications). These non-linearity costs are bounded between 0 (linearity) and 1 (maximal non-linearity). The \textit{linear adjustment} assigns a cost of zero to respondents who reported wanting to make spacing equal between sliders. The \textit{effort adjustment} corrects for inattention or low effort. Across specifications, the SD-based cost lies in a relatively narrow range of around 0.15–0.20.
\end{minipage}
\end{table}

 In the full sample (Panel A), average costs range from 0.149 to 0.198. Our preferred specification — restricting to respondents who pass the attention and comprehension checks and applying the linear and effort adjustment protocols — yields an estimate of $C=0.174$ (95\% CI: $0.161$–$0.187$). Estimates are slightly higher in the US than in the UK (Panel B vs. Panel C), although these differences are not statistically significant at conventional levels.

Furthermore, average costs are remarkably stable across sample restrictions and adjustment procedures, suggesting that neither selection nor correction materially affects the magnitude of non-linearity. Finally, all costs are statistically different from zero at conventional levels, allowing us to reject the hypothesis of linear scale use.

\subsection{Complementary surveys and approaches}
\label{sec:complement}
\vspace{-.5em}

Our main approach has the advantage of providing a direct measure of respondents’ life satisfaction scale use. However, it is a cognitively demanding task and, more generally, it is unclear to what extent the results are driven by the specifics of our design. We therefore compare our findings to three alternative procedures that rely on different protocols and sources of identifying variation, and hence rest on different assumptions. 

First, we use a `linear prompting' design. Here, respondents answer a standard life satisfaction question, with a random subset explicitly instructed to treat the response scale as linear. Under the assumptions that respondents comply with the prompt and that the distribution of underlying satisfaction is comparable across groups -- which follows from random assignment -- we can identify non-linear scale use in the absence of such prompting.\footnote{Specifically, for each discrete value $k \in \{1, \ldots, K\}$ in the unprompted group we find the value $r_k^{*}$ in the prompted group's continuous response distribution satisfying $F_{un}(k) = F_{lin}(r_k^{*})$, where $F_{un}$ and $F_{lin}$ denote the respective CDFs. If unprompted respondents were already using the scale linearly, the relationship between $k$ and $r_k^{*}$ would itself be linear, and the differences $r_{k+1}^{*} - r_k^{*}$ form a gap vector from which we can then compute $C$.}

Second, we use the data and results of \cite{jump}, who implement a `jump' elicitation in which respondents directly report the perceived gap between each pair of adjacent categories. After being shown the full 0-10 scale, respondents report how large they perceive each gap (between 0 and 1, between 1 and 2, and so on) on a continuous scale. This yields, for each respondent $i$, a vector of adjacent-category distances from which we can compute $C_i$. As with our slider task, identification relies on respondents' ability to consistently report relative differences across categories. See \cite{jump} for further details.

Last, we build on the approach of \cite{oswald2008curvature}, who uses objective measures to recover scale use. Here, respondents subjectively rate quantities (i.e. height and weight) on a Likert scale (e.g. `How tall are you?' on a 0-10 scale) and subsequently report their actual values. Within each response category $k$, we then compute mean actual height (or weight) $\bar{h}_k$. The differences $\bar{h}_{k+1} - \bar{h}_k$ give a gap vector from which we compute $C$. Here, identification requires that scale use reflects a stable individual trait rather than being domain-specific. This assumption is supported by recent evidence  \citep{benjamin2023adjusting}.

\begin{table}[!t]
\def\sym#1{\ifmmode^{#1}\else\(^{#1}\)\fi}
\caption{Alternative approaches to elicit average deviations from linear scale use}
\label{tab:complementary_cost}
\centering
    \begin{tabular}{l*{7}{c}}
    
\toprule
\textit{Linear prompt approach} & SD-based cost  \\
\quad Sept 24 survey (N=584) & 0.119 [0.084-0.153]  \\
\textit{Objective-subjective approach}   \\
\quad Height, Apr 26 survey (N=976) & 0.140  [0.103-0.188] \\
\quad Height, Sept 24 survey (N=1,147)  & 0.137 [0.101-0.179] \\
\quad Weight, Apr 26 survey (N=976) & 0.235 [0.180-0.294] \\
\quad Weight, Sept 24 survey (N=1,147) & 0.171  [0.115-0.233]  \\
\textit{Jump approach} \\
\quad Correa-Mackliff (N=508) &  0.124  [0.116-0.132]  \\
\bottomrule
\end{tabular}

\begin{minipage}{.94\textwidth}
\footnotesize
\noindent \textbf{Notes:} This table reports on alternative ways of estimating the average SD-based cost $C$ of non-linearity (bootstrapped CIs in brackets). Results for alternative cost functions are shown in Table \ref{tab:complementary_cost2}.
\end{minipage}

\end{table}

The linear prompting design and the objective–subjective questions are fielded in a survey we conducted in September 2024 on a representative UK sample via Prolific, with the latter also implemented in our April 2026 survey. The jump approach is implemented in an independent UK survey collected by \cite{jump}. For each approach, we compute the average non-linearity cost $C$ and report bootstrapped 95\% confidence intervals. The results are presented in Table \ref{tab:complementary_cost} and descriptive statistics are reported in Table \ref{tab:descriptives_prolific}.

Despite relying on different identifying assumptions, all methods yield estimates of similar magnitude, and in almost all cases smaller than those obtained with the interactive slider approach. The point estimates typically lie between 0.12 and 0.17. Results are also stable across survey implementations. For the objective–subjective approach, estimates based on height are nearly identical across the September 2024 and April 2026 surveys, while those based on weight are somewhat larger. All confidence intervals exclude zero, providing consistent evidence against linear scale use across identification strategies. 

\subsection{How does scale use vary between respondents?}
\label{sec:individual_heterogeneity}
\vspace{-.5em}

While our analysis so far has focused on average departures from linear scale use, respondents need not interpret response scales in a homogeneous way. Examining heterogeneity is therefore informative for assessing whether a single summary measure is a reasonable approximation. In particular, if scale-use costs varied systematically with observable characteristics, focusing on an average $C$ could mask important differences across groups. Conversely, if heterogeneity is limited or largely idiosyncratic, the average cost provides a meaningful and parsimonious summary of scale use in the population. Because individual-level measures are subject to noise, we restrict attention to respondents who pass both the attention and comprehension checks, and we report both unadjusted and effort-adjusted measures of $C_i$.

\begin{figure}[!t]
    \caption{CDF and determinants of individual non-linear scale use costs}
    \centering
    \label{fig:cdf_c}
    \subcaptionbox*{Panel (A)}{\includegraphics[width=0.495\textwidth]{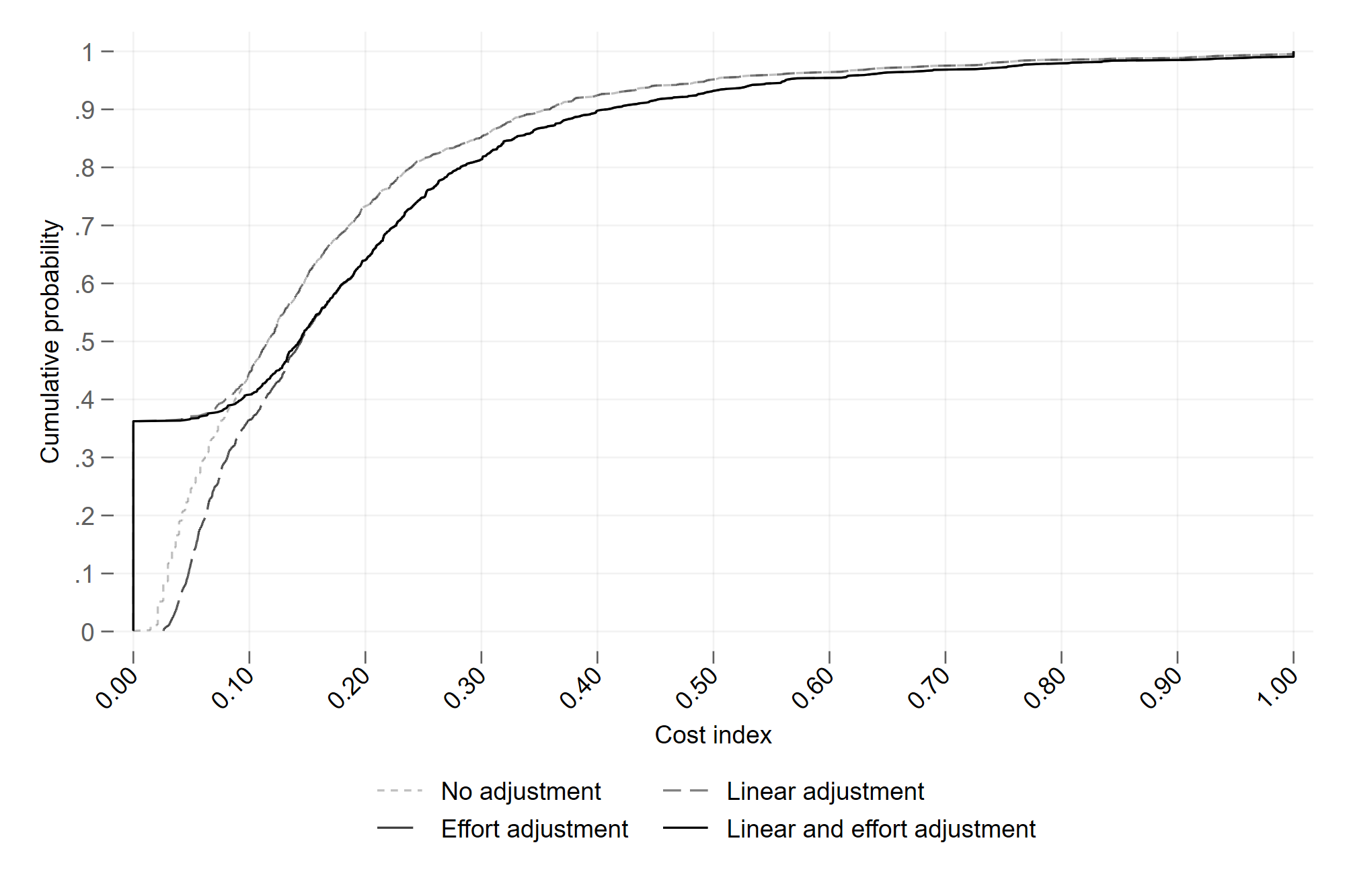}}
    \subcaptionbox*{Panel (B)}{\includegraphics[width=0.495\textwidth]{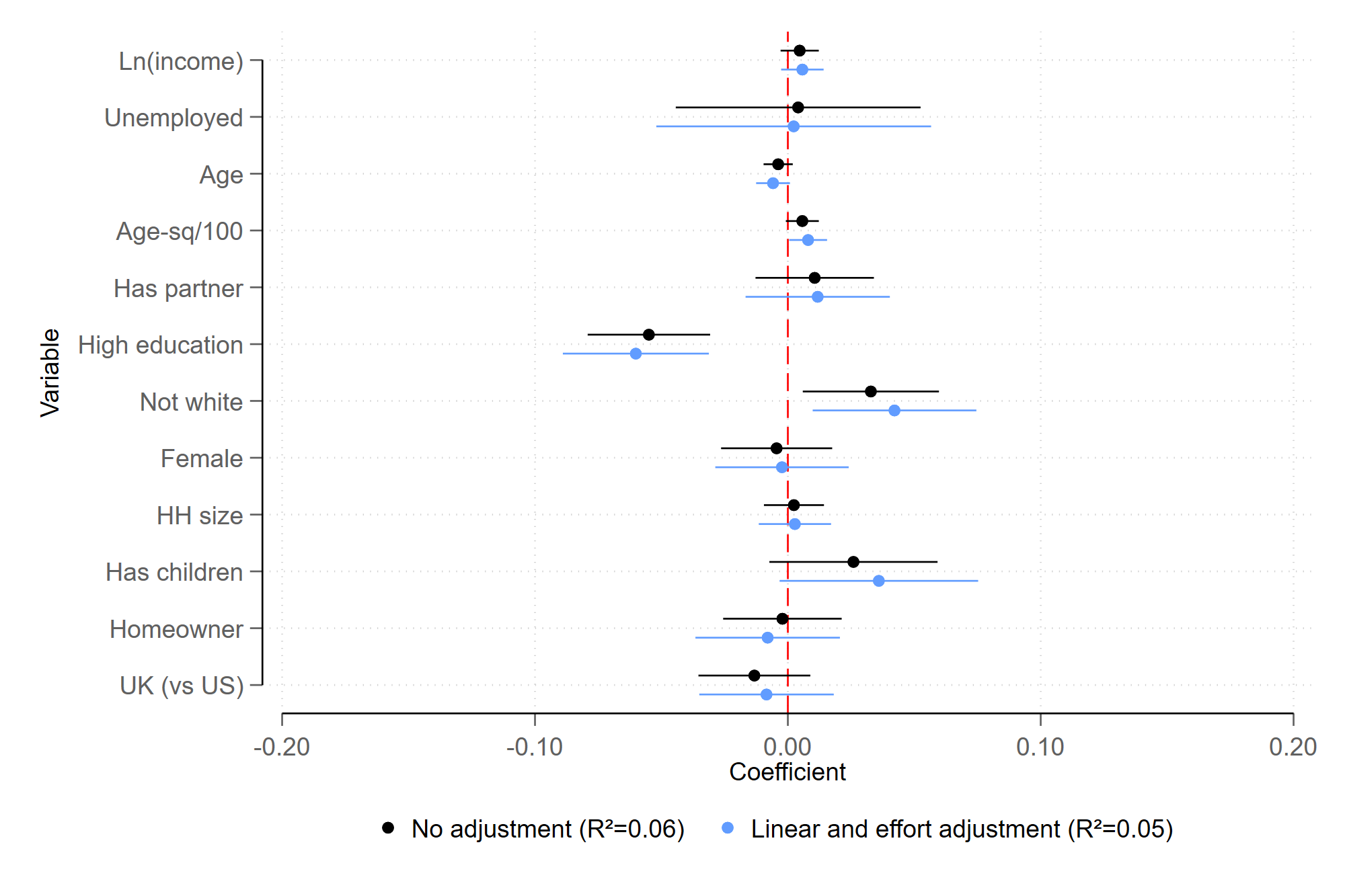}} 
    \begin{minipage}{0.975\textwidth}
    \footnotesize
\noindent \textbf{Notes:} Panel (A) plots the cumulative distribution function of the individual SD-based cost $C$ under different adjustment procedures. The horizontal axis reports the cost index, where $C=0$ corresponds to perfectly linear scale use and higher values indicate greater non-linearity; the vertical axis reports the cumulative share of respondents. The distribution is fairly concentrated at low values of $C$, indicating that most respondents exhibit either linear or only mildly non-linear scale use. Panel (B) reports coefficient estimates from regressions of individual cost measures on respondent characteristics. For brevity, results are shown for the unadjusted and fully adjusted measures of $C$, although intermediate adjustments yield very similar estimates. Dots represent point estimates and horizontal lines denote 95\% confidence intervals. The vertical axis in Panel (B) is restricted to the interval $[-0.2,0.2]$, which corresponds approximately to one standard deviation of the individual cost distribution. Positive coefficients indicate greater non-linearity in scale use. The small and insignificant coefficients, together with the very low explanatory power of these regressions, suggest that heterogeneity in scale use is largely idiosyncratic. \end{minipage}
\end{figure}

Figure \ref{fig:cdf_c}, Panel (A), plots the cumulative distribution functions of $C_i$ under different adjustments. First, there is a large mass at zero after accounting for the linear adjustment: around 36\% of respondents declared wanting to report perfectly linear scale use. Second, among those exhibiting non-linearity, deviations are typically small. Roughly another third of respondents have $C_i<0.2$, and the distribution rises steeply at low values. Large deviations are rare. The right tail is thin, with only a small minority of respondents displaying $C_i>0.5$. Adjusting for either linear rescaling or effort shifts the distribution slightly leftward but leaves its overall shape unchanged. The same pattern holds in both the UK and US samples.

To assess whether these differences in scale use reflect systematic variation across observable characteristics, we regress the individual cost measures on a standard set of socio-demographic variables and report the coefficients in Panel (B). For brevity, we present results for the unadjusted and fully adjusted costs (i.e. $C_i$ and $C_i^+$), but estimates are very similar when using intermediate adjustment procedures. Across specifications, the results are remarkably stable. Higher education and identifying as white are associated with slightly smaller deviations from linearity, with the education gradient in line with qualitative evidence reported by \cite{fabiancognitive}.\footnote{A Cragg hurdle specification, separating the probability of $C_i^+=0$ from the conditional  level of non-linearity, shows that ethnicity operates mainly at the extensive margin, while education affects both margins.} Beyond these two variables, coefficients are small and statistically insignificant. Most importantly, the explanatory power of these regressions is extremely limited, with $R^2$ below 0.06 across all specifications.\footnote{Appendix \ref{sec:evidence_indiv_appendix} further shows that monotonicity is unlikely to be violated, and that differences in spacing between adjacent response categories, like overall non-linearity, are idiosyncratic across individuals.} 

Hence, while heterogeneity in scale use is present, it is concentrated within a fairly narrow range. Most respondents either use the scale linearly or deviate only mildly. Extreme non-linear interpretations are rare and individual differences in scale use are largely idiosyncratic rather than systematically related to socio-demographic traits.

\section{Systematic evidence from \textit{WellBase}}
\label{sec:wellbase_evidence}
\vspace{-.5em}

Having introduced a measure of departures from linear scale use, an optimisation framework for identifying the minimal non-linearity required to overturn results, and experimental benchmarks for the range of plausible departures, we now bring this framework to the published evidence.

This section provides the first systematic assessment of the robustness of the empirical economics of subjective wellbeing. Subjective wellbeing is becoming increasingly central to policy. Among constructs measured using ordered response scales, it also is the main focus of methodological critiques.  

We therefore constructed \textit{WellbBase}, a systematic database of wellbeing research published in top economics journals. \textit{WellBase} includes 73 papers, 1,610 regressions, 28,522 coefficients, and all the underlying data needed to reproduce these. 

We use these replications to quantify three kinds of risks that can arise when analysts assume the response scale to be linear: (I) the risk that a coefficient’s \emph{sign} changes after a positive monotonic transformation of the scale, (II) the risk that its \emph{statistical significance} changes, and (III) the extent to which such transformations can alter the \emph{relative magnitudes} of point estimates. Because the same Likert-style measurement issues may affect other constructs in economics, we also benchmark wellbeing against scales for, among others, risk, trust and political preferences. There, we reproduce 23,104 coefficients across 16 papers. 

\vspace{-1em}
\subsection{Data}
\vspace{-.5em}

Our goal was to reproduce the universe of empirical research on subjective wellbeing published in top economics journals. We had three inclusion criteria. First, we only included articles published in economics journals ranked among the Top 30 on RePEc (as of June 2022), which typically enforce data and code sharing, making reproduction more feasible. Second, we only included papers published between January 2010 and May 2025. Third, we focus on papers that use a cognitive measure of subjective wellbeing as dependent variable in an individual-level analysis. Our search, conducted via Google Scholar, was based on the following keywords: ``Life Satisfaction'', ``Cantril Ladder'', ``Subjective well-being'', and ``Subjective wellbeing''. The first two capture the most common cognitive wellbeing scales, while we added the latter two to capture any papers using less frequent cognitive wellbeing measures. See Figure~\ref{fig:prisma} for a summary of our selection process.

In total, 97 articles were eligible for inclusion. Because of missing replication files or protected data, we reproduced 73 of these articles. Among these, we successfully reproduced all of the 1,610 relevant regressions in both the main manuscripts and any associated appendices (printed or online). Less than 2\% of these regressions (spread across six articles) were not using a linear estimator, but were using an ordered probit approach instead. Additionally, 3\% of regressions (across two papers) were estimated using probit-adjusted OLS. To make these regressions comparable and to apply the methods of Section \ref{sec:background}, we reproduced these regressions using OLS. In all such cases, the results, in terms of sign and statistical significance, remained the same. About 2\% of the regressions (spread across three articles) used a binary dummy for high wellbeing as dependent variable. We re-estimated these regressions using the underlying full versions of the wellbeing measure.

\begin{figure}[!t]
    \centering
    \caption{Selection Process for \textit{WellBase}.}
    \label{fig:prisma}
    \begin{tikzpicture}[>=latex, font={\sf \small}, scale=0.95, transform shape]
\tikzstyle{bluerect} = [rectangle, rounded corners, minimum width=1.2cm, minimum height=0.6cm, text centered, draw=black, fill=cyan!60!gray!45!white, font={\sffamily\small}]
\tikzstyle{textrect} = [rectangle, minimum width=3.8cm, text width=3.6cm, minimum height=1cm, draw=black, font={\sffamily \scriptsize}]

% Identification column
\node (r1blue) at (0, 3.2cm) [draw, bluerect, minimum height=0.8cm]{Identification};
\node (r1left) at (0, 1.5cm) [draw, textrect, minimum height=1.3cm]
  {Records identified through database search: $n=473$};
\node (r1right) at (0, -1.2cm) [draw, textrect, minimum height=1.8cm]
  {Records removed \textit{before screening}: \\ Retracted ($n=1$)};

% Screening column  
\node (r2blue) at (6.45cm, 3.2cm) [draw, bluerect, minimum height=0.8cm]{Screening};
\node (r2left) at (4.3cm, 1.5cm) [draw, textrect, minimum height=1.3cm]
  {Records screened: $n=472$};
\node (r2right) at (4.3cm, -2cm) [draw, textrect, minimum height=3cm]
  {Records excluded: 
    \begin{itemize}
    \setlength{\itemsep}{0pt}
    \item No SWB scale in empirical analysis ($n=351$)
    \item SWB not the dependent variable ($n=20$)
    \item Empirical analysis not at the individual level ($n=4$)
    \end{itemize}     
  };

% Assessment column
\node (r4left) at (8.6cm, 1.5cm) [draw, textrect, minimum height=1.3cm]
  {Records assessed for reproduction: $n=97$ 
};
\node (r4right) at (8.6cm, -1.2cm) [draw, textrect, minimum height=1.3cm]
  {Records excluded:
    \begin{itemize}
    \item Missing replication package or protected data ($n=24$)
    \end{itemize}     
  };

% Included column
\node (r5blue) at (12.9cm, 3.2cm) [draw, bluerect, minimum height=0.8cm]{Included};
\node (r5left) at (12.9cm, 1.5cm) [draw, textrect, minimum height=1.3cm]
  {Papers reproduced and included in \textit{WellBase}: $n=73$};

% Draw arrows between nodes:
\draw[thick, ->] (r1left.south) -- (r1right.north);
\draw[thick, ->] (r1left.east) -- (r2left.west);
\draw[thick, ->] (r2left.south) -- (r2right.north);
\draw[thick, ->] (r2left.east) -- (r4left.west);
\draw[thick, ->] (r4left.south) -- (r4right.north);
\draw[thick, ->] (r4left.east) -- (r5left.west);
\end{tikzpicture}
    \begin{minipage}{0.95\textwidth}
        \footnotesize
        \vspace{10pt}
        \noindent \textbf{Note:} PRISMA \citep{prisma2021} flowchart summarising our selection process to produce \textit{WellBase}.
    \end{minipage}
\end{figure}

Our replication effort yielded two categories of estimates: (1) published coefficients that form the core of each paper's analysis and (2) unpublished coefficients that typically serve as control variables mentioned only in table/figure notes.  In total, we replicated 5,322 published estimates and 23,200 unpublished estimates. Table \ref{tab:wellbase_description} provides a complete list of all reproduced articles, along with details such as the type of wellbeing scale used.

\begin{table}[!t]
\caption{Descriptive Statistics of \textit{WellBase} at the estimate level.}
\label{tab:descriptives}
\centering
\small  
\renewcommand{\arraystretch}{0.95}  

\begin{tabular}{l*{1}{cccc}}
\toprule
                    &        Mean&          SD&         Min&         Max\\
\midrule
\textbf{About the wellbeing scales:} \\
\quad \textit{Number of response categories:} \\
\quad \quad 3-point scale&        0.01\%&        &        0&        1\\
\quad \quad 4-point scale&        26.88\%&        &        0&        1\\
\quad \quad 5-point scale&        14.17\%&        &        0&        1\\
\quad \quad 6-point scale&        0.41\%&        &        0&        1\\
\quad \quad 7-point scale&        4.00\%&        &        0&        1\\
\quad \quad 10-point scale&        22.17\%&        &        0&        1\\
\quad \quad 11-point scale&        35.77\%&        &        0&        1\\
\quad \quad More than 11-point scale&        0.12\%&        &        0&        1\\
\quad \textit{Type of question:} \\
\quad \quad Life Satisfaction&        77.43\%&        &        0&        1\\
\quad \quad Cantril Ladder&        4.51\%&        &        0&        1\\
\quad \quad Happiness Question&        18.05\%&        &        0&        1\\
\textbf{About the estimation samples:} \\
%\quad Number of observations&   158,089.14&   368,570.79&       59&    2,471,360\\
\quad Number of observations (logged)&        9.98&        2.17&        4.08&       14.72\\
%\quad Only one nationality included&        0.66&        &        0&        1\\
\textbf{About the econometric models:} \\
\quad Number of controls&       34.07&       30.02&        1&      191\\
\quad Individual FE&        14.21\%&        &        0&        1\\
\textbf{About the independent variables:} \\
\quad Printed in manuscript&        6.41\%&     &        0&        1\\
\quad Printed in appendix&        12.25\%&       &        0&        1\\
\quad Not printed&        81.30\%&        &        0&        1\\
% \quad Significant at the ten percent&        0.05&        &        0&        1\\
% \quad Significant at the five percent&        0.08&        &       0&        1\\
% \quad Significant at the one percent&        0.34&       &        0&        1\\
\quad Continuous variable&        24.72\%&        &        0&        1\\
\quad Time-varying variable&        75.13\%&        &        0&        1\\
\quad Two-stage least square&        0.60\%&        &        0&        1\\
\quad Individual-specific&        91.35\%&        &        0&        1\\
\quad Natural experiment, RCT, or policy reform&        3.93\% &        &        0&        1\\
\quad Macroeconomic indicator&        3.60\%&        &        0&        1\\
%\quad Absolute t-statistics&        5.11&       13.76&        0.00&      474.74\\
\quad Absolute t-statistics (logged)&        0.45&        1.53&       -9.08&        6.16\\
\midrule
\multicolumn{5}{l}{\textbf{Total number of estimates/regressions/papers:} \quad \quad \quad \quad 28,522/1,610/73} \\
\bottomrule
\end{tabular}

\vspace{5pt}

\begin{minipage}{0.95\textwidth}
\footnotesize
\noindent \textbf{Note:} These numbers refer to the sample of 28,522 estimates included in \textit{WellBase}.
\end{minipage}

\end{table}

Table \ref{tab:descriptives} provides an overview of the characteristics of the replicated estimates. About 6\% of these estimates can be found directly in the published manuscripts. An additional 12\% are reported in the appendices. The majority, constituting 81\%, are coefficients on unprinted control variables not shown in the printed articles.\footnote{Most of these unprinted control variables are standard sociodemographic characteristics that researchers include in regressions, such as age, gender, race, religion, marital status, family size, employment status, job characteristics, income, health, and childhood characteristics.} About 4\% relate to quasi-natural experiments (e.g., centralisation reforms in Switzerland, the London Olympics, or RCTs), while another 4\% are macroeconomic factors (e.g., economic growth, inflation rates). Approximately 25\% of coefficients relate to time-invariant characteristics (e.g., sex). Likewise, 25\% of estimates are based on a continuous covariate (e.g., income, age). 

Table \ref{tab:of_int_detailed} focuses on the 27 papers in \textit{WellBase} for which at least half of the printed regressions use a wellbeing scale as dependent variable. In these studies, the main objective is to uncover the drivers of subjective wellbeing.\footnote{In the remaining papers in \textit{WellBase}, wellbeing is not the \textit{primary} outcome of interest.} For each of these, Table \ref{tab:of_int_detailed} summarizes the hypotheses tested, and records the sign and significance of the main coefficients. A large number of these studies find that economic resources (e.g. household income or labour earnings) are associated with higher levels of reported wellbeing. Reported wellbeing also systematically declines following major adverse life events, including physical violence \citep{johnston2018victimisation}, exposure to the Chernobyl disaster \citep{danzer2016long}, or falling into poverty \citep{clark2016adaptation}. Several papers examine policy or environmental changes, such as centralisation reforms in Switzerland \citep{fleche2021welfare}, income transparency reforms in Norway \citep{perez2020effects}, or the London Olympic Games \citep{dolan2019quantifying}.

\subsection{Results on wellbeing scales}
\vspace{-.5em}

This section presents a series of results systematically assessing how sensitive the \textit{WellBase} estimates are to the the assumption that scale use is linear. We do so with the help of the SD-based cost function $C$ defined in Section \ref{sec:variance}. Recall that when $C=0$, this corresponds to the near-universally adopted assumption that scale use is linear in underlying wellbeing. As $C$ increases, the transformed scale increasingly departs from this assumption. 

Experimental evidence on scale use presented in Section \ref{sec:scale_use_evidence} provides useful guidance on the range of $C$ that is plausibly relevant in practice. Across US and UK samples of our main survey, average cost values derived from the interactive slider approach lie around $0.15$--$0.20$, and virtually no respondents exhibit cost values above $0.50$. We use these figures as reference points when interpreting the sensitivity results that follow.

\vspace{-1em}
\subsubsection{On sign reversals: Documenting the risk of reversal}
\label{sec:evidence_sign_reversals}
\vspace{-.5em}

\begin{figure}[t]
    \caption{Cumulative sign-reversal percentages for different values of $C$ in \textit{WellBase}.}
    \centering
    \label{fig:sign_rev}
    \subcaptionbox*{Panel (A)}{\includegraphics[width=0.495\textwidth]{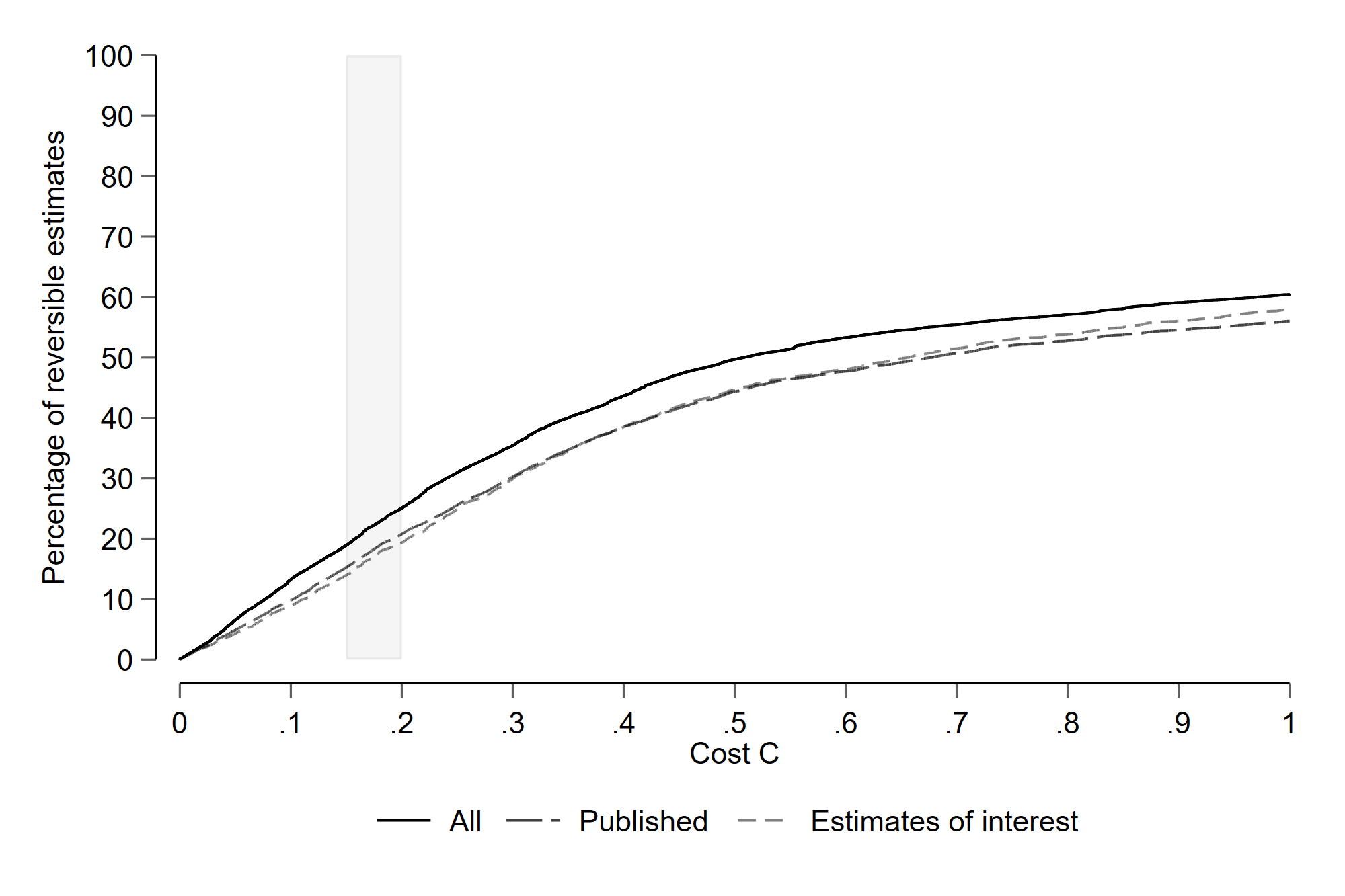}}
    \subcaptionbox*{Panel (B)}{\includegraphics[width=0.495\textwidth]{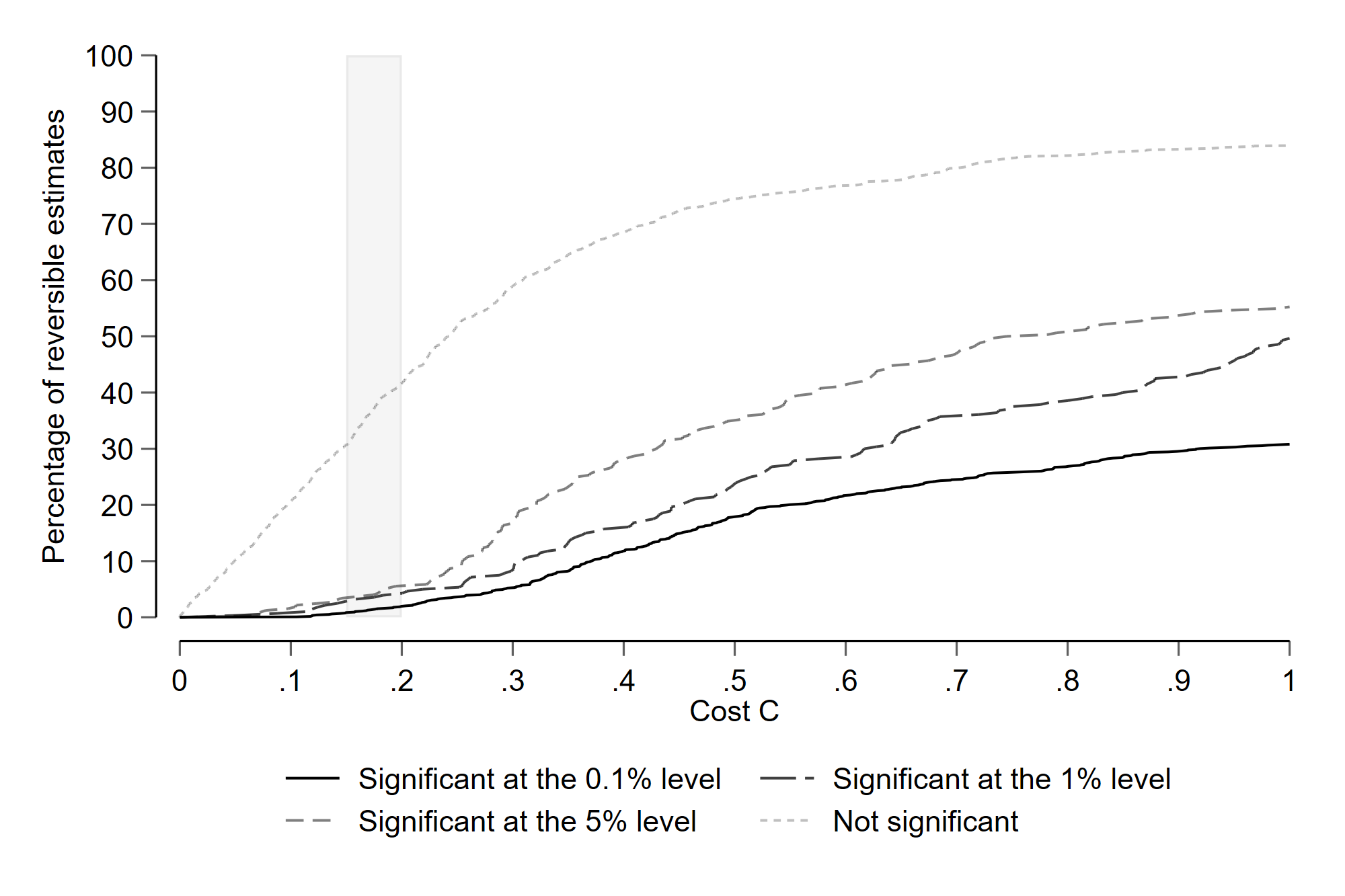}} 
    \begin{minipage}{0.975\textwidth}
    \footnotesize
    \noindent \textbf{Notes:} Most coefficient signs in published wellbeing research are robust to plausible departures from linear scale use. The figure shows the cumulative shares of coefficients included in \textit{WellBase} for which their sign can be reversed by some positive monotonic transformation of the response scale with at most cost $C$. The case of $C=0$ corresponds to assuming that scale-use is linear. When $C$ may take on any value on the unit interval, shown on the far right of the graphs, any monotonic transformation of the original scale is permissible and the assumption of cardinality is replaced by a purely ordinal interpretation. Based on the scale-use evidence presented in Section \ref{sec:scale_use_evidence}, the shaded region indicates the average degrees of non-linearity. Panel (A) shows that at most 60\% of all replicated estimates can be sign-reversed by some positive monotonic transformation of the response scale. This risk drops to 25\% when restricting attention to transformations corresponding to plausible degrees of non-linearity. Panel (B) shows that it is harder to reverse the sign of coefficients that are originally significant at the 5\% level or below.
    \end{minipage}
\end{figure}

Figure \ref{fig:sign_rev} shows the share of point estimates whose sign can be reversed by applying some positive monotonic transformation of the response scale with a cost of at most $C$. 

We report three lines in Panel (A). The solid dark line shows the share of sign reversals among all point estimates in \textit{WellBase}. The remaining two lines present the same statistic for \textit{printed estimates} and for \textit{estimates of interest}. Here, an \textit{estimate of interest} refers to estimates explicitly discussed in the text of the manuscript, and on which the conclusions of the included papers are based. The lines in Panel (A) all exhibit a concave relationship  between the cost $C$ and the percentage of sign reversals. About 60\% of all replicated estimates can be sign-reversed via at least one positive monotonic transformation of the wellbeing scale when allowing for any cost $C$. However, focusing on transformations falling in average degrees of non-linearity only (i.e. $0.15<C<0.20$), the risk of sign reversals drops to 25\% of all point estimates in \textit{WellBase}. \emph{Printed estimates} and \emph{estimates of interest} specifically exhibit even lower risks of sign reversal.

Panel (B) focuses only on \emph{estimates of interest} and displays a further breakdown by estimates' original level of statistical significance. There is a clear gradient between the original level of significance and the possibility of sign reversal: the less significant an estimate at $C=0$, the greater the chance that there is at least one transformation changing its sign. Sign reversals are virtually non-existent under any ``plausible'' transformation among coefficients that meet a 5\% significance threshold. 

Table \ref{tab:of_int_detailed} restricts attention to studies whose main objective is to study the determinants of subjective wellbeing. Its last two columns indicate whether a reversal is possible and, if so, what cost $C$ is required to produce such a reversal. About half of the coefficients reported in Table \ref{tab:of_int_detailed} are reversible. However, the risk is again much lower among the statistically significant coefficients (at the 5\% level): about 33\% of these can be sign-reversed, and in 95\% of cases doing so would require a cost $C>0.20$.

Overall, these results indicate that although sign reversals are often possible in principle, reversals under \textit{plausible} (i.e. $C<0.20$) transformations are not. This is especially true for results that were highly statistically significant in their original form.

\vspace{-1em}
\subsubsection{On sign reversals: Risk of reversal using alternative cost functions}
\label{sec:evidence_sign_reversals_alt}
\vspace{-.5em}

The results presented above rely on the SD-based cost function. As implied by Proposition \ref{prop:plausibility}, however, there are many reasonable ways to quantify departures from linear scale use. In this section, we therefore reassess the risk of sign reversals using alternative members of the class characterised in Proposition \ref{prop:plausibility}, namely a variance- and a Theil-based cost function. 

Using our primary data from the US and the UK, we estimated average non-linear scale use under these alternative cost functions as we did in Section \ref{sec:scale_use_evidence} with the SD-based cost. We report the results in Table \ref{tab:other_cost}. Consistent with the SD-based measure, we reject linearity in scale use across all specifications. For reference, the corresponding plausibility region under the SD-based cost is $C \in [0.15,0.20]$. The plausibility regions implied by the alternative measures differ somewhat: they are lower for the variance-based cost (approximately $C \in [0.06, 0.08]$) and for the Theil-based index (approximately $C \in [0.08, 0.11]$).\footnote{The variance-based cost is just the SD-based cost squared, so $C_{Var}<C_{SD}$ whenever $C_{SD}<1$.} 

\begin{figure}[t]
    \caption{Cumulative sign-reversal percentages for different cost functions in \textit{WellBase}.}
    \centering
    \label{fig:sign_rev_alt}
    \includegraphics[width=0.75\textwidth]{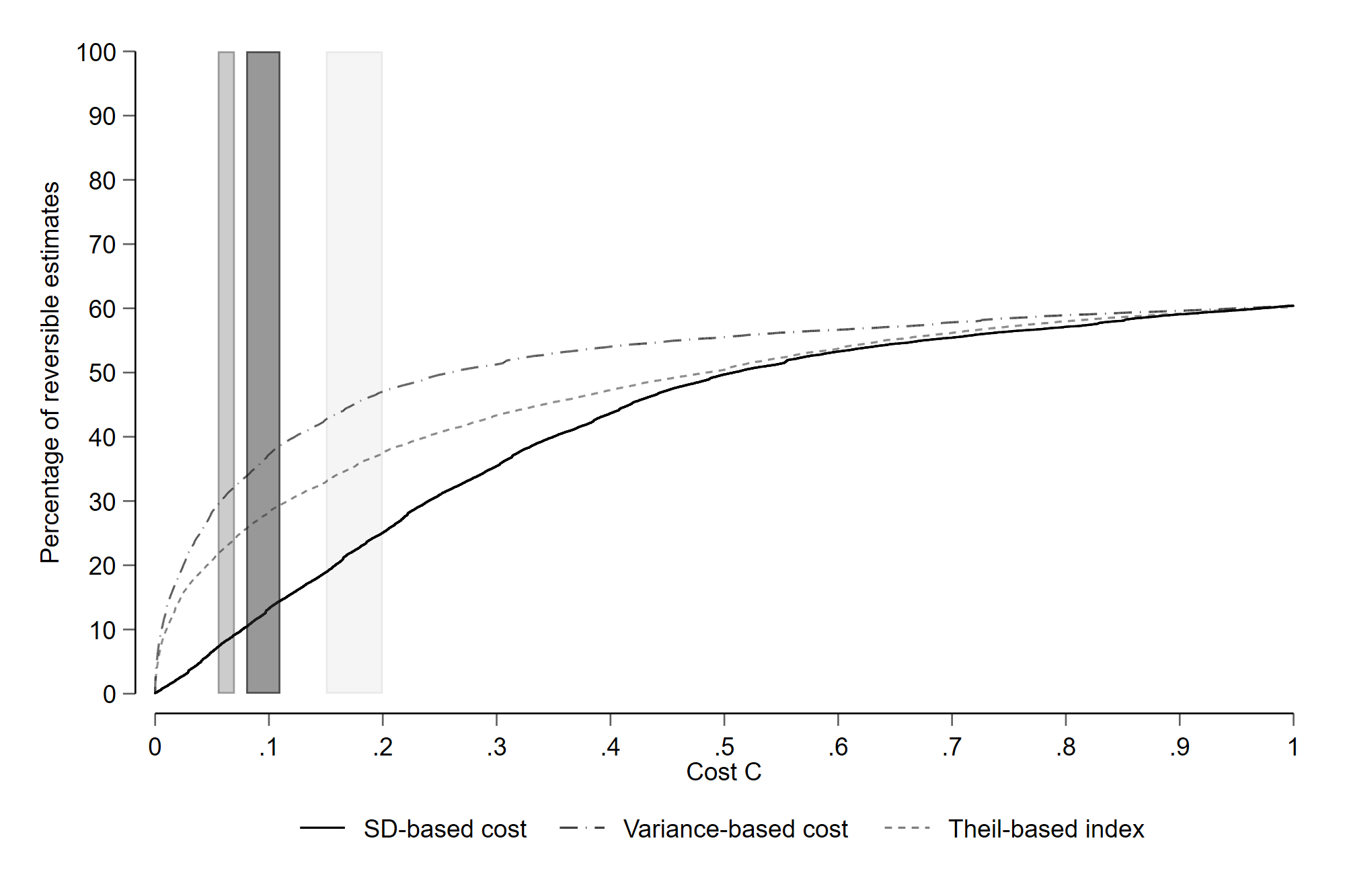}
    \begin{minipage}{0.975\textwidth}
    \footnotesize
    \noindent \textbf{Notes:} The figure shows the cumulative shares of coefficients included in \textit{WellBase} for which their sign can be reversed by some positive monotonic transformation of the response scale with at most cost $C$ using different cost functions. Based on the scale-use evidence presented in Section \ref{sec:scale_use_evidence}, the shaded regions indicate the average degrees of non-linearity, with lighter shading corresponding to the SD-based cost, intermediate shading to the variance-based cost, and darker shading to the Theil-based cost. The risk of sign reversal in the ranges of plausible transformation is similar across cost functions.
    \end{minipage}
\end{figure}

Figure \ref{fig:sign_rev_alt} plots, for each cost function, the cumulative share of coefficients in \textit{WellBase} whose sign can be reversed by some positive monotonic transformation with cost at most $C$. First, all curves mechanically converge to the same upper bound as $C \to 1$. In that limit, the share of reversible coefficients is independent of the particular cost function used. Second, for intermediate values of $C$, the relationship between the cost threshold and the share of reversible coefficients remains concave across all cost functions.

While all lines in the figure are concave, they exhibit different curvature. For a given value of $C$, the variance- and Theil-based measures appear to imply a higher risk of sign reversal than the SD-based cost. At face value, this could suggest that our conclusions depend on the choice of cost function. However, this comparison is not on equal footing: Once attention is restricted to the empirically relevant plausibility regions for each measure (roughly $0.06$--$0.08$ for the variance-based cost, $0.08$--$0.11$ for the Theil index, and $0.15$--$0.20$ for the SD-based measure), the implied risk of sign reversal is roughly comparable across cost functions, ranging from about 25\% under the SD-based cost to 28\% under the Theil index and 30\% under the variance-based measure.

Hence, the substantive conclusions from Section \ref{sec:evidence_sign_reversals} are not an artifact of the specific SD-based cost. Although different functional forms weight deviations from linearity differently, they deliver  similar assessments of reversal risk over plausible ranges of scale use.

\subsubsection{On sign reversals: Predicting the risk of reversal}
\vspace{-.5em}

We now explore whether the risk of sign reversal can be predicted by observable features of the research design. We estimate a linear probability model of the form:
\begin{equation}   
\label{eq:reversal_prob}
    \text{Rev}_{mpr} = \beta_0 +\beta_1\text{ln(\#Observations)}_{pr} + \boldsymbol{\beta}_2\text{\textbf{Model}}_{pr} + \boldsymbol{\beta}_3\text{\textbf{Estimate}}_{mpr} + \boldsymbol{\beta}_4\mathbf{X}_{pr} + \varepsilon_{mpr},
\end{equation}

where the dependent variable, $\text{Rev}_{mpr}$, is a dummy equal to one if there exists at least one positive monotonic transformation of the wellbeing scale capable of reversing the sign of estimate $m$ from regression $r$ in paper $p$, and zero otherwise.\footnote{We also estimate a probit model. Marginal effects are reported in Table \ref{tab:robustness_sign}. Conclusions are the same.}

The term $\text{ln(\#Observations)}_{pr}$ gives the logged number of observations in each regression $r$ in each paper $p$. The vector $\textbf{Model}_{pr}$ includes the logged number of control variables and a dummy for regressions that include individual fixed-effects, reflecting practices through which researchers attempt to limit omitted variable bias. The vector $\textbf{Estimate}_{mpr}$ captures characteristics specific to the covariate $m$.  It includes dummies for whether the covariate is continuous (as opposed to categorical or binary), time-varying, or instrumented via 2SLS. It also includes a categorical variable classifying whether the covariate corresponds to an individual socio-demographic characteristic (the reference category), a natural experiment (e.g., policy reform or RCTs), a placebo, or a macroeconomic indicator. Finally, the vector $\mathbf{X}_{pr}$ comprises control variables: a dummy indicating whether the wellbeing scale includes at least seven categories, and a categorical variable differentiating among life satisfaction questions, the Cantril Ladder, and happiness questions. 

We estimate two versions of Equation (\ref{eq:reversal_prob}): one without and one with the logged t-statistic. We treat the t-statistic differently because, unlike the other variables, which reflect researchers’ design choices, it is an outcome of those choices that is not directly controlled. We include it to test whether the observed negative association between statistical significance and reversal risk (Panel (B), Figure \ref{fig:sign_rev}) continues to hold.

Conditional on the possibility of a sign reversal for a given estimate, we further estimate the following via OLS:
\begin{equation}
\label{eq:reversal_cost}
    \text{Cost}_{mpr} = \beta_0 +\beta_1\text{ln(\#Observations)}_{pr} + \boldsymbol{\beta}_2\text{\textbf{Model}}_{pr} + \boldsymbol{\beta}_3\text{\textbf{Estimate}}_{mpr} + \boldsymbol{\beta}_4\mathbf{X}_{pr} + \varepsilon_{mpr},
\end{equation}

Equation (\ref{eq:reversal_cost}) mirrors Equation (\ref{eq:reversal_prob}) but uses the minimum $C$ needed for a sign reversal as the dependent variable. Comparing Equations (\ref{eq:reversal_prob}) and (\ref{eq:reversal_cost}) enables us to assess whether the probability of reversal and the ease of achieving it share common determinants. In both analyses, we cluster standard errors at the regression–paper ($r \times p$) level. Continuous independent variables are standardised using their means and SDs reported in Table \ref{tab:wellbase_description}. 

\begin{table}[!t]
\caption{Predictors of the Probability and Cost of Sign-reversal.}
\label{tab:predictors_sign}
\centering
\small
\renewcommand{\arraystretch}{0.95}  

\def\sym#1{\ifmmode^{#1}\else\(^{#1}\)\fi}

\begin{tabular}{lcccc}
\toprule
& \multicolumn{2}{c}{P(Sign-reversal)} & \multicolumn{2}{c}{Cost of sign-reversal} \\
\cmidrule(lr){2-3} \cmidrule(lr){4-5}
& (1) & (2) & (3) & (4) \\
\midrule
\textbf{About the estimation sample:} \\
Number of observations (logged) & -0.105\sym{***} & 0.029\sym{***} & 0.016\sym{***} & -0.038\sym{***} \\
& (0.007) & (0.006) & (0.003) & (0.003) \\
\\
\textbf{About the econometric model:} \\
Number of controls & 0.019\sym{*} & 0.001 & -0.009\sym{**} & -0.006 \\
& (0.008) & (0.006) & (0.003) & (0.003) \\
Individual FE & 0.084\sym{***} & 0.004 & -0.045\sym{***} & -0.010 \\
& (0.021) & (0.016) & (0.009) & (0.008) \\
\\
\textbf{About the independent variable:} \\
Continuous variable & -0.080\sym{***} & -0.034\sym{***} & -0.008 & -0.002 \\
& (0.010) & (0.008) & (0.006) & (0.005) \\
Time-varying variable & -0.142\sym{***} & -0.074\sym{***} & 0.065\sym{***} & 0.038\sym{***} \\
& (0.009) & (0.007) & (0.005) & (0.004) \\
Two-stage least square & -0.056 & -0.012 & 0.049\sym{**} & 0.016 \\
& (0.033) & (0.032) & (0.018) & (0.017) \\
Natural experiment & -0.167\sym{***} & -0.090\sym{***} & 0.091\sym{***} & 0.034\sym{***} \\
& (0.017) & (0.014) & (0.013) & (0.009) \\
Macroeconomic indicator & 0.145\sym{***} & -0.026 & -0.009 & 0.024\sym{**} \\
& (0.015) & (0.014) & (0.009) & (0.008) \\
Absolute t-statistics (logged) & & -0.275\sym{***} & & 0.187\sym{***} \\
& & (0.004) & & (0.002) \\
\midrule
Observations & 28,522 & 28,522 & 17,243 & 17,243 \\
R\textsuperscript{2} & 0.163 & 0.411 & 0.105 & 0.512 \\
\bottomrule
\end{tabular}

\vspace{5pt}
\begin{minipage}{0.975\textwidth}
\footnotesize
\textbf{Notes:} The table shows the results from regressions assessing the risk and cost of sign reversal under positive monotonic transformations of the wellbeing scale. Specifically, Columns (1) and (2) report coefficients from an OLS model where the dependent variable equals one if at least one transformation reverses the sign of a coefficient~$m$ from a regression~$r$ reported in paper~$p$. Conditional on a sign reversal being possible, Columns (3) and (4) report coefficients from an OLS model where the dependent variable is the minimum cost~$C$ required for reversal. All regressions control for a dummy indicating whether the wellbeing scale includes at least seven response categories and for the type of well-being measure (life satisfaction, Cantril Ladder, or happiness). Standard errors are clustered at the regression-paper level. Statistical significance is denoted as follows: *~$p<0.05$, **~$p<0.01$, ***~$p<0.001$.

\end{minipage}
\end{table} 

Table \ref{tab:predictors_sign} reports predictors of reversal risk in Columns (1) and (2) and reversal costs in Columns (3) and (4). We highlight three findings. First, the determinants of whether a reversal is possible and how costly it is are largely shared: variables that lower the probability of reversal also increase the cost required to achieve a reversal. Second, the logged t-statistic is the strongest predictor of robustness: estimates with originally larger t-statistics are substantially less prone to reversal and more costly to reverse. This variable alone explains much of the variation, raising the $R^2$ of the model from 17\% to over 41\% in Columns (1) and (2) and from 11\% to over 51\% in Columns (3) and (4). Last, a covariate's source of variation matters: keeping the logged t-statistics constant, the signs of estimates exploiting exogenous sources of variation (e.g., natural experiments) are both less likely to reverse and require larger departures from linearity.\footnote{Some robustness checks are given Table \ref{tab:robustness_sign}. There we re-estimate Columns (2) and (4) of Table \ref{tab:predictors_sign} while adding journal or paper fixed effects, employing a probit model instead of a linear probability model as well as a linear Cragg hurdle model where we jointly model the occurrence and cost of reversibility. Our conclusions are robust across these specifications.}

Overall, the risk and cost of sign reversal are not just random noise. They reflect identifiable features of research design, and are therefore within researchers’ control. Signs of highly significant results are far more likely to persist across scale transformations.

\vspace{-1em}
\subsubsection{On significance reversals: Documenting the risk of reversal}
\vspace{-.5em}

We now quantify the risk of \textit{significance} reversals. To this end, we first divide all \textit{estimates of interest}  in \textit{WellBase} into two groups: those initially significant at the 5\% level, and those not significant at this level. For all estimates, we compute the maximum and minimum attainable p-values under any monotonic transformation. We define significance reversals as cases where some transformation of the wellbeing scale cause the maximum attainable p-value for an originally significant estimate to exceed the 5\% threshold, or conversely, where the minimum attainable p-value for an originally non-significant estimate drops below this threshold. Conditional on a `significance reversal' being possible, we then numerically search for the transformation that produces a reversal with the smallest deviation from linearity. 

\begin{figure}[tbp]
    \caption{\textit{Significance}-reversal shares for different values of $C$ in \textit{WellBase}.}
    \centering
    \label{fig:significance_reversal}
    \subcaptionbox*{Panel (A)}{\includegraphics[width=0.495\textwidth]{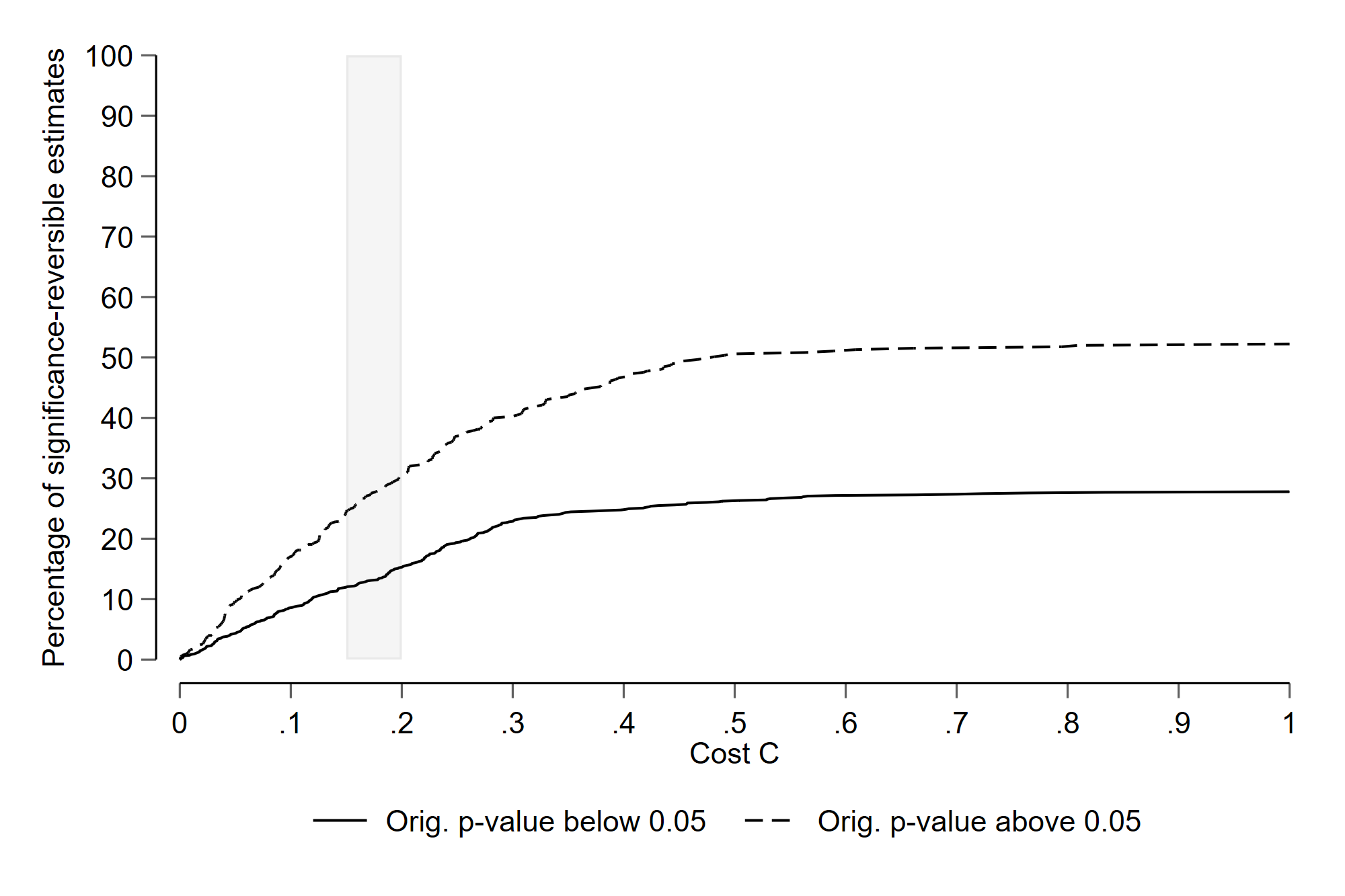}} 
    \subcaptionbox*{Panel (B)}{\includegraphics[width=0.495\textwidth]{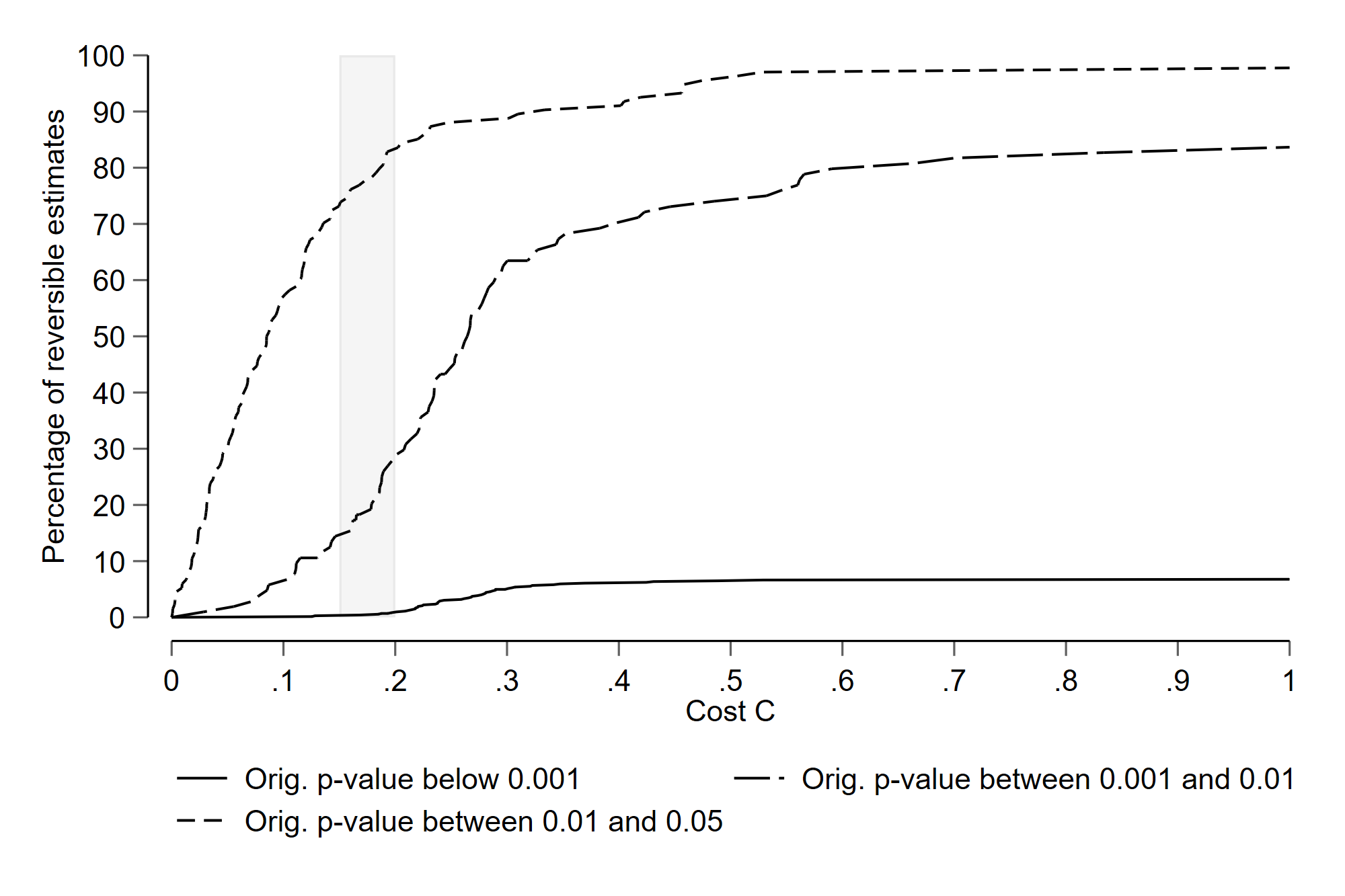}}    
    \begin{minipage}{0.975\textwidth}
    \footnotesize
    \noindent \textbf{Notes:} Cumulative shares of \textit{coefficients of interest} included in \textit{WellBase} for which `statistical significance' can be reversed by at least one positive monotonic transformation of the response scale with at most cost $C$. See notes of Figure \ref{fig:sign_rev} for more details about $C$ and the shaded regions. Panel (A) shows that up to 24\% of originally significant estimates lose significance under at least one positive monotonic transformation, while approximately 60\% of originally insignificant coefficients can be rendered significant. Panel (B) shows that originally highly significant coefficients ($p<0.001$) are extremely robust, whereas marginally significant ones ($0.01<p\leq0.05$) are fragile even under ``plausible'' transformations.
    \end{minipage}
\end{figure}

Figure \ref{fig:significance_reversal} plots the share of significance reversal against the cost $C$. The solid black curve traces this share for coefficients originally significant at the 5\% level. The dotted grey curve shows the corresponding share for originally insignificant coefficients crossing the significance threshold. The relationship between the cost $C$ and the probability of significance reversals is, again, concave. The `hazard' of gaining significance is always greater than that of losing it: 60\% of previously insignificant estimates can become significant with some positive monotonic transformation of the response scale. Only 24\% of significant coefficients can be turned insignificant. Restricting attention to ``plausible'' transformations ($0.15 < C<0.20$) reduces these figures to 30\% and 15\%, respectively. Panel (B) restricts attention to initially significant coefficients. About 85\% of coefficients with originally $0.01 < p \leq 0.05$ lose significance under transformations with a cost lying between 0.15 and 0.20. In contrast, coefficients that were already highly significant ($p<0.001$) are almost immovable: 95\% stay below a p-value of 0.05 under any positive monotonic transformation. 

These results mirror those for sign reversals: significance reversals are a real concern, but their occurrence appears limited when restricting attention to `plausible' departures from linearity. This is especially true for highly significant estimates, which almost never become insignificant regardless of the transformation considered. However, as is intuitive, estimates just below the 5\% threshold easily lose significance even under `mild' transformations.

\vspace{-1em}
\subsubsection{On relative magnitudes: The case of unemployment and income}
\label{sec:rel_magnitudes}
\vspace{-.5em}

Turning to relative magnitudes, we now focus on unemployment and the income–wellbeing relationship. The latter is especially central to policy‑oriented work, because income is used as the numéraire in monetary valuations based on subjective wellbeing data \citep[e.g.][]{dolan2019quantifying}. Our analysis draws on the subset of nine studies in \textit{WellBase} that include both unemployment and household income in their regressions. To facilitate comparability across studies, we standardise each study's wellbeing variable to mean zero and unit SD. 

\begin{figure}[t]
    \caption{Forest plot showing the sensitivity of relative estimate magnitudes to transformations of the wellbeing scale.}
    \centering
    \label{fig:magnitudes}
    \includegraphics[width=0.75\textwidth]{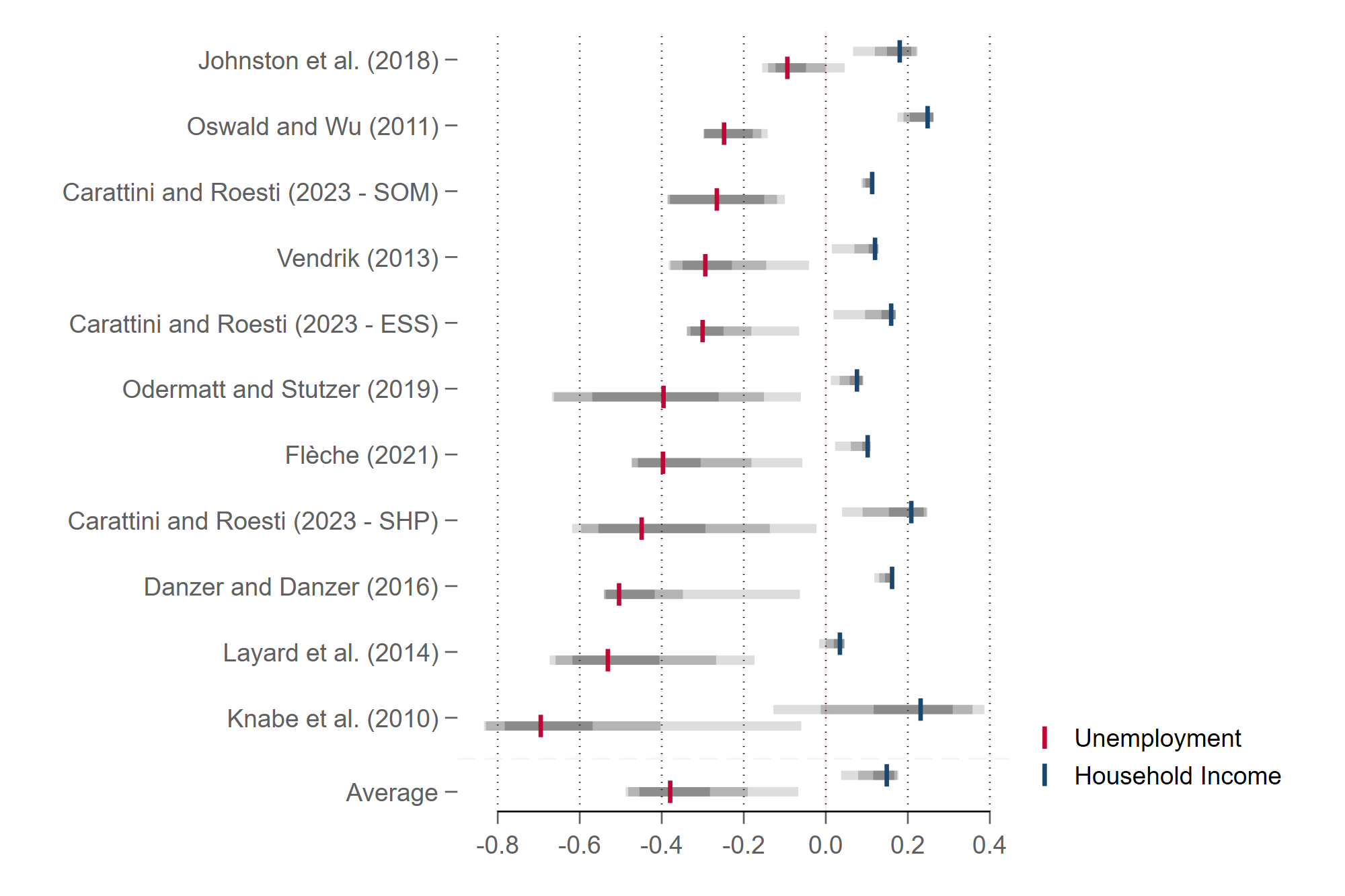}
    
    \begin{minipage}{0.975\textwidth}
    \footnotesize
    \noindent \textbf{Notes:} Standardised average point estimates for unemployment and the log of household income among papers included in \textit{WellBase}. Papers are ranked by the average effect size of the unemployment coefficient. The overall average, weighted by the inverse of the standard error of the individual estimates \citep{borenstein2010basic}, is displayed at the bottom. Wellbeing scales are standardised (mean of zero and standard deviation of one). Grey bars indicate the possible range of point estimates after applying positive monotonic transformations of the wellbeing scales. There are three shades of grey, with the darkest corresponding to $C<0.15$, the middle to $0.15 \leq C<0.30$, and the lightest to $0.30 \leq C$.
    \end{minipage}
\end{figure}

We first compute a paper-specific average point estimate for unemployment and for income weighted by the inverse of the standard error of the estimates.\footnote{The only exception is \cite{carattini2023trust}, who used three distinct datasets, where we treat each dataset from their paper as a unique observation.} The vertical markers in Figure \ref{fig:magnitudes} present such point estimates under the assumption of linearity (i.e. $C=0$). Unemployment is indicated in red. The blue markers show income. On average, unemployment is associated with a decrease in wellbeing of roughly 0.39 standard deviations. A unit increase in log income is, on average, associated with a 0.16 SD increase in wellbeing. 

The grey bars in Figure \ref{fig:magnitudes} show how these estimates may vary as we depart from linear scale use. When taking the meta-analytic average across all studies, allowing for $C<0.15$ (for $C<0.30$), unemployment decreases wellbeing between 0.28 (0.19) and 0.45 (0.50) SDs. A unit increase in log income is correspondingly associated with an average increase between 0.11 (0.04) and 0.17 (0.18) SDs. Thus, the magnitudes of estimates vary widely, even under `plausible' transformations. 

\begin{figure}[tbp]
    \caption{Ranges of unemployment-income ratios.}
    \centering
    \label{fig:ratios}
    \includegraphics[width=0.75\textwidth]{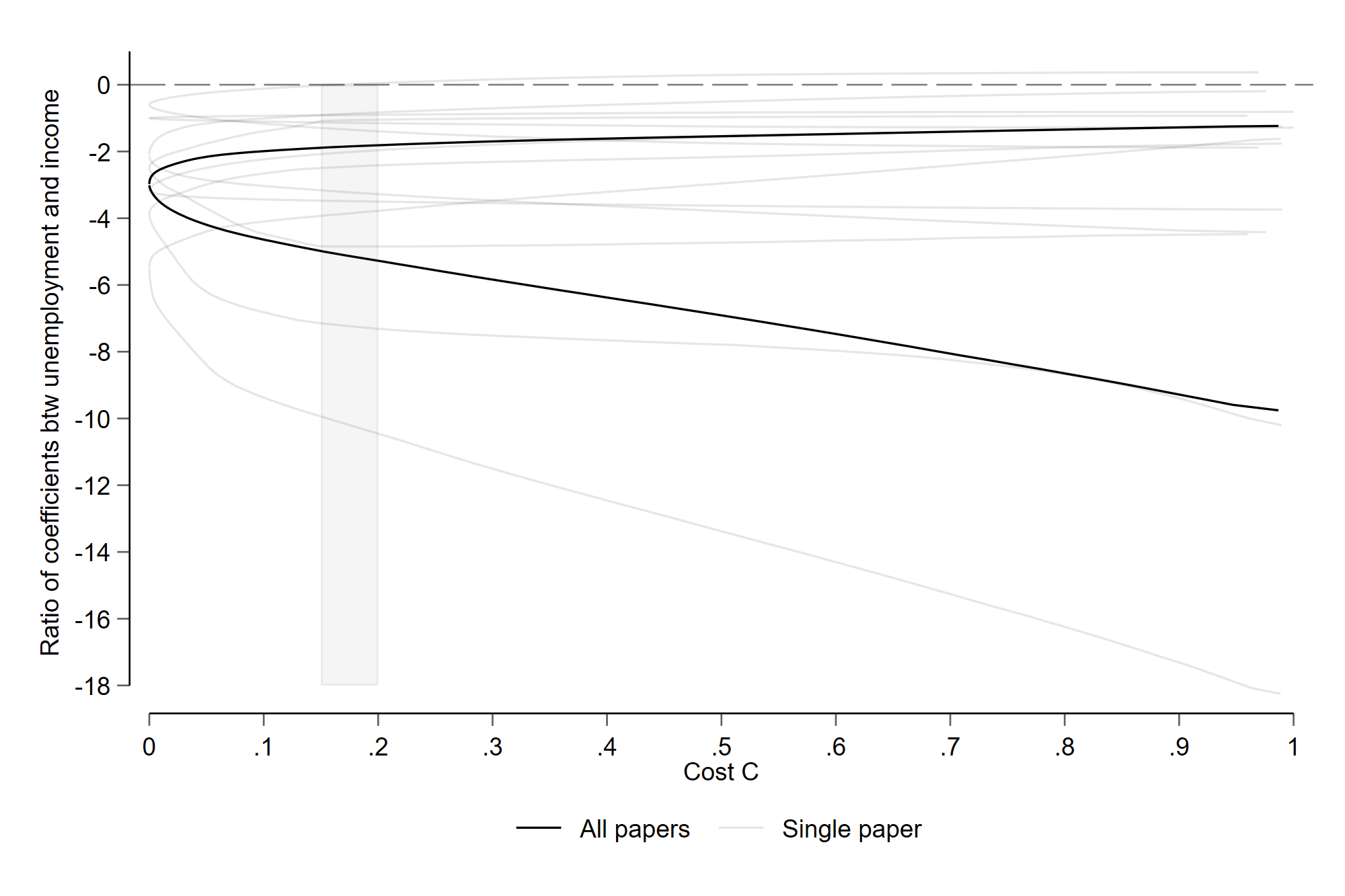}
    %\subcaptionbox*{Panel (B)}{\includegraphics[width=0.495\textwidth]{figures/wellbase_MRS_panelB.png}}    
    \begin{minipage}{0.975\textwidth}
    \footnotesize{\noindent \textbf{Notes:} Ranges of marginal rate of substitution (MRS) between unemployment and log household income by paper (grey) and their average (black) across values of $C$. See notes of Figure \ref{fig:sign_rev} for more details about $C$ and the shaded region. The plotted MRS between unemployment and log household income reveals a wide range across papers: from just positive to –18.25.}  
    \end{minipage}
\end{figure}

However, given that there is no natural absolute scale for wellbeing (linear or not), the absolute magnitudes of coefficients are not entirely meaningful. Ratios of coefficients, in contrast, do provide a meaningful relative measure. When interpreted as causal estimates, such ratios can be read as marginal rates of substitution (MRS) between two variables. Figure \ref{fig:ratios} therefore shows ratios of the coefficient on unemployment to the coefficient on ln(HH income) across different levels for $C$.\footnote{For the computations to follow, we exclude regressions where the coefficient on income was reversible. According to Proposition \ref{prop:ratios}, ratios are unbounded in such a case. We excluded four regressions on that basis: one in \cite{knabe2010dissatisfied} and three in \cite{layard2014predicts} (cf. Figure \ref{fig:ratios}).} Each grey line corresponds to a different paper. The black line shows the average ratio across all papers. We observe that this mean MRS can range from -1 to as low as –10. Under `plausible' transformations, the ranges are narrower, ranging from -2 to -5.

Thus, although the risk of sign and significance reversals appeared relatively small under `plausible' transformations of the scale, the same cannot be said about the \textit{magnitudes} of estimates. In the case of unemployment and income, two key drivers of wellbeing, both absolute and relative magnitudes are highly sensitive to even mild departures from linearity.

\subsection{Likert scales for attitudes, preferences and perceptions}
\vspace{-.5em}

Our focus has so far been on wellbeing scales. But these are not the only constructs in economics measured using ordered response scales. Concepts such as risk aversion, trust, or political preferences are also routinely captured with such scales, and are broadly accepted within the discipline. To gauge whether concerns about the cardinal vs ordinal nature of Likert-style measurement ought to be unique to wellbeing, we now compare the reversal risks between these different types of measures. 

To do so systematically, we screened every article that appeared between January 2010 and May 2025 in the five leading economics journals\footnote{We count the following journals as part of the `top five': \textit{Quarterly Journal of Economics, American Economic Review, Journal of Political Economy, Review of Economic Studies, Econometrica}.} and retained those whose full text contained the term ``Likert scale'' or whose title included at least one of the following expressions: ``attitudes'', ``risk aversion,'' ``risk preferences,'' ``trust,'' or ``preferences for''. This strategy is unlikely to cover all Likert-scale based research published in top-five economics journals, but a true census of all such published research is beyond the scope of this study. 

\begin{figure}[t]
    \caption{Comparing the risk of sign reversal between wellbeing scales and other Likert scales in Top-five Economics journals.}
    \centering
    \label{fig:likert}
    \includegraphics[width=0.75\textwidth]{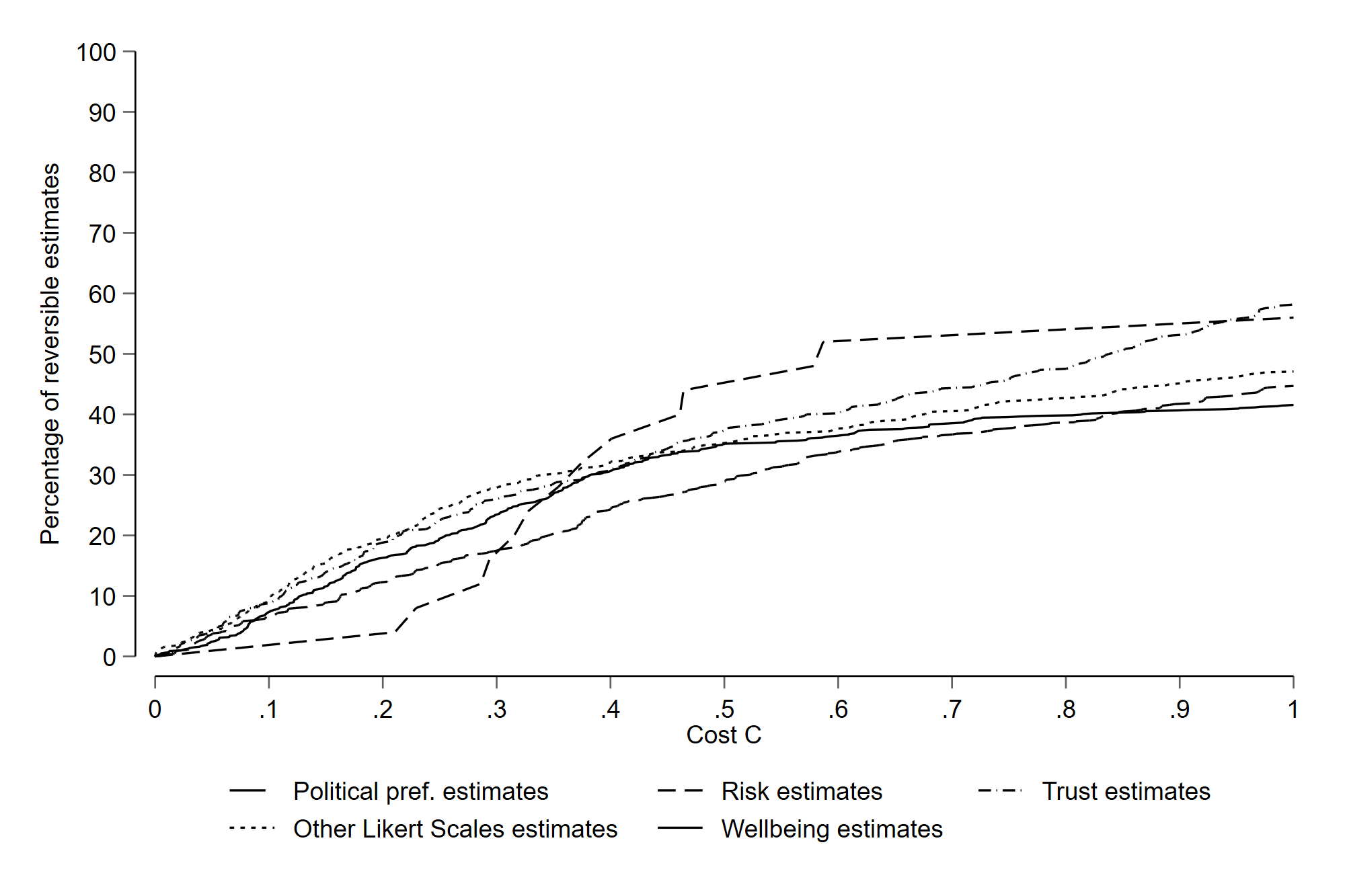}
    \begin{minipage}{0.975\textwidth}
    \footnotesize
    \footnotesize{\noindent \textbf{Notes:} 
    Cumulative shares of coefficients for which the sign can be reversed by at least one positive monotonic transformation of the response scale with at most cost $C$. The figure shows that the risk of reversal is not unique to wellbeing. In many cases, results on constructs such as risk (56\%), trust (58\%), political preferences (44\%), and `other' constructs (47\%) are sign reversible.}  
    \end{minipage}
\end{figure}

As shown in Figure \ref{fig:prisma_likert}, we reproduced 16 articles for a total of 511 regressions and 23,104 estimates (8.61\% of which are printed coefficients). Of the included papers, three contained a Likert-scale measure of trust \citep{acemoglu2020trust,algan2010inherited,falk2018global}, four contained a Likert-scale measure of political preferences \citep{kuziemko2015elastic,alesina2018intergenerational,jha2019valuing,dechezlepretre2025fighting}, two contained a Likert-scale measure of risk aversion \citep{dohmen2010risk,jha2019valuing}, and nine contained a Likert-scale measure of other concepts including hiring interest, optimism, fear, political correctness attitudes, and work morale \citep{cohn2015evidence,jha2019valuing,kessler2019incentivized,exley2022gender,spenkuch2023ideology,braghieri2024political,engelmann2024anticipatory,englmaier2024effect,gagnon2025effect}.

Results are shown in Figure \ref{fig:likert}. There we compare the sign reversal risk for estimates based on wellbeing scales published in top-five economics journals (solid line) with the corresponding risk for estimates based on other Likert scales. The risk of sign reversal for wellbeing estimates in this subsample is around 41\%. This is lower than that for risk (56\%), trust (58\%), political preferences (44\%), and the `other' concepts (47\%).

Finally, to explore whether the risks of sign reversal vary across concepts, we also replicated the analysis of Table \ref{tab:predictors_sign}. Figure \ref{fig:swb_likert_sign_rev} shows that the predictors of reversal risk are similar across types of measures: most notably, larger t-statistics robustly reduce reversal risk. Thus, neither the level nor the determinants of reversal risk are unique to wellbeing. Any concept measured with Likert-type scales is similarly vulnerable.

\vspace{-1em}

\section{Practical guidelines and empirical illustration}
\label{sec:guidelines}
\vspace{-.5em}

In this section, we show how our approach can be adopted for use in regression analyses of the kind systematically reviewed in the previous section. We suggest the following steps.

\begin{enumerate}
\setlength{\itemsep}{0pt}
    \item Estimate baseline regressions as usual, coding the ordered response scale linearly.
    \item For each coefficient of interest, report: 
    \begin{itemize}
    \setlength{\itemsep}{0pt}
        \item  Whether a sign reversal is possible under any monotonic transformation. Conditional on reversibility, the cost required to reverse the sign. 
        \item Whether statistical significance (at a chosen level) can be reversed. Conditional on reversibility, the cost required to reverse statistical significance.
        \item When relevant report ranges for marginal rates of substitution.
    \end{itemize}
\end{enumerate}

To illustrate, we consider a standard applied setting using secondary survey data. We estimate an OLS regression of life evaluation (Cantril Ladder, 0--10 scale) using U.S. Gallup Daily data, covering the period 2008-2017. The specification includes log income, employment status, age and age squared, partnership status, education, race, gender, household size, and the presence of children. The sample contains 1,233,716 observations.

Table \ref{tab:gallup} shows how the proposed reporting protocol can be implemented in practice. Alongside conventional coefficient estimates, we report whether sign and significance reversals are possible and, conditional on reversibility, the minimal cost required to trigger them. The final three columns relate to coefficient ratios, using log income as the denominator. 

\begin{table}[!t]
\def\sym#1{\ifmmode^{#1}\else\(^{#1}\)\fi}
\caption{Reversal risks and required cost: An illustration in a standard welbeing regression}

\label{tab:gallup}

\centering
\small
\renewcommand{\arraystretch}{0.95}  

\resizebox{1\textwidth}{!}{
\begin{tabular}{lcccccccc}
\toprule
& $\beta$ & \multicolumn{2}{c}{Sign-reversal} & \multicolumn{2}{c}{Significance-reversal} & \multicolumn{3}{c}{Ratio wrt Ln(income)} \\
\cmidrule(lr){3-4} \cmidrule(lr){5-6} \cmidrule(lr){7-9}
& & Risk & Cost  & Risk & Cost & C=0 & C=0.20 & C=1 \\
\midrule
Ln(income) & 0.402\sym{***} & No & . & Yes & 0.588 & - & - & - \\
& (0.002)  \\
Unemployed & -0.662\sym{***} & No & .  & No & . & -1.647 & (-1.913; -1.494) & (-1460; -0.912) \\
& (0.009)  \\
Age & -0.066\sym{***} & No & .  & No & . & -0.164 &  (-0.230; -0.133) & (-314.5; -0.097)   \\
& (0.001)  \\
Age-sq/100 & 0.071\sym{***} & No & .  & No & . & 0.176  & (0.131; 0.263) & (0.087; 378.1)  \\
& (0.000) \\
Has partner & 0.414\sym{***}  & No & . & No & . & 1.029 & (0.704; 1.456) & (0.307; 1381)\\
& (0.004) \\
High education & 0.340\sym{***}  & Yes & 0.341  & Yes & 0.262 & 0.846 & (0.450; 1.039) & (-1091; 1.232)\\
& (0.004)  \\
Not white & 0.152\sym{***}  & Yes & 0.461  & Yes & 0.021 & 0.378 & (0.144; 0.862)  & (-0.259; 2546)\\
& (0.006)  \\
Female & 0.285\sym{***} & No & .  & Yes & 0.658 & 0.711 & (0.442; 1.136) & (0.069; 1502)\\
& (0.003)  \\
HH size & -0.035\sym{***}  & Yes & 0.325  & Yes & 0.228 & -0.087 &  (-0.107; -0.049)  & (-0.113; 110.6) \\
&  (0.002)  \\
Has children & 0.053\sym{***} & No & .  & No & . & 0.133  & (0.105; 0.163)  & (0.020; 138.0)\\
& (0.005) \\
\midrule
Observations & 1,233,716 \\
\bottomrule
\end{tabular}
}

\begin{minipage}{0.975\textwidth}
\footnotesize
\noindent \textbf{Notes:} This table illustrates how positive monotonic transformations of the response scale may affect regression results in a representative specification. Gallup Daily data covering the period 2008 to 2017 is used. Column (1) reports OLS coefficients from a regression of life satisfaction on the listed covariates. Standard errors in parentheses; *~$p<0.05$, **~$p<0.01$, ***~$p<0.001$. Columns (2)–(3) indicate whether a sign reversal is possible under some positive monotonic transformation of the response scale and, if so, report the minimum cost $C$ required to obtain such a reversal. Columns (4)–(5) analogously report whether statistical significance at the 5\% level can be reversed and the corresponding minimum cost. Columns (6)–(8) report on the ratio of each coefficient to the coefficient on log income. Column (6) reports the ratio under linear scale use ($C=0$). Columns (7)–(8) report the min. and max. values this ratio can attain when transformations are restricted to have cost $C = 0.15$ and $C = 0.20$, respectively. The cost $C$ is the SD-based measure, bounded between 0 (perfect linearity) and 1 (maximal non-linearity).
\end{minipage}
\end{table}

% \begin{tabular}{lcccccccc}
% \toprule
% & $\beta$ & \multicolumn{2}{c}{Sign-reversal} & \multicolumn{2}{c}{Significance-reversal} & \multicolumn{3}{c}{Ratio wrt Ln(income)} \\
% \cmidrule(lr){3-4} \cmidrule(lr){5-6} \cmidrule(lr){7-9}
% & & Risk & Cost  & Risk & Cost & C=0 & C=0.15 & C=0.20 \\
% \midrule
% Ln(income) & 0.402\sym{***} & No & . & CK & CK & - & - & - \\
% & (0.002)  \\
% Unemployed & -0.662\sym{***} & No & .  & CK & CK & -1.647 & (-1.530; -1.821) & (-1.494; -1.913) \\
% & (0.009)  \\
% Age & -0.066\sym{***} & No & .  & CK & CK& -0.164 & (-0.207; -0.139) &  (-0.230; -0.133)\\
% & (0.001)  \\
% Age-sq/100 & 0.071\sym{***} & No & .  & CK & CK & 0.176 & (0.141; 0.232)  & (0.131; 0.263)\\
% & (0.000) \\
% Has partner & 0.414\sym{***}  & No & . & CK & CK & 1.029 & (0.788; 1.320) & (0.704; 1.456)\\
% & (0.004) \\
% High education & 0.340\sym{***}  & Yes & 0.341  & CK & CK & 0.846 & (0.592; 0.997) & (0.450; 1.039)\\
% & (0.004)  \\
% Not white & 0.152\sym{***}  & Yes & 0.461  & CK & CK & 0.378 & (0.191; 0.688) & (0.144; 0.862) \\
% & (0.006)  \\
% Female & 0.285\sym{***} & No & .  & CK & CK & 0.711 & (0.506; 0.992) & (0.442; 1.136)\\
% & (0.003)  \\
% HH size & -0.035\sym{***}  & Yes & 0.325  & CK & CK & -0.087 & (-0.105; -0.063) &  (-0.107; -0.049)\\
% &  (0.002)  \\
% Has children & 0.053\sym{***} & No & .  & CK & CK & 0.133 & (0.113; 0.154)  & (0.105; 0.163)\\
% & (0.005) \\
% \midrule
% Observations & 1,233,716 \\
% \bottomrule
% \end{tabular}

The first column reports conventional OLS estimates and closely replicates the standard patterns in the wellbeing literature. Life evaluation increases with log income, follows a U-shaped profile over age, and is substantially lower among the unemployed. Individuals with a partner report higher wellbeing.

The second and third columns report on sign-reversal risks due to potentially non-linear scale use. The coefficient on employment status (``unemployed'') cannot be sign-reversed under any monotonic transformation. The same holds for log income, partnership status and gender. Other coefficients admit sign reversals, including the coefficients on education and household size. However, in each of these cases, the required `amount' of non-linearity far exceeds the plausible range from Section \ref{sec:scale_use_evidence}.

The coefficients on ln(income) and being female admit \textit{significance} reversals (at the 5\% level) even though a \textit{sign} reversal is impossible for these. Generally, results on significance can be overturned at a slightly lower cost, which is near the plausible range. High education, for instance, loses significance at a cost of $C=0.262$,  and household size at $C=0.226$. The most extreme case is on the ethnicity dummy (`Not White'), where a transformation with $C=0.021$ is sufficient to render the associated coefficient insignificant at the 5\% level.

Column (6) reports coefficient ratios with respect to log income under linear scale use. The final two columns then evaluate how these trade-offs vary when allowing for departures from linearity. For plausible $C$, the implied ranges are relatively stable. However, even seemingly moderate changes in these ratios can have large economic implications. Consider unemployment. Under linear scale use, the estimated MRS of $1.647$ implies that offsetting the wellbeing loss associated with unemployment requires an increase in income of roughly four times baseline income ($\exp(1.647)-1 \approx 4.19$). Evaluated at the median income in our data (\$54,000), this corresponds to an annual compensation of about \$226,000, or \$18,900 per month; a figure in line with estimates in \citet{clark2002simple}. However, merely allowing for \textit{plausible} non-linearity ($C = 0.2$) shifts this to a range of about \$186,000 to \$312,000 per year (approx. \$15,500 to \$26,000 per month). A modest shift in coefficients thus translates into annual differences exceeding \$100,000 in implied compensation.

\section{Discussion}
\label{sec:discussion}
\vspace{-.5em}

Economists increasingly rely on bounded survey scales to measure latent constructs like risk preferences, trust, political attitudes, and wellbeing. Standard practice treats these scales as cardinal measures, assuming without evidence that psychological distances between adjacent response categories remain constant across the entire scale. Our theoretical framework formalizes when this assumption matters and introduces a cost function $C$ to quantify the minimal deviation from linearity required to reverse the sign, to reverse significance, or to change the relative magnitude of estimated coefficients.

We gathered original experimental data to assess how individuals use response scales. Across a series of elicitation strategies, we find that respondents, on average, use such scales in a way that mildly deviates from linearity. Our estimates imply a rough upper bound on the average cost of deviation from linearity of $C = 0.20$. We use this value as an empirical anchor for judging the plausibility of reversals. Importantly, the distribution of individual-level scale use is highly concentrated: a third of respondents reports using the scale in a linear way, another third exhibits deviations well below $C=0.20$, and we observe few individuals with deviations exceeding $C=0.50$. Thus, while perfect linearity is rejected on average, extreme non-linear interpretations of response scales find little empirical support in our data. Moreover, individual differences in scale use appear largely idiosyncratic. We find no systematic relationship between non-linearity and observable respondent characteristics. In our view, rather than recovering a unique ``true'' reporting function, our experimental data allows to gauge a set of admissible transformations, and to thereby provide an empirical benchmark for judging the plausibility of reversals in the broader literature.

We then ask to what extent wellbeing research published in top-ranked economics journals depends on the assumption of linear scale use. To do so, we constructed \textit{WellBase}, a database comprising the universe of replicable regressions using cognitive wellbeing as a dependent variable in the top 30 economics journals between 2010 and 2025 (more than 28,000 coefficients). For each estimate, we assess whether its sign can be reversed by at least one positive monotonic transformation of the wellbeing scale and, if so, compute the minimal cost of such a transformation. We further examine whether research practices exist that are systematically associated with a lower risk of sign reversal. Finally, we use \textit{WellBase} to document the rate of significance reversals and changes in coefficient ratios under positive monotonic transformations.

Sign reversal risks are concave in the cost of deviating from linearity. Plausible transformations of the wellbeing scale can reverse the sign of about 25\% of the wellbeing research published in top-ranked economics journals. If linearity is entirely abandoned, this share increases to about 60\%. Restricting attention to wellbeing studies in top-five journals, the risk is lower: around 40\% under a purely ordinal interpretation. The corresponding values for our sample of non-wellbeing Likert scales lie between 44\% and 58\%. Among wellbeing-based coefficients with $p$-values below 0.05, the ones typically emphasised in published texts, the risk is negligible if we consider plausible transformations only. More generally, the risk of sign reversal is not random: it can be predicted by observable features of the research design. One key finding is that estimates relying on arguably exogenous variation, such as natural experiments or macroeconomic shocks, are systematically less prone to reversals.

Regarding significance reversals, we again find a concave relationship: the marginal effect of relaxing linearity on the risk of significance reversal diminishes with cost. If the linearity assumption were fully abandoned, roughly 85\% of the estimates originally significant at the 1\% level would remain robust at the 5\% level. However, for estimates with $p$-values between 0.01 and 0.05, the risk of significance reversal is much larger, even under empirically `plausible' transformations of the wellbeing scale. Hence, the bar for statistical inference is higher than in the absence of concerns over non-linear scale use.  

To assess the sensitivity of coefficient magnitudes and ratios, we restricted the analysis to papers that include both unemployment and income as covariates. Even small potential deviations from linear scale use substantially affect the absolute size of these coefficients and easily alter their ratio. Thus, while the direction of estimates tends to be stable, their relative sizes are highly sensitive to scale assumptions.

Overall, some of our conclusions are encouraging. The overall risk of sign reversal is limited under plausible deviations from linearity, and partially predictable based on research design. Likewise, the risk of significance reversal is small for estimates with small original p-values. But other conclusions are more concerning. First, our results are not unique to wellbeing data: estimates based on other widely used Likert-type scales in economics, such as trust, risk preferences, or political attitudes, face similar risks. Potentially non-linear scale use is therefore a concern for a much broader segment of economic research than is widely recognised.  Second, the risk of significance reversal is high for estimates with $p$-values between 0.01 and 0.05. Finally, estimated magnitudes and coefficient ratios are highly unstable. Here, too, minimal non-linearities suffice to reverse substantive conclusions.

These results have practical implications. What is most needed is a broader evidence-base on scale use. Our own tests, while indicative, focus on a single type of wellbeing scale and could not pin down the precise functional form by which scale use departs from linearity. Moreover, future work might fruitfully combine our analysis with current work on correcting for interpersonal differences in scale use \citep{prati2026possible, benjamin2023adjusting}. Further work may also draw on the partial identification literature to, for instance, provide confidence regions for identified coefficient ranges given some cost $C$ \citep[cf.][]{imbens2004confidence,tamer2010partial}. For current practice, we believe that analysts should routinely report robustness to monotonic transformations of the outcome variable, following the guidelines in Section \ref{sec:guidelines}. Our \href{https://github.com/casparwarwick/reversals}{\color{blue}{Stata}} routines make this straightforward.

\newpage
\clearpage

\appendix

\setcounter{figure}{0}
\setcounter{table}{0}
\setcounter{appendixproposition}{0}

\renewcommand{\thetable}{A\arabic{table}}
\renewcommand{\thefigure}{A\arabic{figure}}
\renewcommand{\theappendixproposition}{A\arabic{appendixproposition}}

\begin{center}
\section*{Appendix}
\end{center}

\section{Proof of Proposition \ref{prop:plausibility}}
\label{sec:proofs_theorem}

\subsubsection*{Step 1: Establish identical component functions.}

Desideratum 5 states that $\Pi(\Delta) = g\left(\sum_{k=1}^{K-1} \varphi_k(\Delta_k)\right)$. By Desideratum 2 (Permutation invariance), for any permutation $\sigma$, we have $g\left(\sum_{k=1}^{K-1} \varphi_k(\Delta_k)\right) = g\left(\sum_{k=1}^{K-1} \varphi_k(\Delta_{\sigma(k)})\right)$. Since $g$ is monotonic (Desideratum 5), we also need: $\sum_{k=1}^{K-1} \varphi_k(\Delta_k) = \sum_{k=1}^{K-1} \varphi_k(\Delta_{\sigma(k)})$. Now consider the specific case where $\Delta = (x, y, 0, ..., 0)$ with $x + y = 1$. Under the transposition $\sigma$ that swaps positions 1 and 2: $\varphi_1(x) + \varphi_2(y) = \varphi_1(y) + \varphi_2(x) \iff \varphi_1(x) - \varphi_2(x) = \varphi_1(y) - \varphi_2(y)$. Since this holds for all $x, y$, the difference $\varphi_1(x) - \varphi_2(x)$ must be constant. By similar arguments for all pairs of indices, all $\varphi_k$ differ only by constants. Since adding constants to all $\varphi_k$ can be absorbed into $g$, we can without loss of generality set all $\varphi_k = \psi$ for some common function $\psi$. We thus get: $\Pi(\Delta) = g\left(\sum_{k=1}^{K-1} \psi(\Delta_k)\right)$.

\subsubsection*{Step 2: Determine convexity of $\psi$ and direction of $g$}

Define the inner sum as $S(\Delta) = \sum_{k=1}^{K-1} \psi(\Delta_k)$. Desideratum 3 states that for any mean-preserving spread $\Delta$ of $\Delta'$, we have $\Pi(\Delta) > \Pi(\Delta')$. This implies that $\Pi$ is strictly Schur-convex \citep{marshall1979inequalities}. Since $\Pi(\Delta) = g(S(\Delta))$ where $g$ is monotonic (Desideratum 5), and $\Pi$ is strictly Schur-convex, we know that $S(\Delta)$ is strictly Schur-convex (if $g$ is increasing) or strictly Schur-concave (if $g$ is decreasing). A sum of the form $S(\Delta) = \sum_{k=1}^{K-1} \psi(\Delta_k)$ is strictly Schur-convex (Schur-concave) iff $\psi$ is strictly convex (concave). Therefore, we have two possible cases. Either (Case A) $\psi$ is strictly convex and $g$ is strictly increasing, or (Case B) $\psi$ is strictly concave and $g$ is strictly decreasing. These cases are duals: If $(\psi,g)$ is a valid concave/decreasing pair, we can define $\psi'=-\psi$ (strictly convex) and $g'(x)=g(-x)$ (increasing), which represents the same plausibility measure: $\Pi(\Delta)=g(S(\Delta))=g(\sum\psi(\Delta_k))=g(-\sum[-\psi(\Delta_k)])=g'(\sum\psi'(\Delta_k))$. By convention, we choose Case A without loss of generality. Therefore, $\psi$ is strictly convex and $g$ is increasing.

\subsubsection*{Step 3: Note that $g$ and $\psi$ are continuous}

By Desideratum 4, $\Pi$ is continuous. Since $g$ is monotonic and therefore cannot smooth out discontinuities, both $S(\Delta) = \sum \psi(\Delta_k)$ and $g$ must be continuous. Continuity of $S(\Delta)$ requires $\psi(\Delta_k)$ to be continuous.

\subsubsection*{Step 4: Derive the unique minimum of $\psi$}

By Desideratum 1, $\Pi(\Delta) = 0$ if all $\Delta_k = 1/(K-1)$ and $\Pi(\Delta) = 0$ is the minimum of $\Pi(\Delta)$. Let $\mathbf{u} = (1/(K-1), ..., 1/(K-1))$. Then: $\Pi(\mathbf{u}) = g\left((K-1) \cdot \psi(1/(K-1))\right) = 0$.

From Step 2, we know that $g$ is monotonically increasing. Since $\Pi(\Delta)$ is minimised at $\Pi(\mathbf{u}) = 0$, it follows that the input to $g$ must also be minimised at $\Delta = \mathbf{u}$. Therefore, $\sum_{k=1}^{K-1} \psi(\Delta_k)$ achieves a minimum when $\Delta_k = 1/(K-1)$ for all $k$. Because $\psi(\Delta_k)$ is strictly convex, this minimum is unique. We can set $\psi(1/(K-1)) = 0$ without loss of generality. If instead $\psi(1/(K-1)) = c$ for some constant $c$, we could define $g'(x) = g(x - c(K-1))$ to achieve the same plausibility measure. Setting the minimum to zero simplifies the notation.

\subsubsection*{Step 5: Show that $\Pi$ is maximised at single-jump distributions}

Since $\psi$ is strictly convex (Step 2) the sum $\sum_{k=1}^{K-1} \psi(\Delta_k)$ over $\mathcal{D} = \{\Delta : \sum \Delta_k = 1, \Delta_k \geq 0\}$ is maximised at the extreme points of $\mathcal{D}$. The single-jump distributions where $\Delta_j = 1$ for some $j$ and $\Delta_k = 0$ for $k \neq j$ are the extreme points of $\mathcal{D}$. Finally, by symmetry, all single-jump distributions yield the same maximum value $M = \psi(1) + (K-2) \cdot \psi(0)$. Since $g$ is monotonically increasing, $\Pi$ is also maximised at single-jump distributions.

\subsubsection*{Step 6: Normalize}

In step 4, we established that $\min_{\Delta \in \mathcal{D}} S(\Delta)=\sum_{k=1}^{K-1} \psi(\Delta_k)=0$. In step 5, we showed that $\sum_{k=1}^{K-1} \psi(\Delta_k)$ obtains its maximum $M = \psi(1) + (K-2) \cdot \psi(0)$ at single-jump distributions. By Desideratum 1 we must have: $\min_{\Delta \in \mathcal{D}} \Pi(\Delta) = g(0) = 0$ and $\max_{\Delta \in \mathcal{D}} \Pi(\Delta) = g(M) = 1$. Therefore, $g: [0, M] \rightarrow [0, 1]$, with $g$ continuous (by step 3) and monotonically increasing (by step 2). Define $h:[0,1] \rightarrow [0,1]$ by $h(x) = g(M \cdot x)$. Then $h$ is continuous and increasing, with $h(0) = g(0) = 0$ and $h(1) = g(M) = 1$. Now, for any $\Delta$: $\Pi(\Delta) = g\left(\sum_{k=1}^{K-1} \psi(\Delta_k)\right) = g\left(M \cdot \frac{\sum_{k=1}^{K-1} \psi(\Delta_k)}{M}\right) = h\left(\frac{\sum_{k=1}^{K-1} \psi(\Delta_k)}{M}\right)$. Therefore: $\Pi(\Delta) = h\left(\frac{\sum_{k=1}^{K-1} \psi(\Delta_k)}{\max_{\Delta' \in \mathcal{D}} \sum_{k=1}^{K-1} \psi(\Delta'_k)}\right)$. 

\section{Derivations of decomposition results and proof of bounds on coefficient ratios}
\label{sec:proofs}

\subsection{Decomposition of $\hat{\beta}_m^{(\tilde{r})}$}
\label{sec:decomposition}

We here derive the decomposition of Equation \eqref{eq:decomp}. A version of this derivation originally appeared in \cite{kaiser2023how}. Here, we reproduce it in our notation. The supplementary appendix derives similar decompositions for the 2SLS, fixed effects, and case where $r_i$ is continuous.

Let $l_k$ be the value assigned to the $k^{th}$ response category of the untransformed variable $r_i$, and let $\tilde{l}_k$ denote the label assigned to category $k$ under some positive monotonic transformation $f$. Then:
\begin{equation*}
\tilde{r}_i = \sum_{k=1}^{K} \tilde{l}_k \mathds{1}(r_i = k) = \sum_{k=1}^{K-1} (\tilde{l}_k - \tilde{l}_{k+1})\mathds{1}(r_i \leq k) + \tilde{l}_K = \sum_{k=1}^{K-1} (\tilde{l}_k - \tilde{l}_{k+1})d_{k,i} + \tilde{l}_K.
\end{equation*}

Stacking over individuals: $\tilde{\mathbf{r}} = \sum_{k=1}^{K-1} (\tilde{l}_k - \tilde{l}_{k+1})\mathbf{d}_k + \tilde{l}_K\mathbf{1}$. Now notice that:
\begin{align*}
\hat{\boldsymbol{\beta}}^{(\tilde{r})} &= (\mathbf{X}'\mathbf{X})^{-1}\mathbf{X}'\tilde{\mathbf{r}} = (\mathbf{X}'\mathbf{X})^{-1}\mathbf{X}'\left(\sum_{k=1}^{K-1} (\tilde{l}_k - \tilde{l}_{k+1})\mathbf{d}_k + \tilde{l}_K\mathbf{1}\right) \\
&= \sum_{k=1}^{K-1} (\tilde{l}_k - \tilde{l}_{k+1})(\mathbf{X}'\mathbf{X})^{-1}\mathbf{X}'\mathbf{d}_k + \tilde{l}_K(\mathbf{X}'\mathbf{X})^{-1}\mathbf{X}'\mathbf{1} \\
&= \sum_{k=1}^{K-1} (\tilde{l}_k - \tilde{l}_{k+1})\hat{\boldsymbol{\beta}}_k^{(d)} + \tilde{l}_K(\mathbf{X}'\mathbf{X})^{-1}\mathbf{X}'\mathbf{1}.
\end{align*}

Since $\mathbf{X}$ includes a constant, the second term affects only the intercept. For any non-intercept coefficient $\hat{\beta}_m^{(\tilde{r})}$, we therefore obtain:

\begin{equation*}
    \hat{\beta}_m^{(\tilde{r})} = \sum_{k=1}^{K-1} (\tilde{l}_k - \tilde{l}_{k+1})\,\hat{\beta}_{km}^{(d)}.
\end{equation*}

As $\tilde{l}_k - \tilde{l}_{k+1} < 0$ for all positive monotonic transformations, if all $\hat{\beta}_{km}^{(d)}$ share the same sign, no positive monotonic transformation can reverse the sign of $\hat{\beta}_m^{(\tilde{r})}$. If they differ in sign, a transformation placing large weight on gaps where $\hat{\beta}_{km}^{(d)}$ takes the other sign will reverse $\hat{\beta}_m^{(\tilde{r})}$.

Finally, note that the normalisation $\sum_{k=1}^{K-1}\Delta_k = 1$ adopted elsewhere in the paper affects only the scale of $\hat{\beta}_m^{(\tilde{r})}$, not its sign or the ratio of any two non-intercept coefficients. The decomposition above therefore applies without loss of generality to the normalised case.

\subsection{Residual decomposition}
\label{sec:residual_derivation}

We show that the residuals from a regression of any transformed $\tilde{\mathbf{r}}$ on $\mathbf{X}$ can be expressed as a weighted combination of residuals from the dichotomised regressions. The residuals from the transformed regression are $\tilde{\mathbf{e}} = \tilde{\mathbf{r}} - \mathbf{X}\hat{\boldsymbol{\beta}}^{(\tilde{r})}$. From the decomposition derived in Appendix \ref{sec:decomposition}, we have $\tilde{\mathbf{r}} = \sum_{k=1}^{K-1}(\tilde{l}_k - \tilde{l}_{k+1})\mathbf{d}_k + \tilde{l}_K\mathbf{1}$ and $\hat{\boldsymbol{\beta}}^{(\tilde{r})} = \sum_{k=1}^{K-1}(\tilde{l}_k - \tilde{l}_{k+1})\hat{\boldsymbol{\beta}}_k^{(d)} + (\mathbf{X}'\mathbf{X})^{-1}\mathbf{X}'\tilde{l}_K\mathbf{1}$. Substituting:
\begin{align*}
\tilde{\mathbf{e}} &= \sum_{k=1}^{K-1}(\tilde{l}_k - \tilde{l}_{k+1})\mathbf{d}_k + \tilde{l}_K\mathbf{1} - \mathbf{X}\left(\sum_{k=1}^{K-1}(\tilde{l}_k - \tilde{l}_{k+1})\hat{\boldsymbol{\beta}}_k^{(d)} + (\mathbf{X}'\mathbf{X})^{-1}\mathbf{X}'\tilde{l}_K\mathbf{1}\right)\\
&= \sum_{k=1}^{K-1}(\tilde{l}_k - \tilde{l}_{k+1})(\mathbf{d}_k - \mathbf{X}\hat{\boldsymbol{\beta}}_k^{(d)}) + \tilde{l}_K\mathbf{1} - \tilde{l}_K\mathbf{X}(\mathbf{X}'\mathbf{X})^{-1}\mathbf{X}'\mathbf{1} = \sum_{k=1}^{K-1}(\tilde{l}_k - \tilde{l}_{k+1})\mathbf{e}_{dk}.
\end{align*}
The last equality follows because $\mathbf{X}(\mathbf{X}'\mathbf{X})^{-1}\mathbf{X}'$ is the projection matrix onto the column space of $\mathbf{X}$. Since $\mathbf{X}$ includes a constant, $\mathbf{1}$ lies in its column space, so $\mathbf{X}(\mathbf{X}'\mathbf{X})^{-1}\mathbf{X}'\mathbf{1} = \mathbf{1}$. This decomposition allows us to construct $\hat{\boldsymbol{\Omega}}$, and hence the standard errors and p-values, for any transformation using only results from the $K-1$ dichotomised regressions.

\section{Closed-form solutions for minimum-cost sign reversals and target ratios}
\label{sec:closed_form_proofs}

This section derives closed-form expressions for the minimum-cost transformation to reverse the sign of an estimated regression coefficient (Proposition \ref{prop:sign_reversal_closed_form}), and for the minimum-cost transformation that shifts the ratio of two coefficients to some target value (Proposition \ref{prop:ratio_closed_form}). 

Note that when the monotonicity constraint of the main text binds at the optimum, solutions analogous to Propositions \ref{prop:sign_reversal_closed_form} and \ref{prop:ratio_closed_form} can still be found. However, in that case the solutions do not yield a clear intuition about the ease of reversals, which is why we omit them here. See Supplementary Appendix \ref{sec:algorithm} for these more general cases. 

The same proof strategy is used for both propositions. We show that after some rewriting, the reversal constraint implies, via the Cauchy-Schwarz inequality, a lower bound on the amount of non-linearity in the scale. We then show that this lower bound is always attainable, making it the minimum. 

\subsection{Proof of Proposition \ref{prop:sign_reversal_closed_form}}
\label{sec:proofs_sign}
 
Throughout, $\Delta_k \equiv \tilde{l}_{k+1} - \tilde{l}_k$ denotes the gap between adjacent transformed labels, normalised so that $\sum_{k=1}^{K-1}\Delta_k = 1$. From the decomposition of Equation \eqref{eq:decomp}, the transformed coefficient satisfies $\hat\beta_m^{(\tilde{r})} = -\sum_k \Delta_k \hat\beta_{km}^{(d)}$. Under the linear coding, the original-scale gaps are $(l_K - l_1)/(K-1)$, giving $\hat\beta_m^{(r)} = -(l_K - l_1)\mu_b$, where $\mu_b \equiv \tfrac{1}{K-1}\sum_k \hat\beta_{km}^{(d)}$. Let $V_b \equiv \tfrac{1}{K-1}\sum_k (\hat\beta_{km}^{(d)} - \mu_b)^2$ denote the variance of the dichotomised coefficients.
 
It will be convenient to work with centred gaps $\delta_k \equiv \Delta_k - 1/(K-1)$. The constraint $\sum_k \Delta_k = 1$ then becomes $\sum_k \delta_k = 0$. Using $\text{maxVar}(\Delta) = (K-2)/(K-1)^2$ from Appendix \ref{sec:proof_maxvar}, the squared cost can be written as
\begin{equation}
\label{eq:C2_projection}
C(\boldsymbol{\Delta})^2 = \frac{\text{Var}(\boldsymbol{\Delta})}{\text{maxVar}(\boldsymbol{\Delta})} = \frac{K-1}{K-2}\sum_k \delta_k^2,
\end{equation}
so, by monotonicity, minimising $C$ is equivalent to minimising $\sum_k \delta_k^2$.\footnote{More generally, because the function $h$ in Proposition \ref{prop:plausibility} is positive monotonic, the optimal gap vector will be the same for all cost functions in the family identified by Proposition \ref{prop:plausibility} and that share the same $\psi$.}
 
We thus need to find the gap vector $\boldsymbol{\Delta}$ that minimises $\sum_k \delta_k^2$ subject to (a) the transformed coefficient equals zero: $\sum_k \Delta_k \hat\beta_{km}^{(d)} = 0$ (sign-reversal), and (b) the centred gaps sum to zero: $\sum_k \delta_k = 0$ (scale-preservation). From here, we proceed in four steps.
 
\paragraph{Step 1: Rewrite constraint (a) in centred variables.} Substituting $\Delta_k = \delta_k + 1/(K-1)$ into constraint (a):
$$\sum_k \left(\delta_k + \frac{1}{K-1}\right) \hat\beta_{km}^{(d)} = 0 \quad \Longrightarrow \quad \sum_k \delta_k \hat\beta_{km}^{(d)} + \frac{1}{K-1}\sum_k \hat\beta_{km}^{(d)} = 0.$$
The second term equals $\mu_b = -\hat\beta_m^{(r)}/(l_K - l_1)$, so $\sum_k \delta_k \hat\beta_{km}^{(d)} = \hat\beta_m^{(r)}/(l_K - l_1)$. Since $\sum_k \delta_k = 0$ by constraint (b), we can subtract $\mu_b \sum_k \delta_k = 0$ from the left-hand side without changing its value, giving:
\begin{equation}
\label{eq:sign_constraint}
\sum_{k=1}^{K-1} \delta_k(\hat\beta_{km}^{(d)} - \mu_b) = \hat\beta_m^{(r)}/(l_K - l_1).
\end{equation}
 
\paragraph{Step 2: Derive a lower bound on $\sum_k \delta_k^2$ from constraint (a).} By the Cauchy-Schwarz inequality applied to the sequences $(\delta_k)$ and $(\hat\beta_{km}^{(d)} - \mu_b)$:
$$\left(\sum_k \delta_k(\hat\beta_{km}^{(d)} - \mu_b)\right)^2 \leq \left(\sum_k \delta_k^2\right)\left(\sum_k(\hat\beta_{km}^{(d)} - \mu_b)^2\right).$$
The left-hand side equals $(\hat\beta_m^{(r)}/(l_K - l_1))^2$ by \eqref{eq:sign_constraint}, and $\sum_k(\hat\beta_{km}^{(d)} - \mu_b)^2 = (K-1)V_b$. Rearranging:
\begin{equation}
\label{eq:lower_bound}
\sum_k \delta_k^2 \geq \frac{(\hat\beta_m^{(r)})^2}{(K-1)(l_K - l_1)^2 V_b}.
\end{equation}
This lower bound is attained if and only if $\delta_k = \gamma(\hat\beta_{km}^{(d)} - \mu_b)$ for some scalar $\gamma$ (i.e. the equality condition of Cauchy-Schwarz). 
 
\paragraph{Step 3: Verify that the lower bound implied by Cauchy-Schwarz is feasible for constraint (b).} Constraint (b) requires $\sum_k \delta_k = 0$. Under the Cauchy-Schwarz equality case, $\delta_k = \gamma(\hat\beta_{km}^{(d)} - \mu_b)$, so:
$$\sum_k \delta_k = \gamma \sum_k (\hat\beta_{km}^{(d)} - \mu_b) = 0,$$
where the last equality holds because $(\hat\beta_{km}^{(d)} - \mu_b)$ are deviations from their own mean and therefore sum to zero by construction. Hence, the solution that attains the Cauchy-Schwarz lower bound automatically satisfies constraint (b).\footnote{Substantively this implies that the scale normalisation constraint is never binding at the optimum.} Since the lower bound \eqref{eq:lower_bound} is attainable while satisfying both constraints, it is the minimum.
 
\paragraph{Step 4: Find $\gamma$ and obtain $C^\star$.} Substituting $\delta_k = \gamma(\hat\beta_{km}^{(d)} - \mu_b)$ into \eqref{eq:sign_constraint}:
$$\gamma \sum_k (\hat\beta_{km}^{(d)} - \mu_b)^2 = \hat\beta_m^{(r)}/(l_K - l_1) \quad \Longrightarrow \quad \gamma = \frac{\hat\beta_m^{(r)}}{(K-1)(l_K - l_1)V_b}.$$
Translating back to gaps:
$$\Delta_k^\star = \frac{1}{K-1} + \gamma(\hat\beta_{km}^{(d)} - \mu_b) = \frac{1}{K-1} + \frac{\hat\beta_m^{(r)}(\hat\beta_{km}^{(d)} - \mu_b)}{(K-1)(l_K - l_1)V_b},$$
which is the expression given in the proposition. Substituting $\sum_k \delta_k^2 = (\hat\beta_m^{(r)})^2/[(K-1)(l_K - l_1)^2 V_b]$ into \eqref{eq:C2_projection} and taking the square-root yields $C^\star = |\hat\beta_m^{(r)}|/[(l_K - l_1)\sqrt{(K-2)V_b}]$.

\subsection{Proof of Proposition \ref{prop:ratio_closed_form}}
\label{sec:proofs_ratios_closed_form}
 
The ratio constraint $\hat\beta_m^{(\tilde{r})}/\hat\beta_n^{(\tilde{r})} = \rho$ rearranges to $\hat\beta_m^{(\tilde{r})} - \rho \, \hat\beta_n^{(\tilde{r})} = 0$. Using the decomposition of Equation \eqref{eq:decomp}:
$$\hat\beta_m^{(\tilde{r})} - \rho \, \hat\beta_n^{(\tilde{r})} = -\sum_k \Delta_k \, c_k,$$
where $c_k \equiv \hat\beta_{km}^{(d)} - \rho  \hat\beta_{kn}^{(d)}$. This is structurally identical to the sign-reversal problem with $\hat\beta_{km}^{(d)}$ replaced by $c_k$. The analogue of $\mu_b$ is $\mu_c \equiv \tfrac{1}{K-1}\sum_k c_k$, which by linearity equals $-(\hat\beta_m^{(r)} - \rho \, \hat\beta_n^{(r)})/(l_K - l_1)$, and the analogue of $V_b$ is $V_c \equiv \tfrac{1}{K-1}\sum_k(c_k - \mu_c)^2$. Substituting into the sign-reversal formula of Proposition \ref{prop:sign_reversal_closed_form} yields $C^\star_\rho = |\hat\beta_m^{(r)} - \rho \, \hat\beta_n^{(r)}| / [(l_K - l_1)\sqrt{(K-2)V_c}]$.
 
%\newpage
%\clearpage

\setlength{\bibsep}{1.25pt}
\bibliography{references}

\newpage
\clearpage

\appendix

\setcounter{figure}{0}
\setcounter{table}{0}
\setcounter{appendixproposition}{0}

\renewcommand{\thetable}{S\arabic{table}}
\renewcommand{\thefigure}{S\arabic{figure}}
\renewcommand{\theappendixproposition}{S\arabic{appendixproposition}}
\setcounter{section}{0}
\renewcommand{\thesection}{S\arabic{section}}

\begin{center}
\section*{Supplementary Appendix}
\end{center}

\section{Properties of the SD-based cost function}
\label{sec:sd_properties}

\subsection{The SD-based cost satisfies Proposition \ref{prop:plausibility}}
\label{sec:sd_satisfies_prop}

Set $\psi(x) = (x - 1/(K-1))^2$ and $h(x) = \sqrt{x}$ in Proposition \ref{prop:plausibility}. Then $\psi$ is continuous and strictly convex with a unique minimum at $1/(K-1)$, and $h:[0,1]\to[0,1]$ is continuous and increasing with $h(0)=0$, $h(1)=1$. Since $\sum_{k=1}^{K-1}\Delta_k = 1$, we have $\sum_{k=1}^{K-1}(\Delta_k - 1/(K-1))^2 = (K-1)\,\mathrm{Var}(\Delta)$. The same identity applied to the denominator gives $\max_{\Delta'}\sum(\Delta'_k - 1/(K-1))^2 = (K-1)\,\mathrm{maxVar}(\Delta)$. The $(K-1)$ factors cancel, so $C(\Delta) = h\bigl(\mathrm{Var}(\Delta)/\mathrm{maxVar}(\Delta)\bigr) = \mathrm{SD}(\Delta)/\mathrm{maxSD}(\Delta)$.

\subsection{Derivation of $\text{maxSD}(\Delta)$}
\label{sec:proof_maxvar}

Since $\sum_{k=1}^{K-1}\Delta_k = 1$, the mean gap is $\bar{\Delta} = 1/(K-1)$ and the variance of the gap vector is $\mathrm{Var}(\Delta) = \frac{1}{K-1}\sum_{k=1}^{K-1}(\Delta_k - \bar{\Delta})^2$. Subject to $\Delta_k \geq 0$ and $\sum_k \Delta_k = 1$, this is maximised when $\Delta_j = 1$ for some $j$ and $\Delta_k = 0$ for all $k \neq j$. Hence:
\begin{align*}
\mathrm{maxVar}(\Delta) &= \frac{1}{K-1}\left[\left(1 - \frac{1}{K-1}\right)^2 + (K-2)\left(\frac{1}{K-1}\right)^2\right] \\
&= \frac{1}{K-1}\left[\frac{(K-2)^2}{(K-1)^2} + \frac{K-2}{(K-1)^2}\right] = \frac{K-2}{(K-1)^3}\left[(K-2) + 1\right] = \frac{K-2}{(K-1)^2}.
\end{align*}
Therefore $\mathrm{maxSD}(\Delta) = \sqrt{K-2}\,/\,(K-1)$.

\subsection{Linear homogeneity}
\label{sec:homogen}

Let $\mathbf{u} = (1/(K-1), \ldots, 1/(K-1))$ denote the uniform gap vector and define $\Delta^{(\lambda)} = \lambda\Delta + (1-\lambda)\mathbf{u}$ for $\lambda \in [0,1]$. Then $C(\Delta^{(\lambda)}) = \lambda \, C(\Delta)$. To see this, note that $\Delta^{(\lambda)}_k - 1/(K-1) = \lambda(\Delta_k - 1/(K-1))$, so:
$$C(\Delta^{(\lambda)}) = \sqrt{\frac{\lambda^2 \sum_{k=1}^{K-1} (\Delta_k - 1/(K-1))^2}{\max_{\Delta'} \sum_{k=1}^{K-1} (\Delta'_k - 1/(K-1))^2}} = \lambda \, C(\Delta).$$
Thus, mixing any gap vector with the uniform vector by weight $\lambda$ scales the cost by $\lambda$.

More generally, the same property holds for the normalised $L^p$ family of cost functions,
$$C(\Delta) = \left(\frac{\frac{1}{K-1}\sum_{k=1}^{K-1}|\Delta_k - 1/(K-1)|^p}{\max_{\Delta'}\frac{1}{K-1}\sum_{k=1}^{K-1}|\Delta'_k - 1/(K-1)|^p}\right)^{1/p}, \quad p > 1,$$
of which our SD-based cost is the special case $p=2$. The argument remains the same since $|\Delta_k^{(\lambda)} - 1/(K-1)|^p = \lambda^p\,|\Delta_k - 1/(K-1)|^p$, the normalised ratio scales as $\lambda^p$, and the outer exponent $1/p$ yields $C(\Delta^{(\lambda)}) = \lambda\,C(\Delta)$.

\section{Decompositions for fixed effects, 2SLS, and continuous outcomes}
\label{sec:analytical}

Appendix \ref{sec:decomposition} established the decomposition
\begin{equation*}
\hat{\beta}_m^{(\tilde{r})} = \sum_{k=1}^{K-1} (\tilde{l}_k - \tilde{l}_{k+1})\,\hat{\beta}_{km}^{(d)}
\end{equation*}
for OLS estimators applied to a discrete ordered outcome. In what follows, we show that an analogous decomposition holds for fixed-effects and 2SLS estimators, as well as for the case in which the underlying response variable $r_i$ is continuous. As a result, the methods of Section \ref{sec:optimisation} extend immediately to all of these cases.

\subsection{Fixed-effects estimator}
\label{sec:proof_fixed_effects}

Suppose we have panel data on respondents $i$ observed in periods $t = 1, \ldots, T_i$. Let $\bar{\mathbf{X}}$, $\bar{\tilde{\mathbf{r}}}$, and $\bar{\mathbf{d}}_k$ collect the within-person means of $\mathbf{X}$, $\tilde{\mathbf{r}}$, and $\mathbf{d}_k$, respectively, and let $\dot{\mathbf{X}} \equiv \mathbf{X} - \bar{\mathbf{X}}$, $\dot{\tilde{\mathbf{r}}} \equiv \tilde{\mathbf{r}} - \bar{\tilde{\mathbf{r}}}$, and $\dot{\mathbf{d}}_k \equiv \mathbf{d}_k - \bar{\mathbf{d}}_k$ denote the corresponding demeaned quantities. The fixed-effects estimator is $\hat{\boldsymbol{\beta}}^{(\tilde{r}),FE} = (\dot{\mathbf{X}}'\dot{\mathbf{X}})^{-1}\dot{\mathbf{X}}'\dot{\tilde{\mathbf{r}}}$. Because the transformation $f$ is applied identically to $r_{it}$ at every $(i,t)$, the identity from Appendix \ref{sec:decomposition} holds observation by observation:
\begin{equation*}
\tilde{r}_{it} = \sum_{k=1}^{K-1}(\tilde{l}_k - \tilde{l}_{k+1})\,d_{k,it} + \tilde{l}_K.
\end{equation*}
Taking within-person means and differencing, the constant $\tilde{l}_K$ cancels:
\begin{equation*}
\dot{\tilde{r}}_{it} = \sum_{k=1}^{K-1}(\tilde{l}_k - \tilde{l}_{k+1})\,\dot{d}_{k,it}.
\end{equation*}
Stacking observations and substituting into the FE estimator:
\begin{equation*}
\hat{\boldsymbol{\beta}}^{(\tilde{r}),FE} = (\dot{\mathbf{X}}'\dot{\mathbf{X}})^{-1}\dot{\mathbf{X}}'\sum_{k=1}^{K-1}(\tilde{l}_k - \tilde{l}_{k+1})\,\dot{\mathbf{d}}_k = \sum_{k=1}^{K-1}(\tilde{l}_k - \tilde{l}_{k+1})\,\hat{\boldsymbol{\beta}}_k^{(d),FE},
\end{equation*}
where $\hat{\boldsymbol{\beta}}_k^{(d),FE} \equiv (\dot{\mathbf{X}}'\dot{\mathbf{X}})^{-1}\dot{\mathbf{X}}'\dot{\mathbf{d}}_k$ is the FE estimate from a regression of $d_{k,it}$ on $\mathbf{X}_{it}$ with individual fixed effects. The same decomposition therefore holds element-by-element for $\hat{\beta}_m^{(\tilde{r}),FE}$.

\subsection{2SLS estimator}
\label{sec:proof_2sls}
 
Let $\mathbf{Z}$ collect the instruments, with the constant in $\mathbf{X}$ included as its own instrument. In the just-identified case, $\hat{\boldsymbol{\beta}}^{(\tilde{r}),IV} = (\mathbf{Z}'\mathbf{X})^{-1}\mathbf{Z}'\tilde{\mathbf{r}}$ and $\hat{\boldsymbol{\beta}}_k^{(d),IV} = (\mathbf{Z}'\mathbf{X})^{-1}\mathbf{Z}'\mathbf{d}_k$. Analogously to the OLS case:
\begin{align*}
\hat{\boldsymbol{\beta}}^{(\tilde{r}),IV} &= (\mathbf{Z}'\mathbf{X})^{-1}\mathbf{Z}'\tilde{\mathbf{r}} = (\mathbf{Z}'\mathbf{X})^{-1}\mathbf{Z}'\left(\sum_{k=1}^{K-1}(\tilde{l}_k - \tilde{l}_{k+1})\mathbf{d}_k + \tilde{l}_K\mathbf{1}\right) \\
&= \sum_{k=1}^{K-1}(\tilde{l}_k - \tilde{l}_{k+1})\,\hat{\boldsymbol{\beta}}_k^{(d),IV} + \tilde{l}_K(\mathbf{Z}'\mathbf{X})^{-1}\mathbf{Z}'\mathbf{1}.
\end{align*}
The term $\tilde{l}_K(\mathbf{Z}'\mathbf{X})^{-1}\mathbf{Z}'\mathbf{1}$ is the IV estimate from a regression of the constant $\tilde{l}_K\mathbf{1}$ on $\mathbf{X}$ using $\mathbf{Z}$, which affects only the intercept. Hence, all non-intercept elements satisfy $\hat{\beta}_m^{(\tilde{r}),IV} = \sum_{k=1}^{K-1}(\tilde{l}_k - \tilde{l}_{k+1})\,\hat{\beta}_{km}^{(d),IV}$. 

\subsection{Continuous outcomes}
\label{sec:proof_continuous}
 
Suppose now that $r_i$ is continuous with support $[r_{\min}, r_{\max}]$, and let $f:[r_{\min}, r_{\max}]\to\mathbb{R}$ be any continuously differentiable, strictly increasing function. Define $\tilde{r}_i = f(r_i)$. By the fundamental theorem of calculus,
\begin{equation*}
f(r_i) = f(r_{\max}) - \int_{r_i}^{r_{\max}} f'(t)\,dt = f(r_{\max}) - \int_{r_{\min}}^{r_{\max}} f'(t)\,\mathds{1}\{r_i \leq t\}\,dt.
\end{equation*}
Stacking observations,
\begin{equation*}
\tilde{\mathbf{r}} = f(r_{\max})\,\mathbf{1} - \int_{r_{\min}}^{r_{\max}} f'(t)\,\mathbf{d}(t)\,dt,
\end{equation*}
where $\mathbf{d}(t)$ is the vector with $i$-th entry $\mathds{1}\{r_i \leq t\}$. Substituting into the OLS estimator:
\begin{equation*}
\hat{\boldsymbol{\beta}}^{(\tilde{r})} = (\mathbf{X}'\mathbf{X})^{-1}\mathbf{X}'\tilde{\mathbf{r}} = f(r_{\max})\,(\mathbf{X}'\mathbf{X})^{-1}\mathbf{X}'\mathbf{1} - \int_{r_{\min}}^{r_{\max}} f'(t)\,\hat{\boldsymbol{\beta}}^{(d)}(t)\,dt,
\end{equation*}
where $\hat{\boldsymbol{\beta}}^{(d)}(t) \equiv (\mathbf{X}'\mathbf{X})^{-1}\mathbf{X}'\mathbf{d}(t)$ is the OLS coefficient vector from the regression of $\mathds{1}\{r_i \leq t\}$ on $\mathbf{X}_i$. As before, the first term affects only the intercept. For any non-intercept coefficient,
\begin{equation*}
\hat{\beta}_m^{(\tilde{r})} = -\int_{r_{\min}}^{r_{\max}} f'(t)\,\hat{\beta}_m^{(d)}(t)\,dt.
\end{equation*}
This is the continuous analogue of the discrete decomposition: $\hat{\beta}_m^{(\tilde{r})}$ is a weighted integral of the dichotomised coefficients $\hat{\beta}_m^{(d)}(t)$, with weights given by the slope $f'(t)$ of the transformation. In practice, since $\hat{\beta}_m^{(d)}(t)$ is piecewise constant in $t$ and changes only at observed values of $r_i$, it suffices to run one dichotomised regression per observed value, which reduces the integral to a finite sum.

\section{Bounds on coefficient ratios}
\label{sec:proof_ratios}
In Section \ref{sec:ratio_method} of the main text, we made reference to the following: 

\begin{appendixproposition}[Bounded coefficient ratios]
\label{prop:ratios}
If and only if $\hat{\beta}_n^{(\tilde{r})}$ in the denominator is not reversible across all positive monotonic transformations of $r_i$, the ratio $\hat{\beta}_m^{(\tilde{r})}/\hat{\beta}_n^{(\tilde{r})}$ is bounded by the minimum and maximum values of $\hat{\beta}_{km}^{(d)}/\hat{\beta}_{kn}^{(d)}$ across all $k = 1, \ldots, K-1$.
\end{appendixproposition}

To prove this, we begin with the decomposition of Equation \eqref{eq:decomp}: $\hat{\beta}_m^{(\tilde{r})} = \sum_{k=1}^{K-1} (\tilde{l}_k - \tilde{l}_{k+1})\hat{\beta}_{km}^{(d)}$. For any two covariates $m$ and $n$, the ratio of their coefficients is $\frac{\hat{\beta}_m^{(\tilde{r})}}{\hat{\beta}_n^{(\tilde{r})}} = \frac{\sum_{k=1}^{K-1} (\tilde{l}_k - \tilde{l}_{k+1})\hat{\beta}_{km}^{(d)}}{\sum_{k=1}^{K-1} (\tilde{l}_k - \tilde{l}_{k+1})\hat{\beta}_{kn}^{(d)}}$.

There are two cases, depending on whether the coefficient in the denominator is reversible.

\subsubsection*{Case 1: $\hat{\beta}_n^{(\tilde{r})}$ is not reversible}

If $\hat{\beta}_n^{(\tilde{r})}$ is not reversible across all positive monotonic transformations, then by the non-reversibility condition established in the preceding section, all $\hat{\beta}_{kn}^{(d)}$ share the same sign. First assume that all $\hat{\beta}_{kn}^{(d)}>0$. Let $\tilde{w}_k = -(\tilde{l}_k - \tilde{l}_{k+1})$. Given that $\tilde{l}_k - \tilde{l}_{k+1} < 0$ for all positive monotonic transformations, we have $\tilde{w}_k > 0$. Then:
\begin{align*}
\frac{\hat{\beta}_m^{(\tilde{r})}}{\hat{\beta}_n^{(\tilde{r})}} 
&= 
\frac{\sum_{k=1}^{K-1} \tilde{w}_k\hat{\beta}_{km}^{(d)}}{\sum_{k=1}^{K-1} \tilde{w}_k\hat{\beta}_{kn}^{(d)}}
=
\frac{\sum_{k=1}^{K-1} \tilde{w}_k\hat{\beta}_{kn}^{(d)} \frac{\hat{\beta}_{km}^{(d)}}{\hat{\beta}_{kn}^{(d)}}}{\sum_{k=1}^{K-1} \tilde{w}_k\hat{\beta}_{kn}^{(d)}}
=
\frac{1}{\sum_{k=1}^{K-1} \tilde{w}_k\hat{\beta}_{kn}^{(d)}}\sum_{k=1}^{K-1}
\tilde{w}_k\hat{\beta}_{kn}^{(d)}
\frac{\hat{\beta}_{km}^{(d)}}{\hat{\beta}_{kn}^{(d)}}\\
&=
\sum_{k=1}^{K-1}
\frac{\tilde{w}_k\hat{\beta}_{kn}^{(d)}}{\sum_{j=1}^{K-1} \tilde{w}_j\hat{\beta}_{jn}^{(d)}}\frac{\hat{\beta}_{km}^{(d)}}{\hat{\beta}_{kn}^{(d)}}
=
\sum_{k=1}^{K-1}
\alpha_k \frac{\hat{\beta}_{km}^{(d)}}{\hat{\beta}_{kn}^{(d)}}
\end{align*}

Since $\tilde{w}_k > 0$ and $\hat{\beta}_{kn}^{(d)} > 0$ for all $k$ (by assumption), we have $\alpha_k > 0$ for all $k$. Additionally, $\sum_{k=1}^{K-1} \alpha_k = 1$. Therefore, the ratio $\hat{\beta}_m^{(\tilde{r})}/\hat{\beta}_n^{(\tilde{r})}$ is a convex combination of the ratios $\hat{\beta}_{km}^{(d)}/\hat{\beta}_{kn}^{(d)}$, where the weights are given by $\alpha_k \equiv \frac{\tilde{w}_k\hat{\beta}_{kn}^{(d)}}{\sum_{j=1}^{K-1} \tilde{w}_j\hat{\beta}_{jn}^{(d)}}$. Thus, the ratio $\hat{\beta}_m^{(\tilde{r})}/\hat{\beta}_n^{(\tilde{r})}$ must lie between the minimum and maximum values of $\hat{\beta}_{km}^{(d)}/\hat{\beta}_{kn}^{(d)}$:
$$\min_{k} \frac{\hat{\beta}_{km}^{(d)}}{\hat{\beta}_{kn}^{(d)}} < \frac{\hat{\beta}_m^{(\tilde{r})}}{\hat{\beta}_n^{(\tilde{r})}} < \max_{k} \frac{\hat{\beta}_{km}^{(d)}}{\hat{\beta}_{kn}^{(d)}}.$$

By choosing appropriate values for $\tilde{w}_k$ (which corresponds to choosing an appropriate positive monotonic transformation), we can make the ratio $\hat{\beta}_m^{(\tilde{r})}/\hat{\beta}_n^{(\tilde{r})}$ arbitrarily close to either bound. For example, to approach the maximum value $\max_{k} \frac{\hat{\beta}_{km}^{(d)}}{\hat{\beta}_{kn}^{(d)}}$, we could choose a transformation where $\tilde{w}_k$ is very large for the $k$ that maximises $\frac{\hat{\beta}_{km}^{(d)}}{\hat{\beta}_{kn}^{(d)}}$ and very small for all other values of $k$. Finally, when the signs of the $\hat{\beta}_{kn}^{(d)}$ are all negative, the same argument applies, except that the inequalities are reversed due to the negative sign in the denominator. However, the bounds remain the same.

\subsubsection*{Case 2: $\hat{\beta}_n^{(\tilde{r})}$ is reversible}

Now consider the case where $\hat{\beta}_n^{(\tilde{r})}$ can be reversed by some positive monotonic transformation. In that case, the ratio $\frac{\hat{\beta}_m^{(\tilde{r})}}{\hat{\beta}_n^{(\tilde{r})}}$ is not bounded. To see this, note that since $\hat{\beta}_n^{(\tilde{r})}$ is reversible, we can find a transformation such that $\hat{\beta}_n^{(\tilde{r})} = \varepsilon$ for some arbitrarily small $\varepsilon > 0$. Depending on the sign of $\hat{\beta}_m^{(\tilde{r})}$ for that transformation, this will cause the ratio $\frac{\hat{\beta}_m^{(\tilde{r})}}{\hat{\beta}_n^{(\tilde{r})}}$ to be arbitrarily large negative (for $\hat{\beta}_m^{(\tilde{r})}<0$) or positive (for $\hat{\beta}_m^{(\tilde{r})} \geq 0$). By the same argument, we can always find another transformation such that $\hat{\beta}_n^{(\tilde{r})} = \epsilon$ for some arbitrarily small $\epsilon < 0$, and obtain an arbitrarily large positive (for $\hat{\beta}_m^{(\tilde{r})}<0$) or large negative (for $\hat{\beta}_m^{(\tilde{r})} \geq 0$) ratio.\footnote{We exclude the degenerate case where $\hat{\beta}_m^{(\tilde{r})}$ switches sign for exactly the same transformation as $\hat{\beta}_n^{(\tilde{r})}$.}

\section{Computing minimum costs when monotonicity binds}
\label{sec:algorithm}

Propositions \ref{prop:sign_reversal_closed_form} and \ref{prop:ratio_closed_form} give closed-form expressions for the minimum-cost sign-reversing and ratio-targeting transformations \emph{under the assumption that the monotonicity constraint (Constraint 2) is non-binding at the optimum}. This appendix describes how to handle the case when the monotonicity constraint is binding. 

The gap vector $\Delta$ lies in the simplex $\mathcal{D} = \{\Delta \in \mathbb{R}^{K-1}_+ : \sum_k \Delta_k = 1\}$. When monotonicity binds, the optimal gap vector lies on the boundary of $\mathcal{D}$, with $\Delta_k = 0$ for some indices $k$. We refer to the subset of indices for which $\Delta_k = 0$ as the \emph{active set} $A \subseteq \{1, \ldots, K-1\}$, and to the corresponding subset of $\mathcal{D}$ as a \emph{face} of the simplex.

\subsection{Closed-form solution on a single face}

For any $A \subseteq \{1, \ldots, K-1\}$, let $I \equiv \{1, \ldots, K-1\} \setminus A$ and $M \equiv |I|$. We refer to indices in $A$ as \textit{active} and to indices in $I$ as \textit{free}. Let
\begin{equation*}
\mu_I \;\equiv\; \frac{1}{M}\sum_{k \in I} \hat\beta_{km}^{(d)}, \qquad
V_I  \;\equiv\; \frac{1}{M}\sum_{k \in I} \bigl(\hat\beta_{km}^{(d)} - \mu_I\bigr)^{2}
\end{equation*}
denote the mean and variance of the dichotomised coefficients restricted to the free indices. We assume throughout that $M \geq 2$ and $V_I > 0$. Faces violating either condition are referred to as \emph{degenerate} and need not be considered.

\begin{appendixproposition}[Minimum-cost sign reversal on a single face]
\label{prop:face_solution}
Suppose the face $A$ is non-degenerate, and suppose the monotonicity restriction on the free indices is non-binding at the optimum of the face problem. Then the minimum-cost gap vector on the face $A$ under the SD-based cost function is
\begin{equation*}
\Delta^\star_{A,k} \;=\;
\begin{cases}
\dfrac{1}{M} \;-\; \dfrac{\mu_I\,\bigl(\hat\beta_{km}^{(d)} - \mu_I\bigr)}{M\,V_I}, & k \in I, \\[8pt]
0, & k \in A,
\end{cases}
\end{equation*}
with associated cost $C^\star_A = C(\Delta^\star_A)$.
\end{appendixproposition}

\noindent \textit{Proof}. By definition of the active set, $\Delta_k = 0$ for $k \in A$. The remaining problem is to choose $(\Delta_k)_{k \in I}$ to minimise $C(\Delta)$ subject to $\sum_{k\in I}\Delta_k = 1$ and $\sum_{k \in I}\Delta_k\,\hat\beta_{km}^{(d)} = 0$. This is structurally identical to the problem solved in Appendix \ref{sec:proofs_sign}, with $K-1$ replaced by $M$, $\mu_b$ replaced by $\mu_I$, and $V_b$ replaced by $V_I$. Following Steps 1-4 of that proof on the free indices yields the stated expression for $\Delta^\star_{A,k}$ on $I$. $C^\star_A$ then follows from evaluating $C$ at $\Delta^\star_A$.

The closed-form $\Delta^\star_A$ is the minimum on the face \emph{ignoring monotonicity on the free indices}. The full problem additionally requires $\Delta^\star_{A,k} \geq 0$ for all $k \in I$. We say the face $A$ is \emph{feasible} if this condition holds at $\Delta^\star_A$. The global minimum is then
\begin{equation}
\label{eq:global_min}
C^\star \;=\; \min\bigl\{\, C^\star_A \,:\, A \subseteq \{1,\ldots,K-1\},\; A \text{ is feasible and non-degenerate} \,\bigr\}.
\end{equation}

\subsection{Pruning the search over faces}

In principle, $C^\star$ can be obtained by enumerating all $2^{K-1}$ subsets $A \subseteq \{1, \ldots, K-1\}$, solving each face via Proposition \ref{prop:face_solution}, and retaining the minimum cost among those that are feasible and non-degenerate. This approach becomes intractable for large $K$.

To make the search more efficient, we note that for any non-degenerate faces $A \subseteq A'$, we have $C^\star_{A} \leq C^\star_{A'}$. To see this, note that solving the problem on $A'$ differs from solving it on $A$ only by the addition of equality constraints $\Delta_k = 0$ for $k \in A' \setminus A$, and adding equality constraints to a minimisation problem weakly enlarges the optimal value. Therefore, once any feasible point with cost $C^\circ$ has been discovered, every face $A$ with $C^\star_A \geq C^\circ$ can be discarded together with all its supersets, since no superset can achieve a lower cost. This procedure is summarised in Algorithm \ref{alg:main}. There, $\textsc{SolveFace}(A)$ denotes evaluation of Proposition \ref{prop:face_solution} on the active set $A$, returning either the pair $(\Delta^\star_A, C^\star_A)$ or a \textsc{degenerate} flag. The algorithm returns the global minimum $C^\star$ of \eqref{eq:global_min} if it exists.\footnote{Our Stata implementation \texttt{coeff\_reverser} additionally uses a simple heuristic to obtain an initial feasible point before the breadth-first search begins, which further facilitates pruning.}

\begin{algorithm}[H]
\caption{Minimum-cost sign-reversing transformation.}
\label{alg:main}
\begin{algorithmic}[1]
\Require Dichotomised coefficients $\bigl(\hat\beta_{km}^{(d)}\bigr)_{k=1}^{K-1}$.
\Ensure $(\Delta^\star, C^\star)$, or \textsc{none} if no feasible reversal exists.
\Statex
\State $(\Delta^\star, C^\star) \gets (\bot, +\infty)$
\Statex
\State \textit{// Level 0: try the unconstrained interior solution.}
\State $\textsc{res} \gets \Call{SolveFace}{\emptyset}$
\If{$\textsc{res} = \textsc{degenerate}$} \Return \textsc{none} \EndIf
\State $(\Delta_0, C_0) \gets \textsc{res}$
\If{$\min_j \Delta_{0,j} \geq 0$} \Return $(\Delta_0, C_0)$ \Comment{Unconstrained solution is feasible.}
\EndIf
\Statex
\State \textit{// Otherwise, breadth-first search over faces with cost-based pruning.}
\State $\mathcal{F} \gets \{(\emptyset, C_0)\}$
\For{$L = 1, \ldots, K-2$}
  \State $\mathcal{F} \gets \bigl\{(A,C) \in \mathcal{F} : C < C^\star\bigr\}$ \Comment{Re-prune frontier with latest $C^\star$.}
  \If{$\mathcal{F} = \emptyset$} \textbf{break} \EndIf
  \State $\mathcal{C} \gets$ unique $A \cup \{k\}$ across $(A,\cdot) \in \mathcal{F}$ and $k \notin A$
  \State $\mathcal{F}' \gets \emptyset$
  \For{$A \in \mathcal{C}$}
    \State $\textsc{res} \gets \Call{SolveFace}{A}$
    \If{$\textsc{res} = \textsc{degenerate}$} \textbf{continue} \EndIf
    \State $(\Delta, C) \gets \textsc{res}$
    \If{$C \geq C^\star$} \textbf{continue} \Comment{Cost-based pruning.}
    \EndIf
    \If{$\min_j \Delta_j \geq 0$}
      \State $(\Delta^\star, C^\star) \gets (\Delta, C)$ \Comment{New feasible incumbent.}
    \Else
      \State $\mathcal{F}' \gets \mathcal{F}' \cup \bigl\{(A, C)\bigr\}$
    \EndIf
  \EndFor
  \State $\mathcal{F} \gets \mathcal{F}'$
\EndFor
\State \Return $(\Delta^\star, C^\star)$
\end{algorithmic}
\end{algorithm}

\subsection{Extension to coefficient ratios}
\label{sec:mrs_full_analytic}

An analogous result holds for the ratio-targeting problem. Recall from Appendix \ref{sec:proofs_ratios_closed_form} that, defining $c_k \equiv \hat\beta_{km}^{(d)} - \rho\,\hat\beta_{kn}^{(d)}$, the ratio-targeting problem has the same structure as the sign-reversal problem with the dichotomised coefficients $\hat\beta_{km}^{(d)}$ replaced by $c_k$. Applying this substitution to Proposition \ref{prop:face_solution} yields the analogue for binding monotonicity.

\begin{appendixproposition}[Minimum-cost ratio targeting on a single face]
\label{prop:face_solution_ratio}
Let $c_k \equiv \hat\beta_{km}^{(d)} - \rho\,\hat\beta_{kn}^{(d)}$, and let
\begin{equation*}
\mu_I^{(c)} \;\equiv\; \frac{1}{M}\sum_{k \in I} c_k, \qquad
V_I^{(c)}  \;\equiv\; \frac{1}{M}\sum_{k \in I} \bigl(c_k - \mu_I^{(c)}\bigr)^{2}.
\end{equation*}
Suppose the face $A$ is non-degenerate, and suppose the monotonicity restriction on the free indices is non-binding at the optimum of the face. Then the minimum-cost gap vector for attaining the target ratio $\rho$ on the face $A$ is
\begin{equation*}
\Delta^\star_{A,k} \;=\;
\begin{cases}
\dfrac{1}{M} \;-\; \dfrac{\mu_I^{(c)}\bigl(c_k - \mu_I^{(c)}\bigr)}{M\,V_I^{(c)}}, & k \in I, \\[8pt]
0, & k \in A,
\end{cases}
\end{equation*}
with associated cost $C^{\star,\rho}_A = C(\Delta^\star_A)$.
\end{appendixproposition}

Algorithm \ref{alg:main} therefore applies with the only modification that $\textsc{SolveFace}$ is evaluated using $c_k$ in place of $\hat\beta_{km}^{(d)}$. This is the procedure implemented in our Stata routine \texttt{mrs\_reverser}.

\section{Effort adjustment in the slider task}
\label{sec:correction}

As sliders are initialised at random positions, respondents must actively move them to report their intended scale use. If adjustment is effortful, final positions will be partially anchored to their starting values. We model this and exploit the randomisation to correct for it.

Recall that each respondent $i$ positions $K-2$ sliders to indicate their perceived labels for the response categories. Let $l_{i,k}^+$ denote respondent $i$'s intended label for category $k$, $l_{i,0,k}$ the randomised initial position of the corresponding slider, and $l_{i,k}$ the reported position. We assume that $i$ chooses $l_{i,k}$ to minimise a loss $L_i$ that trades off accuracy against effort:
\begin{equation}
\label{eq:slider_loss}
L_i(l_{i,k}) = (l_{i,k}^+ - l_{i,k})^2 + \gamma_i(l_{i,0,k} - l_{i,k})^2, \qquad \gamma_i > 0.
\end{equation}
The first-order condition yields:
\begin{equation}
\label{eq:shrinkage}
l_{i,k} = (1 - \lambda_i)\,l_{i,k}^+ + \lambda_i\,l_{i,0,k},
\end{equation}
where $\lambda_i \equiv \gamma_i/(1 + \gamma_i) \in [0,1)$. When $\lambda_i = 0$ the respondent fully reveals their intended scale. As $\lambda_i$ increases, the reported position shrinks toward the random starting value.

To estimate $\lambda_i$, note that \eqref{eq:shrinkage} implies a linear relationship between $l_{i,k}$ and $l_{i,0,k}$ with an individual-specific slope equal to $\lambda_i$. We make use of this by estimating via OLS:
\begin{equation}
\label{eq:slider_regression}
l_{i,k} = \alpha_k + \mu_i + \hat{\lambda}_i l_{i,0,k} + \varepsilon_{i,k}.
\end{equation}

Here, $\alpha_k$ are slider fixed effects, $\mu_i$ are individual fixed effects, and $\hat{\lambda}_i$ is an individual-specific slope on the initial position. To see that $\hat{\lambda}_i$ is unbiased for $\lambda_i$, rewrite \eqref{eq:shrinkage} as $l_{i,k} = \lambda_i l_{i,0,k} + u_{i,k}$, where $u_{i,k} \equiv (1 - \lambda_i)\,l_{i,k}^+$. The error $u_{i,k}$ thus contains the intended labels $l_{i,k}^+$, which are correlated with $l_{i,0,k}$ in levels (both increase in $k$). However, the slider fixed effects $\alpha_k$ absorb $\bar{l}_{\cdot,0,k}$, i.e. the cross-respondent mean of the initial position at each slider. The identifying variation in $l_{i,0,k}$ therefore reduces to $\tilde{l}_{i,0,k} \equiv l_{i,0,k} - \bar{l}_{\cdot,0,k}$, which reflects only the respondent-specific draw from the randomisation. By design, $\tilde{l}_{i,0,k} \perp l_{i,k}^+$, so $\tilde{l}_{i,0,k} \perp u_{i,k}$ and $\hat{\lambda}_i$ is unbiased for $\lambda_i$ (but noisy, given that we rely on just $9$ observations per respondent).

The individual fixed effects $\mu_i$ improve efficiency. To see this, decompose the error into its within-respondent mean and a deviation: $u_{i,k} = (1-\lambda_i)\,\bar{l}_{i,\cdot}^+ + (1-\lambda_i)(l_{i,k}^+ - \bar{l}_{i,\cdot}^+)$. The first term is constant across $k$ and therefore uncorrelated with $\tilde{l}_{i,0,k}$ (which has mean zero across $k$ for each $i$), so it does not bias $\hat{\lambda}_i$, but it inflates the residual variance. 

Given $\hat{\lambda}_i$, corrected labels are recovered as $\hat{l}_{i,k}^+ = \frac{l_{i,k} - \hat{\lambda}_i  l_{i,0,k}}{1 - \hat{\lambda}_i}$. From these corrected labels, we compute the gap vector $\Delta_{i,k} = \hat{l}_{i,k+1}^+ - \hat{l}_{i,k}^+$ and the individual cost $C_i^+$ in the usual way.

\section{Further evidence on individual scale use}
\label{sec:evidence_indiv_appendix}

This Appendix provides additional evidence on individual-level scale use using the interactive slider elicitation described in Section \ref{sec:scale_use_evidence}.

\subsection{Heterogeneous shapes of scale use}

The main text shows that the overall cost measure $C_i^+$ is only weakly related to standard socio-demographic characteristics. However, this summary index may conceal heterogeneity in the \emph{shape} of scale use.

To investigate this, we estimate separate regressions for each adjacent-category distance:
\begin{equation*}
\Delta_{ik} = X_i' \gamma_k + u_{ik},
\end{equation*} 
for all $k$.

\begin{figure}[t]
    \caption{Predictors of adjacent-category distances}
    \centering
    \label{fig:indiv_c_2}
    \includegraphics[width=0.9\textwidth]{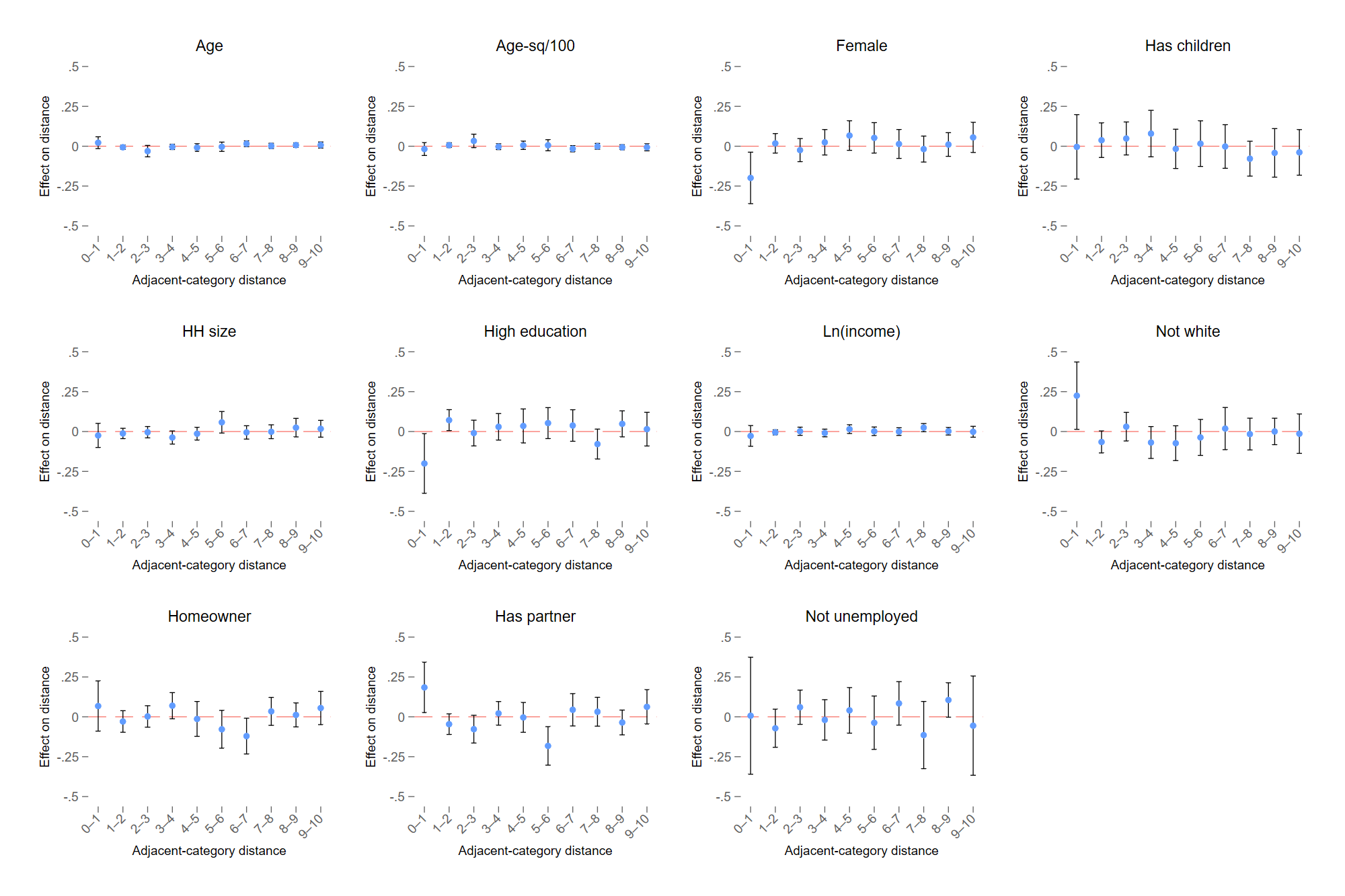}
    \begin{minipage}{0.975\textwidth}
    \small
    \noindent \footnotesize{\textbf{Notes:} This figure reports coefficient estimates from regressions of adjacent-category distances on respondent characteristics. Each panel corresponds to a separate covariate, and each point represents the estimated coefficient from a regression where the dependent variable is the distance between a given pair of adjacent categories (e.g. 0–1, 1–2, …, 9–10). Dots indicate point estimates and vertical lines show 95\% confidence intervals. A flat profile across adjacent categories indicates no systematic variation in local scale use, while deviations across the scale reflect heterogeneous non-linearities.} 
    \end{minipage}
\end{figure}

\begin{figure}[t]
    \caption{Predictors of adjusted adjacent-category distances}
    \centering
    \label{fig:indiv_c_2_star}
    \includegraphics[width=0.9\textwidth]{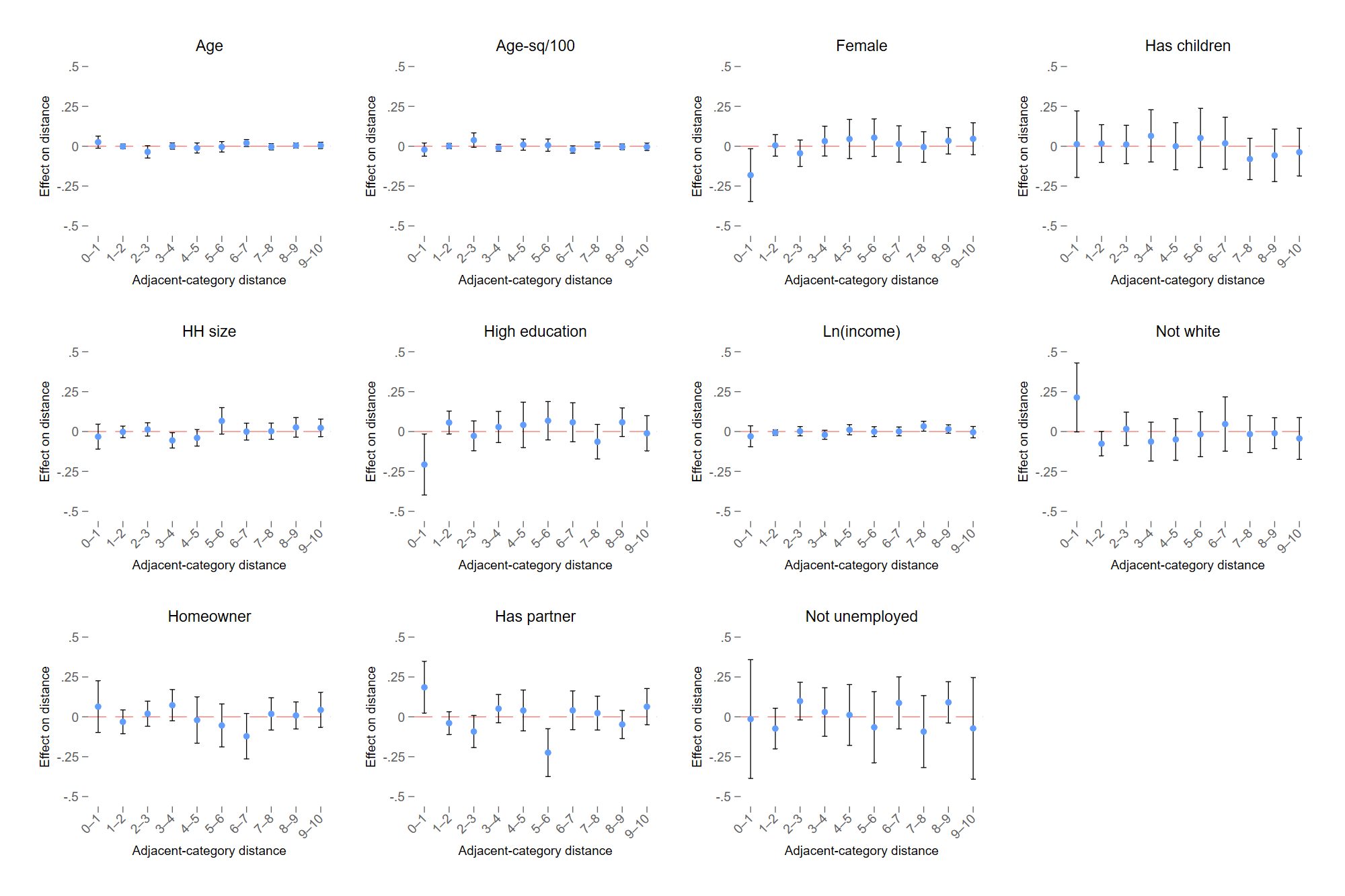}
    \begin{minipage}{0.975\textwidth}
    \small
    \noindent \footnotesize{\textbf{Notes:} This figure reports coefficient estimates from regressions of adjusted adjacent-category distances on respondent characteristics. Each panel corresponds to a separate covariate, and each point represents the estimated coefficient from a regression where the dependent variable is the distance between a given pair of adjacent categories (e.g. 0–1, 1–2, …, 9–10). The adjustment corrects for the effort component induced by the initial slider positions and sets all adjacent distances to one for respondents who indicated that they intended to report a linear scale. Dots indicate point estimates and vertical lines show 95\% confidence intervals. A flat profile across adjacent categories indicates no systematic variation in local scale use, while deviations across the scale reflect heterogeneous non-linearities.
    } 
    \end{minipage}
\end{figure}

Figures \ref{fig:indiv_c_2} and \ref{fig:indiv_c_2_star} report the resulting coefficients and 95\% confidence intervals. Most estimates are small, statistically insignificant, and display no systematic pattern across the scale. The earlier association between education and overall non-linearity can nevertheless be traced to specific segments of the scale: highly educated respondents tend to report slightly smaller distances at the lower end of the scale. Given the large number of coefficients estimated, isolated significant effects cannot be taken as evidence of systematic heterogeneity, as they may simply reflect sampling variability and would not survive conventional adjustments for multiple testing.

Overall, the evidence suggests that scale-use heterogeneity is largely idiosyncratic rather than systematically structured along observable socio-demographic dimensions.

These results should be interpreted with caution. The absence of strong predictors does not imply that scale use is perfectly random. Small effects may remain undetected, and standard socio-demographic variables may not capture the psychological traits governing scale interpretation. Future work could explore alternative sources of heterogeneity, including numeracy, cultural background, or survey framing.

\subsection{Evidence on monotonicity from continuous follow-up responses}
\label{sec:monotonicity_continuous}

Our optimisation problem imposes that admissible transformations of the response scale are positive monotonic. This is Constraint 2 in Section \ref{sec:optimisation}. The interactive slider task also imposes monotonicity by construction, since respondents are asked to indicate the relative position of ordered response categories.

We therefore provide an additional check using an exercise that does not impose monotonicity. After respondents answered the standard discrete life-satisfaction question (and before the interactive slider task), we asked them to report the exact value they would have chosen had they not been constrained to use the discrete response scale. This gives us, for each respondent, both a discrete response and a continuous follow-up measure. Hence, the continuous follow-up question allows responses that are inconsistent with a monotonic mapping from discrete to continuous answers: for example, a respondent who selected 5 on the discrete scale could report a continuous value of 6.2.

\begin{figure}[t]
    \caption{Difference between continuous and discrete life satisfaction responses by discrete LS category}
    \centering
    \label{fig:mean_ci}
    \includegraphics[width=0.9\textwidth]{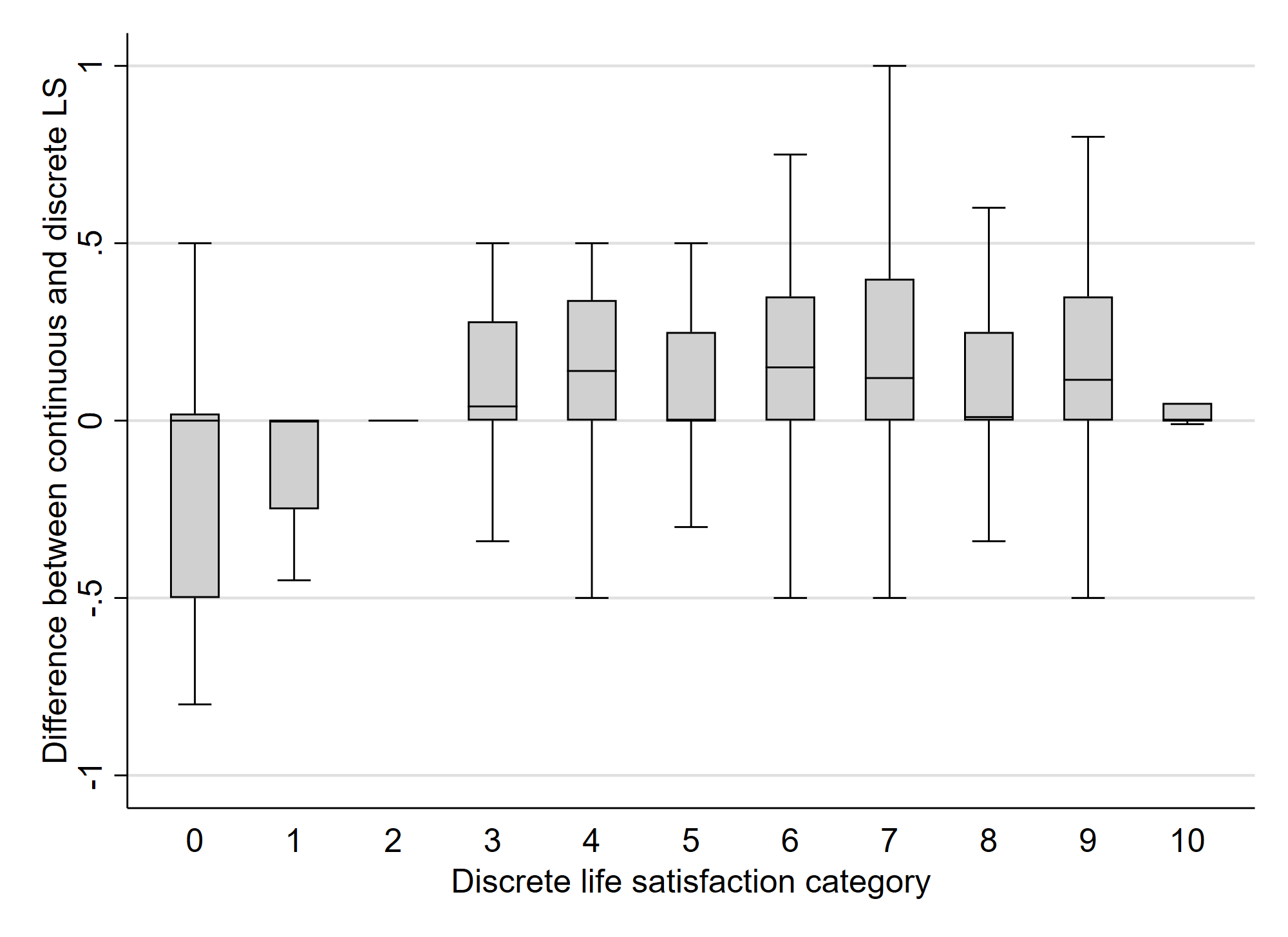}
    \begin{minipage}{0.975\textwidth}
    \small
    \noindent \footnotesize{\textbf{Notes:} This figure plots the distribution of the difference between respondents' continuous and discrete life-satisfaction responses, separately by the discrete response category initially selected. Each box corresponds to one discrete life-satisfaction category, from 0 to 10. The central line denotes the median, the box the interquartile range, and whiskers extend to the most extreme non-outlier observations. Values close to zero indicate that respondents' continuous follow-up responses are close to their initial discrete answers. Values above $1$ or below $-1$ would indicate that respondents frequently locate their continuous response more than one category away from their discrete response, which would raise concerns about violations of the ordinal interpretation of the scale. The distributions lie mostly between $-0.5$ and $0.5$, providing little evidence of such violations.}
    \end{minipage}
\end{figure}

Figure \ref{fig:mean_ci} plots the difference between the continuous and discrete life-satisfaction measures, separately by the respondent's discrete response category. A violation of monotonicity would be apparent if these differences were systematically larger than 1 or smaller than $-1$. We report the differences separately by discrete category because violations of the ordinal interpretation may be more likely around focal response values \citep{barrington2024econometrics}. The figure provides little evidence of such violations. Most responses lie between $-0.5$ and $0.5$, and none of the box plots extends beyond the interval from $-1$ to $1$. This supports the view that the monotonicity restriction used in our optimisation exercise is empirically innocuous in this setting.

\clearpage

\section{Assessing potential biases from discretisation}
\label{sec:discreteness}
 
Even when scale use is linear, individuals reporting the same response category may differ in their underlying state $s_i$ in ways that systematically correlate with covariates. Under sufficiently adverse within-category distributions, this can flip the sign of estimated coefficients. We here show that, in practice, this seems to be a minor concern.
 
If discreteness were generating large biases, regressions using a near-continuous measure of satisfaction would yield different coefficients than regressions using the standard discrete measure. We can therefore quantify the bias from discreteness, for each covariate $m$, as $\hat{\gamma}_m \equiv \hat{\beta}_m^{cont} - \hat{\beta}_m^{disc}$, where $\hat{\beta}_m^{cont}$ and $\hat{\beta}_m^{disc}$ are the OLS coefficients on $X_{im}$ from regressions using, respectively, a continuous and a discrete measure of satisfaction.
 
To assess this possibility, we again make use of our data where respondents first reported their life satisfaction on the standard 0-10 discrete scale, and then indicated, within their chosen category, where exactly they would place themselves on a continuous slider. This yields, for each respondent, both $r_i^{disc}$ and a (quasi-)continuous response $r_i^{cont}$. We then regress each measure on a standard set of socio-economic covariates and compute $\hat{\gamma}_m$.
 
\begin{figure}[!b]
    \caption{Discrete and continuous measures of satisfaction yield near-identical estimates.}
    \centering
    \label{fig:gamma_linear_prolific}
    \includegraphics[width=0.8\textwidth]{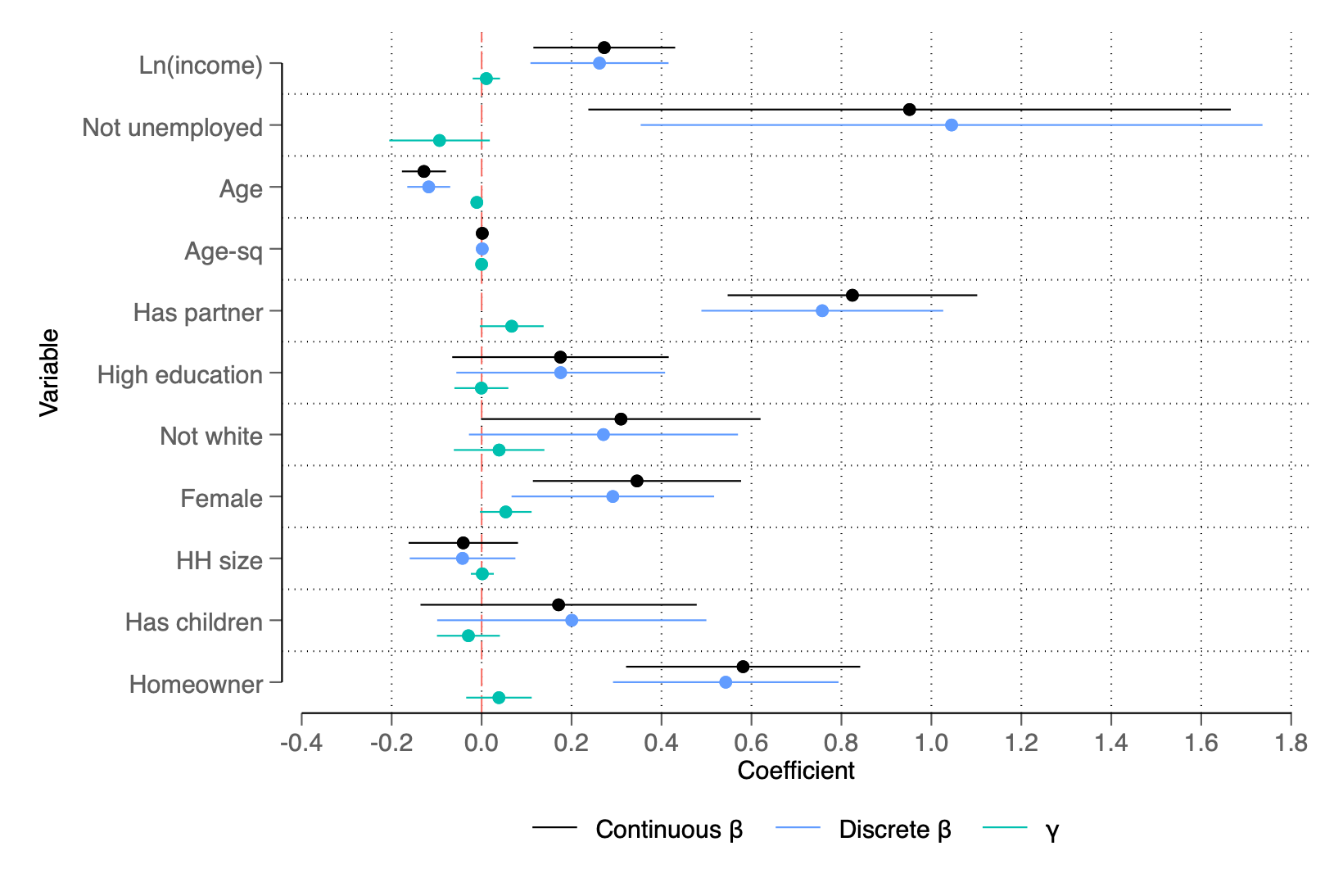}
    \begin{minipage}{0.975\textwidth}
    \footnotesize
    \noindent \textbf{Notes:} Comparison of regression coefficients using either a continuous (black dots) or a discrete (blue dots) 11-point measure of life satisfaction. The differences between these estimates ($\hat{\gamma}_m$; teal dots) represent the bias from discretisation. Whiskers indicate 95\% confidence intervals. Across all covariates, the $\hat{\gamma}_m$ estimates are small and statistically insignificant at the 5\% level. Data from September 2024 survey.
    \end{minipage}
\end{figure}
 
Figure \ref{fig:gamma_linear_prolific} shows the results. Across all covariates, $\hat{\gamma}_m$ is small and statistically insignificant at the 5\% level. The coefficient patterns themselves align closely between the two specifications and with prior literature: life satisfaction follows a U-shape with age, unemployment strongly reduces it, higher household income increases it, and having a partner is beneficial. Taken together, the evidence suggests that the within-category-heterogeneity concern does not generate empirically meaningful biases in standard Likert-scale regressions. 

%====================================
%====================================

\begin{landscape}
\clearpage
\section{WellBase Details}
{\scriptsize
\setstretch{1} 
\begin{longtable}{>{\raggedright\arraybackslash}p{3.9cm}p{10.6cm}>{\centering\arraybackslash}p{1.25cm}>{\centering\arraybackslash}p{2cm}>{\centering\arraybackslash}p{1.25cm}p{3.55cm}}
\caption{Description of Papers included in WellBase\label{tab:wellbase_description}} \\

\midrule
Reference & Measure of $r$ & \makecell{Response \\ options} & \makecell{Unemployment \\ estimates} & \makecell{Income \\ estimates} &Comments \\
\endfirsthead

\multicolumn{4}{c}{\tablename\ \thetable\ -- \textit{Continued from previous page}} \\
\toprule
Reference & Measure of $r$ & \makecell{Response \\ options} & \makecell{Unemployment \\ estimates} & \makecell{Income \\ estimates} &Comments \\
\midrule
\endhead

\bottomrule
\multicolumn{4}{r}{\textit{Continued on next page}} \\
\endfoot

\bottomrule
\endlastfoot
\midrule

\href{https://doi.org/10.1111/j.1468-0297.2010.02395.x}{Banks et al. (2010)} & I am satisfied with my life & 4 & . & . & None \\
\href{https://doi.org/10.1111/j.1468-0297.2010.02359.x}{Clark and Senik (2010)} & Taking all things together, how happy would you say you are? \newline All things considered, how satisfied are you with your life as a whole nowadays? \newline How satisfied are you with how your life has turned out so far? &   \phantom{X}\hspace{0.05cm}11    \newline  \phantom{X}\hspace{0.05cm}11  \newline11 & . &  $\checkmark$ & None \\
\href{https://doi.org/10.1111/j.1468-0297.2009.02347.x}{Knabe et al. (2010)} & All things considered, how satisfied are you with your life as a whole these days? & 11 & $\checkmark$ & $\checkmark$ & None \\
\href{https://doi.org/10.1162/REST_a_00133}{Oswald and Wu (2011)} & In general, how satisfied are you with your life? & 6 & $\checkmark$ & $\checkmark$ & Income logged-transformed and change of LFS reference category for Section \ref{sec:rel_magnitudes} \\
\href{https://doi.org/10.1162/REST_a_00133}{Bertrand (2013)} & Taken all together, how would say things are these days - would you say that you are very happy, pretty happy or not too happy? & 3 & . & . &  None \\
\href{https://doi.org/10.1016/j.jpubeco.2013.06.009}{Vendrik (2013)} & All things considered, how satisfied are you with your life as a whole these days? & 11 & $\checkmark$ & $\checkmark$  & None\\
\href{https://doi.org/10.1257/aer.104.7.2210}{Ashraf et al. (2014)}  & How satisfied are you with your life as a whole these days?& 5 & . & .  & None\\
\href{https://doi.org/10.1111/ecoj.12085}{Frijters et al. (2014)} & Here is a scale from 0 to 10, where "0" dissatisfied and "10" means that you are completely satisfied. Please enter the number which corresponds with how satisfied or dissatisfied you are with the way life has turned out so far. & 11 & $\checkmark$ & .  & None\\  
\href{https://doi.org/10.1162/REST_a_00353}{Kesternich et al. (2014)} & On a scale from 0 to 10 where 0 means completely dissatisfied and 10 means completely satisfied, how satisfied are you with your life? & 11 & . & .  & None\\
\href{https://doi.org/10.1111/ecoj.12170}{Layard et al. (2014)} & Here is a scale from 0 to 10. On it, ``0'' means that you are completely dissatisfied and ``10'' means that you are completely satisfied. Please tick the box with the number above it which shows how dissatisfied or satisfied you are about the way your life has turned out so far. & 11 & $\checkmark$ & $\checkmark$  & Income unstandardized for Section \ref{sec:rel_magnitudes}\\
\href{https://doi.org/10.1093/qje/qju032}{Bloom et al. (2015)} & How satisfied are you with your life as a whole these days? & 7 & . & .  & None\\
\href{https://doi.org/10.1093/qje/qjv002}{Campante and Yanagizawa-Drott (2015)} & Taking all things together, would you say you are: not at all happy, not very happy, quite happy, very happy?  \newline How satisfied are you with your life as a whole these days?&   \phantom{X}\hspace{0.05cm}4  \newline   \newline  10 & . & .  & Converted binary $r$ to original continuous $r$\\
\href{https://doi.org/10.1016/j.jdeveco.2014.12.009}{Dinkelman and Schulhofer-Wohl (2015)} & Taking everything into account, how satisfied is the household with the way it lives these days? & 5 & . & .  & Original measure of $r$ in log; delogged for WellBase\\
\href{https://doi.org/10.1086/681096}{Oswald et al. (2015)} & How would you rate your happiness at the moment? & 6 & . & .  & Ordered probit replaced by OLS\\
\href{https://doi.org/10.1257/aer.20150338}{Aghion et al. (2016)} & Please imagine a ladder with steps numbered from zero at the bottom to 10 at the top. The top of the ladder represents the best possible life for you and the bottom of the ladder represents the worst possible life for you. On which step of the ladder would you say you personally feel you stand at this time? \newline In general, how satisfied are you with your life?
 &  \phantom{X}\hspace{0.2cm} 11 \newline \newline \newline \newline 4 & . & $\checkmark$  & None\\
\href{https://doi.org/10.1162/REST_a_00544}{Clark et al. (2016)} & How satisfied are you with your life, all things considered? & 11 & $\checkmark$ & .  & None\\
\href{https://doi.org/10.1016/j.jpubeco.2016.01.001}{Danzer and Danzer (2016)} & To what extent are you satisfied with your life in general at the present time? & 5 & $\checkmark$ & $\checkmark$  & None \\
\href{https://doi.org/10.1016/j.jpubeco.2016.10.006}{Gerritsen (2016)} & How dissatisfied or satisfied are you with your life overall? &7  & .  & $ \checkmark$ & None \\
\href{https://doi.org/10.1086/684044}{Glaeser et al. (2016)} & In general, how satisfied are you with your life? & 4 & . & .  & Some regressions based on propriety data missing \\
\href{https://doi.org/10.1093/qje/qjw025}{Haushofer and Shapiro (2016)}  & Taking all things together, would you say you are `very happy' (1), `quite happy' (2), `not very happy' (3), or `not at all happy' (4)?"  \newline  All things considered, how satisfied are you with your life as a whole these days? &\phantom{X}\hspace{0.05cm}4   \newline  \newline   11  & . & .  & None\\
\href{https://doi.org/10.1111/ecoj.12256}{Cheng et al. (2017)} & How satisfied are you with your life, all things considered? \newline How dissatisfied or satisfied are you with your life overall? \newline All things considered, how satisfied are you with your life in general? \newline All things considered, how satisfied are you with your life? & \phantom{X}\hspace{0.05cm}11  \newline \phantom{X}\hspace{0.05cm}7 \newline\phantom{X}\hspace{0.05cm}10  \newline 11 & . & .  & None\\
\href{https://doi.org/10.1257/app.20170173}{Blattman and Dercon (2018)} & Please imagine a ladder with steps numbered from zero at the bottom to 10 at the top. The top of the ladder represents the best possible life for you and the bottom of the ladder represents the worst possible life for you. On which step of the ladder would you say you personally feel you stand at this time? & 11 & . & . & None \\
\href{https://doi.org/10.1257/aer.20171676}{Blumenstock et al. (2018)} & All things considered, how satisfied are you with life as a whole? & 11 & . & . & None \\
\href{https://doi.org/10.1162/REST_a_00697}{De Neve et al. (2018)} & On the whole, are you very satisfied, fairly satisfied, not very satisfied, or not at all satisfied with the life you lead? \newline Please imagine a ladder with steps numbered from zero at the bottom to 10 at the top. The top of the ladder represents the best possible life for you and the bottom of the ladder represents the worst possible life for you. On which step of the ladder would you say you personally feel you stand at this time? \newline In general, how satisfied are you with your life?
 &  \phantom{X}\hspace{0.05cm}4 \newline \newline  \phantom{X}\hspace{0.05cm}11    \newline \newline \newline \newline 4 & . & . & None \\
\href{https://doi.org/10.1111/ecoj.12478}{Johnston et al. (2018)} & All things considered, how satisfied are you with your life? &
 11 & $\checkmark$ & $\checkmark$& Income logged-transformed and change of LFS reference category for Section \ref{sec:rel_magnitudes} \\
\href{https://doi.org/10.1016/j.jpubeco.2019.07.002}{Dolan et al. (2019)} & Overall, how satisfied are you with your life nowadays? & 11 & $\checkmark$ & .& None \\
\href{https://doi.org/10.1093/ej/uez011}{Fisher and Zhu (2019)} & All things considered, how satisfied are you with your life? & 11 & . & .& None \\
\href{https://doi.org/10.1257/jep.33.4.100}{Guriev and Treisman (2019)} & Please imagine a ladder with steps numbered from zero at the bottom to 10 at the top. The top of the ladder represents the best possible life for you and the bottom of the ladder represents the worst possible life for you. On which step of the ladder would you say you personally feel you stand at this time? & 11 & . & .& None \\
\href{https://doi.org/10.1162/rest_a_00748}{Heffetz and Reeves (2019)} & In general, how satisfied are you with your life?& 4 & . & .& None \\
\href{https://doi.org/10.1093/jeea/jvy005}{Odermatt and Stutzer (2019)} & How satisfied are you with your life, all things considered? & 11 & $\checkmark$ & $\checkmark$& None \\
\href{https://doi.org/10.1162/rest_a_00784}{Tur-Prats (2019)} & How satisfied are you with your life as a whole these days? & 10 & . & .& Converted binary $r$ to original continuous $r$ \\
\href{https://doi.org/10.1257/aer.20190658}{Allcott et al. (2020)} & During the past 4 weeks, I was satisfied with my life & 7 & . & .& None \\
\href{https://doi.org/10.1257/aer.20180976}{Blakeslee et al. (2020)} & All things considered, how satisfied are you with your life as a whole these days? & 10 & . & . & None \\
\href{https://doi.org/10.1016/j.jdeveco.2019.102416}{Haushofer et al. (2020)} & Taking all things together, would you say you are `very happy' (1), `quite happy' (2), `not very happy' (3), or `not at all happy' (4)?" \newline All things considered, how satisfied are you with your life as a whole these days? & \phantom{X}\hspace{0.05cm}4 \newline\newline  11 & . & . & None \\
\href{https://doi.org/10.1086/705417}{Lee et al. (2020)} & All things considered, how satisfied are you with your life as a whole these days? & 10 & . & $\checkmark$& None \\
\href{https://doi.org/10.1257/aer.20160256}{Perez-Truglia (2020)} & Will you mostly describe yourself as: Very happy; Quite happy; Not particularly happy; Not at all happy How satisfied are you with your life? \newline How satisfied are you with your life, all things considered? & \phantom{X}\hspace{0.05cm}4 \newline \newline  11 & $\checkmark$ & . & Probit-adjusted OLS replaced by OLS \\
\href{https://doi.org/10.1016/j.jdeveco.2018.10.003}{Singh and Masters (2020)} & How satisfied are you with your life, all things considered? & 6 & . & .& None \\
\href{https://doi.org/10.1016/j.jdeveco.2020.102603}{Aksoy and Tumen (2021)} & All things considered, I am satisfied with my life now & 5 &. & .& None \\
\href{https://doi.org/10.1093/qje/qjab013}{Bessone et al. (2021)} & How happy are you today? \newline All things considered, how satisfied are you with your life as a whole? & \phantom{X}\hspace{0.05cm}5  \newline 10& . & .& None \\
\href{https://doi.org/10.1093/qje/qjaa023}{Bryan et al. (2021)} & How would you describe your satisfaction with life? \newline Taking all things together, would you say you are & \phantom{X}\hspace{0.05cm}4  \newline 10& . & . \\
\href{https://doi.org/10.1016/j.jdeveco.2021.102664}{Chen and Fang (2021)} & Please think about your life-as-a-whole. How satisfied are you with it? & 5  & . & .& None \\
\href{https://doi.org/10.1093/jeea/jvab007}{Dalton et al. (2021)} & How satisfied are you with your life at this point? & 10 & . & . & None\\
\href{https://doi.org/10.1162/rest_a_00894}{Fl\`eche (2021)} & In general, how satisfied are you with your life? & 11& $\checkmark$ & $\checkmark$& Regressions with Municipality FE not reproduced \newline Probit-adjusted OLS replaced by OLS \\
\href{https://doi.org/10.1162/rest_a_00921}{Huang et al. (2021)} & Are you happy? & 11& . & . & None \\
\href{https://doi.org/10.1162/rest_a_00921}{Kab\'atek and Ribar (2021)} & How satisfied are you with the life you lead at the moment? & 11 & . & .& Ordered logit replaced by OLS \\
\href{https://doi.org/10.1093/restud/rdaa016}{Levitt (2021)} & All things considered, how happy are you as a whole right now? & 10 & . & .& None \\
\href{https://doi.org/10.1016/j.jdeveco.2021.102634}{Li (2021)} & How happy are you? \newline How satisfied are you with your life as a whole? & \phantom{X}\hspace{0.05cm} 5 \newline 5 & . & . & None \\
\href{https://doi.org/10.1016/j.jdeveco.2022.102899}{Ajzenman et al. (2022)} & All things considered, I am satisfied with my life now & 5 & . & .& None \\
\href{https://doi.org/10.1162/rest_a_00944}{Binder and Makridis (2022)} & Please imagine a ladder with steps numbered from zero at the bottom to 10 at the top. The top of the ladder represents the best possible life for you and the bottom of the ladder represents the worst possible life for you. On which step of the ladder would you say you personally feel you stand at this time? & 11 & $\checkmark$ & .& None  \\
\href{https://doi.org/10.1093/restud/rdab089}{Dahl et al. (2022)} & Overall, how satisfied are you with your life?& 11 & . & . & None\\
%\citet{dellavigna2022evidence} & & . & . \\
\href{https://doi.org/10.1257/app.20200164}{Meier (2022)} & How satisfied are you with your life, all things considered? & 11 & . & .& None \\
\href{https://doi.org/10.1016/j.jpubeco.2023.104949}{Adhvaryu et al. (2023)} & Please imagine a ladder with steps numbered from zero at the bottom to 10 at the top. The top of the ladder represents the best possible life for you and the bottom of the ladder represents the worst possible life for you. On which step of the ladder would you say you personally feel you stand at this time? & 11 & . & .& None \\
\href{https://doi.org/10.1016/j.jdeveco.2023.103153}{Bha et al. (2023)} & Please imagine a ladder with steps numbered from zero at the bottom to 10 at the top. The top of the ladder represents the best possible life for you and the bottom of the ladder represents the worst possible life for you. On which step of the ladder would you say you personally feel you stand at this time? & 11 & . & .& None \\
\href{https://doi.org/10.1086/721655}{Caria et al. (2023)} & Please imagine a ladder with steps numbered from zero at the bottom to 10 at the top. The top of the ladder represents the best possible life for you and the bottom of the ladder represents the worst possible life for you. On which step of the ladder would you say you personally feel you stand at this time? & 11
 & . & .& None \\
\href{https://doi.org/10.1162/rest_a_01379}{Coville et al. (2023)} & All things considered, how satisfied are you with your life as a whole these days? \newline Taking all things together, would you say you are:
 & \phantom{X}\hspace{0.05cm} 10 \newline 4 & . & . & None \\
\href{https://doi.org/10.1162/rest_a_01074}{Edmonds et al. (2023)} & Please imagine a ladder with steps numbered from zero at the bottom to 10 at the top. The top of the ladder represents the best possible life for you and the bottom of the ladder represents the worst possible life for you. On which step of the ladder would you say you personally feel you stand at this time? & 11 & $\checkmark$ & . & None \\
\href{https://doi.org/10.1016/j.jdeveco.2023.103169}{Gazeaud et al. (2023)} & Please imagine a ladder with steps numbered from zero at the bottom to 10 at the top. The top of the ladder represents the best possible life for you and the bottom of the ladder represents the worst possible life for you. On which step of the ladder would you say you personally feel you stand at this time? & 11 & . & . & None\\
\href{https://doi.org/10.1257/aer.20211051}{Liu and Netzer (2023)} & Taking all together, how would you say things are these days?  Would you say that you are rather happy, neither happy nor unhappy or rather unhappy? & 3 & . & . & Ordered probit replaced by OLS \\
\href{https://doi.org/10.1016/j.jpubeco.2023.105014}{Sarmiento et al. (2023)} & How satisfied are you with your life, all things considered? & 11 & . & . & None \\
\href{https://doi.org/10.1016/j.jdeveco.2022.102985}{Sha (2023)} & How satisfied are you with your life as a whole? & 5 & . & . & None\\
\href{https://doi.org/10.1093/restud/rdac055}{Stango and Zinman (2023)} & How satisfied are you with your life as a whole these days? & 100 & . & . & None\\
\href{https://doi.org/10.1257/aer.20210687}{Angelucci and Bennett (2024)} & I am satisfied with my life
& 10 & . & . & None\\
\href{https://doi.org/10.1093/ej/uead116}{Ciancio and Delavande (2024)}  & How satisfied are you with your life, all things considered? & 6 & . & .& None \\
\href{https://doi.org/10.1093/ej/uead118}{Clark and Zhu (2024)} & All things considered, how satisfied are you with your life?
 & 11 & . & . & None\\
\href{https://doi.org/10.1257/app.20220443}{Giacobino et al. (2024)} & Happiness question - wording not reported \newline Life satisfaction question - wording not reported & \phantom{X}\hspace{0.05cm}4 \newline 10 & . & .& None \\
\href{https://doi.org/10.1016/j.jdeveco.2024.103344}{Grimm et al. (2024)} & Imagine for a moment that you are living the best life you can imagine living.
Now, imagine a situation where your life is as bad as it could possibly be.
Let’s consider a scale from 1 to 6.
Suppose we say that the top of the scale (6) represents the best possible life for you, and the bottom (1) represents the worst possible life for you.
Which step of the scale best represents your current personal situation? &  6 & . & .& None \\
\href{https://doi.org/10.1162/rest_a_01533}{Krekel et al. (2024)} & Overall, how satisfied are you with your life nowadays? & 11 & . & . & Ordered logit replaced by OLS\\
\href{https://doi.org/10.1093/ej/ueae029}{Priebe et al. (2024)} & Life Satisfaction question - not reported & 5 & . & .& None \\
\citet{riley2024resisting} &  Happiness question - not reported \newline Life satisfaction question - not reported & \phantom{X}\hspace{0.05cm}5 \newline 10 & . & . & None\\
\href{https://doi.org/10.1257/app.20210655}{Vlassopoulos et al. (2024)} & Taking all things together, how happy are you these days? \newline How satisfied are you with your life as a whole these days?
 & \phantom{X}\hspace{0.05cm}11 \newline 11 & . & .& None. \\
\href{https://doi.org/10.1257/app.20230227}{Bjorvatn et al. (2025)} & How happy are you with your life? \newline In your opinion, where are you on the ladder of life at the
moment? & \phantom{X}\hspace{0.05cm}11 \newline 11 & . & . & None\\
\href{https://doi.org/10.1162/rest_a_01303}{Carattini and Roesti (2025)} & All things considered, how satisfied are you with your life as a whole nowadays? Ranges from 0 (extremely dissatisfied) to 10 (extremely satisfied) \newline Taking all things together, how happy would you say you are? - ranges from 0 (extremely unhappy) to 10 (extremely happy) In general, how satisfied are you with your life? \newline How satisfied as a whole, 1 (not at all) to 4 (very satisfied)& \phantom{X}\hspace{0.05cm} 11  \newline\newline \phantom{X}\hspace{0.05cm} 11  \newline \newline \newline 4 & $\checkmark$ & $\checkmark$ & SHP, ESS and SOM samples analysed separately in Section \ref{sec:rel_magnitudes} \newline Ordered probit replaced by OLS \\
\href{https://doi.org/10.1093/jeea/jvaf004}{Courtemanche et al. (2025)} & In general, how satisfied are you with your life? & 11 & $\checkmark$ & . & Ordered probit replaced by OLS\\
\end{longtable}

\begin{minipage}{0.95\textwidth}
\small
\noindent \textbf{Note:} This table lists all the papers included in \textit{WellBase}.
\end{minipage}}  % Table with descriptions for each papers in WellBase

\clearpage
{\footnotesize
\setstretch{1} 
\begin{longtable}{>{\raggedright\arraybackslash}p{6.2cm}p{6.7cm}p{4.5cm}>{\centering\arraybackslash}p{.8cm}>{\centering\arraybackslash}p{1.5cm}>{\centering\arraybackslash}p{.5cm}>{\centering\arraybackslash}p{1.9cm}}

\caption{Risk of sign reversal and main conclusions of \textit{WellBase}}\label{tab:of_int_detailed} \\

\toprule
Author & Test(s) of the paper & Source & Sign & Sig. & Reversal & Cost \\
\midrule
\endfirsthead
\multicolumn{7}{c}{\textit{(Continued from previous page)}} \\
\toprule
Author & Test(s) of the paper & Source & Sign & Sig. & Reversal & Cost \\
\midrule
\endhead
\bottomrule
\endfoot

\href{https://doi.org/10.1111/j.1468-0297.2010.02359.x}{Clark and Senik (2010)}  & Income & Table 7 - Column 4 & + & 1\% & No & . \\
         & Important to compare income & Table 7 - Column 4 & - & 1\% & No & . \\
         & Comparison direction: work colleagues & Table 7 - Column 4 & - & 5\% & Yes & 0.715 \\
         & Comparison direction: family members & Table 7 - Column 4 & - & 1\% & Yes & 0.164 \\
         & Comparison direction: others & Table 7 - Column 4 & - & 1\% & Yes & 0.138 \\
         & Comparison direction: don't compare & Table 7 - Column 4 & - & 1\% & Yes & 0.252 \\
\href{https://doi.org/10.1111/j.1468-0297.2009.02347.x}{Knabe et al. (2010)} & Unemployment & Table 5 - Column 3 & - & 1\% & No & . \\
\href{https://doi.org/10.1162/REST_a_00133}{Oswald and Wu (2011)} & US States Fixed effects & Table 2 - Column 4 & Mix & NS to 1\% & 44\% & 0.010 to 0.980 \\
\href{https://doi.org/10.1162/REST_a_00133}{Bertrand (2013)}  & Having a job & Table 1 - Panel A & + & 1\% & No & . \\
            & Being married & Table 1 - Panel A & + & 1\% & No & . \\
            & Having a job and being married & Table 1 - Panel A & - & 5\% & No & . \\
            & Having a job & Table 1 - Panel B & + & 1\% & No & . \\
            & Having kids & Table 1 - Panel B & + & 1\% & No & . \\
            & Having a job and having kids & Table 1 - Panel B & - & 5\% & No & . \\
\href{https://doi.org/10.1016/j.jpubeco.2013.06.009}{Vendrik (2013)} & Current own income & Table 1 - Column 5 & + & 1\% & No & . \\
        & Past own income (one year) & Table 1 - Column 5 & - & NS & Yes & 0.165 \\
        & Past own income (two years) & Table 1 - Column 5 & - & NS & Yes & 0.280 \\
        & Past own income (three years) & Table 1 - Column 5 & + & 10\% & Yes & 0.564 \\
        & Future own income (one year) & Table 1 - Column 5 & + & 1\% & No & . \\
          & Current reference income & Table 1 - Column 5 & - & NS & Yes & 0.321 \\
          & Past reference income (one year) & Table 1 - Column 5 & - & 10\% & Yes & 0.226 \\
         & Future reference income (one year) & Table 1 - Column 5 & + & NS & Yes & 0.053 \\
\href{https://doi.org/10.1111/ecoj.12085}{Frijters et al. (2014)} & Wage & Table 4 - Column 5 & + & 1\% & Yes & 0.226 \\
            & Employment & Table 4 - Column 5 & + & 1\% & Yes & 0.399 \\
            & Unemployment & Table 4 - Column 5 & + & NS & Yes & 0.084 \\
            & Married & Table 4 - Column 5 & + & 1\% & Yes & 0.733 \\
            & Poor Health & Table 4 - Column 5 & - & 1\% & No & . \\
            & Education & Table 4 - Column 5 & + & NS & Yes & 0.278 \\
            & Lagged satisfaction (age 46) & Table 4 - Column 5 & + & 1\% & No & . \\
            & Lagged satisfaction (age 42) & Table 4 - Column 5 & + & 1\% & No & . \\
            & Lagged satisfaction (age 33) & Table 4 - Column 5 & + & 1\% & No & . \\
\href{https://doi.org/10.1111/ecoj.12170}{Layard et al. (2014)}   & Income & Table 1 - Column 3 & + & 1\% & Yes & 0.354 \\
            & Education & Table 1 - Column 3 & + & 1\% & Yes & 0.049 \\
            & Having a job & Table 1 - Column 3 & + & 1\% & No & . \\
            & Good conduct & Table 1 - Column 3 & + & 1\% & No & . \\
            & Having a partner & Table 1 - Column 3 & + & 1\% & Yes & 0.856 \\
            & Self-perceived health & Table 1 - Column 3 & + & 1\% & Yes & 0.786 \\
            & Emotional health & Table 1 - Column 3 & + & 1\% & No & . \\
            & Female & Table 1 - Column 3 & + & 1\% & Yes & 0.15 \\
\href{https://doi.org/10.1093/qje/qjv002}{Campante and Yanagizawa-Drott (2015)} & Ramadan hours & Table 2 - Column 12 & + & 1\% & No & . \\

\href{https://doi.org/10.1257/aer.20150338}{Aghion et al. (2016)} & Job turnover rate & Table 2 - Column 3 - Panel B & + & 5\% & Yes & 0.342 \\
            & Unemployment rate & Table 2 - Column 3 - Panel B  & - & 1\% & No & . \\
            & Job creation rate & Table 3 - Column 2 - Panel B & + & 1\% & Yes & 0.575 \\
            & Job destruction rate & Table 3 - Column 2 - Panel B & - & 1\% & Yes & 0.459 \\
\href{https://doi.org/10.1162/REST_a_00544}{Clark et al. (2016)} & Incidence of poverty & Table 2 - Column 1 & - & 1\% & Yes & 0.593 \\
            & Intensity of poverty & Table 2 - Column 1 & - & 1\% & No & . \\
            & 0 to 1 years of poverty & Table 3 - Column 1 & - & 1\% & No & . \\
            & 1 to 2 years of poverty & Table 3 - Column 1 & - & 1\% & No & . \\
            & 2 to 3 years of poverty & Table 3 - Column 1 & - & 1\% & No & . \\
            & 3 to 4 years of poverty & Table 3 - Column 1 & - & 1\% & Yes & 0.494 \\
            & 4 to 5 years of poverty & Table 3 - Column 1 & - & 1\% & Yes & 0.324 \\
            & 5 years of poverty or more & Table 3 - Column 1 & - & 1\% & Yes & 0.504 \\
\href{https://doi.org/10.1016/j.jpubeco.2016.01.001}{Danzer and Danzer (2016)}  & Radiation & Table 2 - Column 3 & - & 1\% & Yes & 0.962 \\
\href{https://doi.org/10.1016/j.jpubeco.2016.10.006}{Gerritsen (2016)} & Income & Table 1 - Column 1 & + & 1\% & No & . \\
             & Hours of work & Table 1 - Column 1 & + & 5\% & Yes & 0.181 \\
             & Hours of work squared & Table 1 - Column 1 & - & 5\% & Yes & 0.115 \\
\href{https://doi.org/10.1086/684044}{Glaeser et al. (2016)}  & Population size & Table 1 - Column 2 & - & 5\% & No & . \\
\href{https://doi.org/10.1111/ecoj.12256}{Cheng et al. (2017)}  & Age & Figure 2 - Panel A & - & 1\% & No & . \\
        & Age squared & Figure 2 - Panel A & + & 1\% & No & . \\
         & Age & Figure 2 - Panel B & - & 1\% & No & . \\
         & Age squared &Figure 2 - Panel B  & + & 1\% & No & . \\
         & Age & Figure 2 - Panel C & - & 1\% & No & . \\
         & Age squared & Figure 2 - Panel C & + & 1\% & No & . \\
         & Age & Figure 2 - Panel D & - & 1\% & No & . \\
         & Age squared & Figure 2 - Panel D & + & 1\% & No & . \\
\href{https://doi.org/10.1162/REST_a_00697}{De Neve et al. (2018)}  & Economic growth - World Sample & Table 1 - Column 1 & + & 1\% & No & . \\
            & Negative growth - World Sample & Table 1 - Column 2 & - & 1\% & No & . \\
            & Positive growth - World Sample &  Table 1 - Column 2 & + & NS & Yes & 0.482 \\
            & Economic growth - European Sample & Table 1 - Column 3 & + & 1\% & No & . \\
            & Negative growth - European Sample & Table 1 - Column 4 & - & 1\% & No & . \\
            & Positive growth - European Sample & Table 1 - Column 4 & + & 5\% & No & . \\
            & Economic growth - US Sample & Table 1 - Column 5 & + & 1\% & No & . \\
            & Negative growth - US Sample & Table 1 - Column 6 & - & 1\% & No & . \\
            & Positive growth - US Sample & Table 1 - Column 6 & + & 1\% & No & . \\
\href{https://doi.org/10.1111/ecoj.12478}{Johnston et al. (2018)} & Victim of physical violence - Women sample & Table 3 - Column 1 & - & 1\% & No & . \\
           & Victim of physical violence - Men sample & Table 3 - Column 2 & - & 1\% & No & . \\
\href{https://doi.org/10.1016/j.jpubeco.2019.07.002}{Dolan et al. (2019)} & Olympic games in London & Table 2 - Column 6 & + & 1\% & No & . \\
\href{https://doi.org/10.1093/jeea/jvy005}{Odermatt and Stutzer (2019)} & Widowhood (zero to one year) & Table 2 - Column 2 & - & 1\% & No & . \\
           & Widowhood (five to six year) &Table 2 - Column 2   & - & 1\% & Yes & 0.081 \\
            & Unemployment (zero to one year) & Table 2 - Column 4 & - & 1\% & No & . \\
            & Unemployment (five to six year) & Table 2 - Column 4 & - & 1\% & Yes & 0.293 \\
            & Disability (zero to one year) & Table 2 - Column 6 & - & 1\% & Yes & 0.509 \\
            & Disability (five to six year) & Table 2 - Column 6 & - & 1\% & Yes & 0.182 \\
            & Plant closure (zero to one year) & Table 2 - Column 8 & - & 1\% & No & . \\
            & Plant closure (five to six year) & Table 2 - Column 8 & - & 1\% & Yes & 0.198 \\
\href{https://doi.org/10.1257/aer.20160256}{Perez-Truglia (2020)} & Income Rank*2001--201*High Internet & Table 3 - Column 4 & + & 1\% & No & . \\
            & Income Rank*2001--2013*High Internet & Table 3 - Column 6 & + & NS & Yes & 0.428 \\
\href{https://doi.org/10.1162/rest_a_00894}{Fl\`eche (2021)} & Centralization reforms & Table 1 - Column 4  & - & 1\% & No & . \\
\href{https://doi.org/10.1093/restud/rdaa016}{Levitt (2021)}   & All major life decisions after two months & Table 5 - Column 2 - Row 1 & + & 1\% & No & . \\
            & All major life decisions after two months & Table 5 - Column 3 - Row 1 & + & NS & Yes & 0.087 \\
            & All major life decisions after six months & Table 5 - Column 5 - Row 1 & + & 1\% & No & . \\
            & All major life decisions after six months & Table 5 - Column 6 - Row 1 & + & 5\% & No & . \\
\href{https://doi.org/10.1016/j.jdeveco.2021.102634}{Li (2021)}  & First son * Sex ratio & Table 3 - Column 1 & - & 5\% & No & . \\
           & First son * Sex ratio & Table 3 - Column 2 & - & 1\% & No & . \\
\href{https://doi.org/10.1093/restud/rdab089}{Dahl et al. (2022)} & Post-reform*Immigrant & Table 1 - Column 4 - Panel A & - & 1\% & No & . \\
            & Post-reform*Immigrant & Table 1 - Column 4 - Panel B & + & NS & Yes & 0.123 \\
\href{https://doi.org/10.1016/j.jpubeco.2023.105014}{Sarmiento et al. (2023)}  & LEZ introduction & Table 8 - Column 1 & + & 1\% & Yes & 0.310 \\
\href{https://doi.org/10.1162/rest_a_01533}{Krekel et al. (2024)}  & Volunteering in England's NHS & Table 3 - Column 2 & + & 1\% & No & . \\
\href{https://doi.org/10.1162/rest_a_01303}{Carattini and Roesti (2025)}  & Trust & Table 1 - Column 1 & + & 1\% & No & . \\
\href{https://doi.org/10.1093/jeea/jvaf004}{Courtemanche et al. (2025)}  & Chain restaurant calorie posting laws & Table 5 - Column 2 & + & 1\% & Yes & 0.645 \\
\end{longtable}
\begin{minipage}{1.3\textwidth}
\noindent \scriptsize{\textbf{Note:} This table lists the risk of sign reversal in all the papers included in \textit{WellBase} for which at least half of the regressions printed uses a measure of cognitive subjective wellbeing as dependent variable.}
\end{minipage}}  % Table with descriptions for each papers in WellBase
\end{landscape}

\clearpage
\section{Additional Tables and Figures}

\begin{figure}[!h]
    \centering
    \caption{PRISMA Chart - Other Likert scales}
    \label{fig:prisma_likert}
    \begin{tikzpicture}[>=latex, font={\sf \small}, scale=0.95, transform shape]
\tikzstyle{bluerect} = [rectangle, rounded corners, minimum width=1.2cm, minimum height=0.6cm, text centered, draw=black, fill=cyan!60!gray!45!white, font={\sffamily\small}]
\tikzstyle{textrect} = [rectangle, minimum width=3.8cm, text width=3.6cm, minimum height=1cm, draw=black, font={\sffamily \scriptsize}]
% Identification column
\node (r1blue) at (0, 3.2cm) [draw, bluerect, minimum height=0.8cm]{Identification};
\node (r1left) at (0, 1.5cm) [draw, textrect, minimum height=1.3cm]
  {Records identified through database searching: $n=115$};
\node (r1right) at (0, -1.2cm) [draw, textrect, minimum height=1.8cm]
  {Records removed \textit{before screening}: \\ Comment of published records ($n=2$)};
% Screening column  
\node (r2blue) at (6.45cm, 3.2cm) [draw, bluerect, minimum height=0.8cm]{Screening};
\node (r2left) at (4.3cm, 1.5cm) [draw, textrect, minimum height=1.3cm]
  {Records screened: $n=113$};
\node (r2right) at (4.3cm, -2cm) [draw, textrect, minimum height=3cm]
  {Records excluded: 
    \begin{itemize}
    \setlength{\itemsep}{0pt}
    \item No Likert scale in empirical analysis ($n=64$)
    \item Likert scale not the dependent variable ($n=14$)
    \item Empirical analysis not at the individual level ($n=2$)
    \end{itemize}     
  };
% Assessment column
\node (r4left) at (8.6cm, 1.5cm) [draw, textrect, minimum height=1.3cm]
  {Records assessed for reproduction: $n=33$ 
};
\node (r4right) at (8.6cm, -1.2cm) [draw, textrect, minimum height=1.3cm]
  {Records excluded:
    \begin{itemize}
    \item Missing replication package or protected data ($n=17$)
    \end{itemize}     
  };
% Included column
\node (r5blue) at (12.9cm, 3.2cm) [draw, bluerect, minimum height=0.8cm]{Included};
\node (r5left) at (12.9cm, 1.5cm) [draw, textrect, minimum height=1.3cm]
  {Papers reproduced and included in analysis: $n=16$};
% Draw arrows between nodes:
\draw[thick, ->] (r1left.south) -- (r1right.north);
\draw[thick, ->] (r1left.east) -- (r2left.west);
\draw[thick, ->] (r2left.south) -- (r2right.north);
\draw[thick, ->] (r2left.east) -- (r4left.west);
\draw[thick, ->] (r4left.south) -- (r4right.north);
\draw[thick, ->] (r4left.east) -- (r5left.west);
\end{tikzpicture}
    \vspace{5pt}
    \begin{minipage}{0.90\textwidth}
        \small
        \noindent \textbf{Note:} This chart describes the selection of papers included in the Likert scale analysis.
    \end{minipage}
\end{figure}

\begin{figure}[!h]
    \caption{Predictors of the Probability of Sign-reversal - Wellbeing and Likert scales}
    \centering
    \label{fig:swb_likert_sign_rev}
    \includegraphics[width=0.7\textwidth]{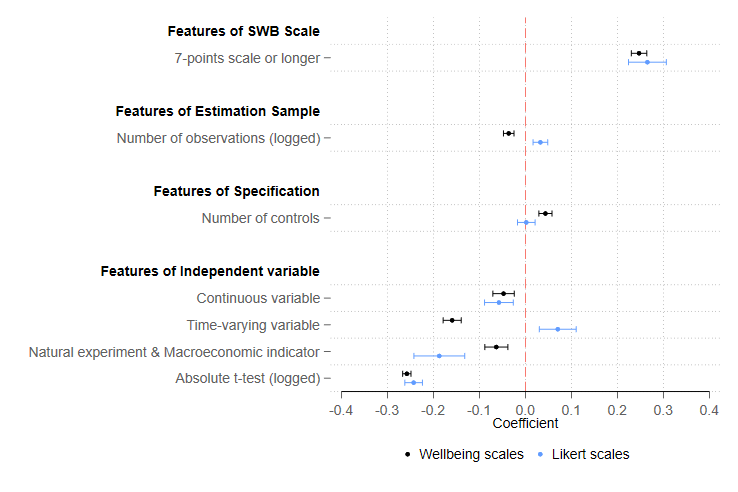}
    \begin{minipage}{0.95\textwidth}
    \small
    \noindent \textbf{Notes:} The figure shows the coefficients from linear probability models estimating the probability of sign reversal for estimates from wellbeing and Likert scale regressions published in Top-5 Economics journals. Standard errors are clustered at the regression–paper level. Whiskers represent 95\% confidence intervals.
    \end{minipage}
\end{figure}

\begin{table}[!htbp]\centering
\caption{Average deviations from linearity with bootstrapped confidence intervals using alternative cost functions}
\label{tab:other_cost}
\resizebox{1.01\textwidth}{!}{
\begin{tabular}{lcccccc}
\toprule
& \multicolumn{6}{c}{Average cost measures} \\   \cmidrule(lr){2-7}
 & (1) & (2) & (3) & (4) & (5) & (6) \\
\midrule
\multicolumn{7}{l}{\textbf{Panel A. US and UK}} \\
% SD-based &     0.184 &     0.173 &     0.165 &     0.149 &     0.198 &     0.174 \\
%  & [0.173,     0.195] & [0.162,     0.183] & [0.154,     0.177] & [0.137,     0.161] & [0.186,     0.210] & [0.161,     0.187] \\
Variance-based &     0.069 &     0.061 &     0.056 &     0.055 &     0.072 &     0.070 \\
 & [0.060,     0.078] & [0.053,     0.069] & [0.047,     0.064] & [0.046,     0.064] & [0.062,     0.082] & [0.060,     0.080] \\
Theil-based &     0.099 &     0.089 &     0.082 &     0.080 &     0.110 &     0.105 \\
 & [0.089,     0.109] & [0.079,     0.099] & [0.072,     0.092] & [0.071,     0.090] & [0.099,     0.121] & [0.094,     0.117] \\
% Cpp &     0.161 &     0.150 &     0.144 &     0.131 &     0.172 &     0.152 \\
%  & [0.151,     0.171] & [0.140,     0.161] & [0.133,     0.155] & [0.119,     0.142] & [0.160,     0.183] & [0.139,     0.164] \\
Observations &      1127 &      1035 &       883 &       883 &       883 &       883 \\
\midrule
\multicolumn{7}{l}{\textbf{Panel B. US}} \\
% SD-based &     0.195 &     0.182 &     0.173 &     0.156 &     0.205 &     0.180 \\
%  & [0.180,     0.210] & [0.166,     0.198] & [0.155,     0.191] & [0.137,     0.175] & [0.187,     0.224] & [0.160,     0.201] \\
Variance-based &     0.078 &     0.068 &     0.063 &     0.062 &     0.080 &     0.078 \\
 & [0.065,     0.091] & [0.054,     0.082] & [0.048,     0.077] & [0.047,     0.076] & [0.063,     0.096] & [0.061,     0.094] \\
Theil-based &     0.109 &     0.097 &     0.089 &     0.087 &     0.117 &     0.112 \\
 & [0.095,     0.124] & [0.082,     0.113] & [0.073,     0.106] & [0.071,     0.104] & [0.098,     0.135] & [0.093,     0.131] \\
% Cpp &     0.172 &     0.160 &     0.152 &     0.138 &     0.180 &     0.159 \\
%  & [0.157,     0.187] & [0.144,     0.175] & [0.134,     0.169] & [0.120,     0.156] & [0.161,     0.198] & [0.139,     0.179] \\
Observations &       578 &       528 &       442 &       442 &       442 &       442 \\
\midrule
\multicolumn{7}{l}{\textbf{Panel C. UK}} \\
% SD-based &     0.172 &     0.163 &     0.158 &     0.142 &     0.190 &     0.167 \\
%  & [0.158,     0.187] & [0.148,     0.177] & [0.143,     0.173] & [0.126,     0.158] & [0.174,     0.206] & [0.149,     0.185] \\
Variance-based &     0.059 &     0.053 &     0.049 &     0.048 &     0.065 &     0.063 \\
 & [0.048,     0.070] & [0.043,     0.064] & [0.039,     0.059] & [0.038,     0.058] & [0.053,     0.076] & [0.051,     0.075] \\
Theil-based &     0.089 &     0.080 &     0.075 &     0.073 &     0.102 &     0.099 \\
 & [0.076,     0.101] & [0.068,     0.093] & [0.063,     0.087] & [0.061,     0.086] & [0.088,     0.117] & [0.084,     0.113] \\
% Cpp &     0.149 &     0.141 &     0.136 &     0.123 &     0.164 &     0.145 \\
%  & [0.136,     0.163] & [0.127,     0.154] & [0.122,     0.150] & [0.109,     0.138] & [0.149,     0.179] & [0.128,     0.161] \\
Observations &       549 &       507 &       441 &       441 &       441 &       441 \\
\midrule
\multicolumn{7}{l}{\textbf{Specification:}} \\
Passed attention check & . & \checkmark & \checkmark & \checkmark & \checkmark  & \checkmark \\
Passed comprehension check & . & . & \checkmark & \checkmark &  \checkmark & \checkmark \\
Linear adjustment & . & . & . & \checkmark & . & \checkmark \\
Effort adjustment & . & . & . & . & \checkmark & \checkmark \\
\bottomrule
\end{tabular}

}

\begin{minipage}{1\textwidth}
\footnotesize
\noindent \textbf{Notes:} This table reports the average non-linear scale-use costs $C$ with bootstrapped confidence intervals (500 replications). These costs are bounded between 0 (perfect linearity) and 1 (maximal non-linearity). The \textit{linear adjustment} assigns a cost of zero to respondents who reported wanting to make spacing equal between sliders. The \textit{effort adjustment} corrects for inattention or low effort, exploiting random initial slider positions to account for respondents who do not actively adjust distances. Linearity ($C=0$) is consistently rejected. Both adjustments have only a modest impact on the estimates, suggesting that the observed non-linearity is not primarily driven by mechanical scale transformations or low effort.
\end{minipage}
\end{table}

\begin{table}[!t]
\def\sym#1{\ifmmode^{#1}\else\(^{#1}\)\fi}
\caption{Alternative approaches to elicit average deviation from linearity}
\label{tab:complementary_cost2}
\centering
    \begin{tabular}{l*{7}{c}}
    
\toprule
 &  Variance-based & Theil-based \\  

 \midrule

% \textit{Main Survey - Interactive sliders:} \\

% \quad All Respondents (N=883) & 0.174 & 0.070 & 0.105 & 0.152 \\
%  & [0.161-0.187]  & [0.060-0.080] & [0.094-0.117] & [0.139-0.164] \\
% \quad US Respondents (N=442) & 0.180 & 0.078 & 0.112 & 0.159 \\
%  & [0.160-0.201]  & [0.061-0.094] & [0.094-0.131] & [0.139-0.179] \\
% \quad UK Respondents (N=441) & 0.167 & 0.063 & 0.099 & 0.145 \\
%  & [0.149-0.185]  & [0.051-0.075] & [0.084-0.113] & [0.128-0.161] \\

% \textit{Complementary Evidence} \\
\textit{Linear prompt approach}   \\
 \quad Sept 24 survey (N=584)  & 0.015 & 0.032   \\
 &  [0.007-0.024] &  [0.014-0.053]    \\

 \textit{Objective-subjective approach}   \\
 \quad \textit{Using height}  \\
 \quad \quad Apr 26 survey (N=976) & 0.020 & 0.037  \\
  & [0.011-0.035] & [0.018-0.067] \\
\quad \quad Sept 24 survey (N=1,147)  &  0.019 & 0.038  \\
 & [0.010-0.032]  &  [0.021-0.061]  \\
\quad \textit{Using weight}  \\
\quad \quad Apr 26 survey (N=976) & 0.057 & 0.115  \\
  & [0.032-0.087] & [0.069-0.176]  \\
\quad \quad Sept 24 survey (N=1,147) & 0.030  & 0.066    \\
 & [0.013-0.054]  &  [0.030-0.113]    \\ 
\textit{Jump approach} \\
\quad Correa Mackliff survey (N=508)  & 0.024 & 0.040 &    \\
 & [0.019-0.028]  &  [0.035-0.045] &    \\
\bottomrule
\end{tabular}

\begin{minipage}{.94\textwidth}
\footnotesize
\noindent \textbf{Notes:} This table reports the average non-linear scale-use cost $C$. These alternative cost functions are reported to assess the robustness of the results reported in Table \ref{tab:complementary_cost} and to quantify the range of average non-linear scale use were another function deemed more adequate.
\end{minipage}

\end{table}

\begin{landscape}
\begin{table}[!h]
\def\sym#1{\ifmmode^{#1}\else\(^{#1}\)\fi}
\caption{Predictors of the Probability and Cost of Sign-reversal: Robustness Checks}
\label{tab:robustness_sign}
\centering
    \begin{tabular}{l*{6}{c}}
\toprule
  &\multicolumn{3}{c}{P(Sign-reversal)}  &\multicolumn{3}{c}{Cost of sign-reversal}                               \\
                           \cmidrule(lr){2-4} \cmidrule(lr){5-7}
                    &\multicolumn{1}{c}{(1)}         &\multicolumn{1}{c}{(2)}         &\multicolumn{1}{c}{(3)}         &\multicolumn{1}{c}{(4)}         &\multicolumn{1}{c}{(5)}    &\multicolumn{1}{c}{(6)}                    \\
\midrule
% 7-points scale or longer&       0.227\sym{***}&       0.219\sym{***}&       0.290\sym{***}&      -0.176\sym{***}&      -0.213\sym{***}\\
%                     &     (0.016)         &     (0.027)         &     (0.011)         &     (0.008)         &     (0.010)         \\

% Cantril Ladder      &       0.006         &       0.198\sym{***}&      -0.047\sym{**} &       0.024\sym{**} &       0.090\sym{*}  \\
%                     &     (0.024)         &     (0.049)         &     (0.021)         &     (0.011)         &     (0.046)         \\

% Happiness question  &      -0.034\sym{***}&      -0.036\sym{**} &      -0.008         &       0.001         &      -0.027\sym{***}\\
%                     &     (0.010)         &     (0.014)         &     (0.009)         &     (0.006)         &     (0.009)         \\
\textbf{About the estimation sample:} \\
\quad Number of observations (logged)&       0.026\sym{***}&       0.019\sym{*}  &       0.048\sym{***}&      -0.029\sym{***}&      -0.026\sym{***}&      -0.060\sym{***}\\
                    &     (0.006)         &     (0.009)         &     (0.006)         &     (0.003)         &     (0.004)         &     (0.004)         \\
%\addlinespace
%\quad Only one nationality included&       0.016         &       0.041         &       0.008         &      -0.004         &       0.141\sym{***}\\
%                    &     (0.017)         &     (0.040)         &     (0.012)         &     (0.007)         &     (0.013)         \\
\textbf{About the econometric model:} \\
\quad Number of controls  &      -0.001         &      -0.015         &       0.004         &      -0.004         &      -0.003         &      -0.009\sym{*} \\
                    &     (0.009)         &     (0.014)         &     (0.006)         &     (0.003)         &     (0.006)         &     (0.004)         \\
%\addlinespace
Individual FE       &       0.002         &       0.120\sym{***}&      -0.000         &      -0.013\sym{*}  &       0.077  &      -0.010         \\
                    &     (0.013)         &     (0.035)         &     (0.015)         &     (0.007)         &     (0.041)         &     (0.012)         \\

\textbf{About the independent variable:} \\
\quad Continuous variable &      -0.032\sym{***}&      -0.030\sym{***}&      -0.034\sym{***}&       0.006         &       0.018\sym{***}&       0.003         \\
                    &     (0.007)         &     (0.007)         &     (0.007)         &     (0.005)         &     (0.005)         &     (0.006)         \\
%\addlinespace
\quad Time-varying variable&      -0.078\sym{***}&      -0.084\sym{***}&      -0.068\sym{***}&       0.057\sym{***}&       0.066\sym{***}&       0.040\sym{***}\\
                    &     (0.008)         &     (0.008)         &     (0.007)         &     (0.004)         &     (0.004)         &     (0.005)         \\
%\addlinespace
\quad Two-stage least squares&    -0.030         &      -0.051         &      -0.035         &       0.021         &       0.036\sym{*} &       0.011         \\
                    &     (0.032)         &     (0.031)         &     (0.029)         &     (0.018)         &     (0.015)         &     (0.021)         \\
%\addlinespace
\quad Natural experiment  &      -0.058\sym{***}&      -0.081\sym{***}&      -0.037\sym{***}&       0.034\sym{***}&       0.050\sym{***}&       0.023\sym{*} \\
                    &     (0.014)         &     (0.014)         &     (0.010)         &     (0.010)         &     (0.010)         &     (0.010)         \\
%\addlinespace
\quad Macroeconomic indicator&      -0.066\sym{***}&      -0.036\sym{*} &      -0.062\sym{***}&       0.037\sym{***}&       0.015         &       0.047\sym{***}\\
                    &     (0.015)         &     (0.014)         &     (0.012)         &     (0.008)         &     (0.010)         &     (0.011)         \\
%\addlinespace
Absolute t-statistics (logged)&-0.285\sym{***}&      -0.282\sym{***}&      -0.315\sym{***}&       0.193\sym{***}&       0.196\sym{***}&       0.273\sym{***}\\
                    &     (0.004)         &     (0.004)         &     (0.004)         &     (0.002)         &     (0.003)         &     (0.004)         \\
\midrule
Observations        &       28,522         &       28,522         &       28,522         &       17,243         &       17,243    & 28,522     \\
Journal FE  &  $\checkmark$ & . & . & $\checkmark$ & .& . \\
Paper FE & . & $\checkmark$ & . & .  & $\checkmark$ & . \\
\bottomrule
\end{tabular}

\begin{minipage}{1.15\textwidth}
\footnotesize
\noindent \textbf{Notes:} Columns (3) reports marginal effects from probit models and Column (6) reports the coefficients of a linear Cragg hurdle model where we assign a cost of reversal of zero for estimates that cannot be reversed. The other Columns report OLS coefficients. All regressions control for a dummy equal to one for wellbeing scales including at least seven categories, a categorical variable indicating whether the wellbeing measure is a life-satisfaction, Cantril Ladder or happiness question. Standard errors are clustered at the regression-paper level. Statistical significance is denoted as follows: *$p < 0.05$, **$p < 0.01$, and ***$p < 0.001$.
  
\end{minipage}

\end{table}  % this will be Table A4
\end{landscape}

\clearpage

\begin{landscape}
\subsection{Additional tables on primary and secondary data}

\begin{table}[!h]
\caption{Description of Primary Datasets}
\label{tab:dataset_description}
\small 
\scriptsize
\begin{tabular}{lp{1.5cm}p{1cm}p{6cm}p{6cm}p{3.5cm}}
\toprule
Short Name & Country & Time & Measure of $r$ & Notes & Reference \\
\midrule
Prolific & US/UK & 2026 & ``Overall, how satisfied are you with your life nowadays?'' & The discrete measure has 11 response options and mirrors the questions used in the UK APS. Continuous measure constructed by asking respondents about their location within a given discrete response option. Sample obtained via Prolific, with the nationally representative option.  & \hyperref[sec:titlepage]{Kaiser and Lepinteur (2025)} \\
Prolific & UK & 2024 & ``Overall, how satisfied are you with your life nowadays?'' & The discrete measure has 11 response options and mirrors the questions used in the UK APS. Continuous measure constructed by asking respondents about their location within a given discrete response option. Sample obtained via Prolific, with the nationally representative option.  & \hyperref[sec:titlepage]{Kaiser and Lepinteur (2025)} \\
% Benjamin et al. & US & 2022 & Discrete measure is Cantril's ladder of life (11 response options). Continuous measure asked: ``How satisfied you are with your life?'' & Continuous and discrete measure obtained with two questions in the same survey. Sample obtained via MTurk. & \cite{benjamin2023adjusting} \\
% Prati \& Kaiser & UK & 2023-2024 & ``All things considered, how satisfied are you with your life nowadays?'' & The discrete measure has 7 response options and mirrors the question used in the UKHLS. Continuous and discrete measure obtained with two questions in the same survey. Sample obtained via Prolific & \cite{kaiser2025from}  \\
% LISS & NL & 2011 & ``Taking all things together, how happy would you say you are?” & The discrete measure has 10 response options.  In both measures, extremes are labelled “completely unhappy” and “completely happy''. Continuous and discrete measures obtained via two surveys administered one month apart. Sample based on long-standing \hyperlink{LISS}{https://www.lissdata.nl/} panel. & \cite{studer2012does}. Also used in \cite{kaiser2023how} \\
\bottomrule
\end{tabular}
\vspace{5pt}
\begin{minipage}{0.95\textwidth}
\small
\noindent \textbf{Note:} Description of datasets used in Section \ref{sec:scale_use_evidence}.
\end{minipage}
\end{table}  % Table with descriptions for each of the datasets used to estimates gamma etc
\end{landscape}

\begin{table}[!h]
\caption{Descriptive statistics for April 2026 Prolific data}
\label{tab:descriptives_prolific26}
\centering
\begin{tabular}{l*{1}{ccccc}}
\toprule
            &       N &        Mean &          SD &         Min &         Max \\
\midrule
\textbf{Satisfaction measure} & & & & & \\
Life satisfaction (discrete)&        1127&        6.37&        2.14&        0.00&       10.00\\
Life satisfaction (continuous)&        1106&        6.50&        2.13&        0.00&       10.00\\
\midrule
\textbf{Height \& weight} & & & & & \\
Height(cm)  &        1107&      171.00&       10.17&      129.54&      198.12\\
Weight(kg)  &        1108&       79.05&       22.08&       40.37&      190.51\\
\midrule
\textbf{Slider values} & & & & & \\
Slider 1    &        1127&        1.10&        1.33&        0.00&       10.00\\
Slider 2    &        1127&        1.90&        1.32&        0.00&       10.00\\
Slider 3    &        1127&        2.83&        1.37&        0.00&       10.00\\
Slider 4    &        1127&        3.82&        1.37&        0.00&       10.00\\
Slider 5    &        1127&        4.88&        1.31&        0.00&       10.00\\
Slider 6    &        1127&        5.92&        1.28&        0.00&       10.00\\
Slider 7    &        1127&        7.04&        1.16&        0.00&       10.00\\
Slider 8    &        1127&        8.06&        1.08&        0.00&       10.00\\
Slider 9    &        1127&        9.00&        0.94&        0.00&       10.00\\
\midrule
\textbf{Demographics} & & & & & \\
Ln(Income)  &        1127&        7.21&        1.81&        0.00&        9.39\\
Not Unemployed  &        1127&        0.93&        0.25&        0.00&        1.00\\
Age         &        1122&       45.54&       15.56&       18.00&       81.00\\
Age Squared &        1122&       23.16&       14.42&        3.24&       65.61\\
Has partner &        1127&        0.61&        0.49&        0.00&        1.00\\
Higher education&        1127&        0.60&        0.49&        0.00&        1.00\\
Non-White   &        1127&        0.25&        0.43&        0.00&        1.00\\
Female      &        1111&        0.51&        0.50&        0.00&        1.00\\
Household Size&        1127&        2.59&        1.28&        1.00&        7.00\\
Has Children&        1127&        0.27&        0.45&        0.00&        1.00\\
Homeowner   &        1127&        0.62&        0.49&        0.00&        1.00\\
\bottomrule
\end{tabular}

\vspace{5pt}

\begin{minipage}{0.95\textwidth}
\small
\noindent \textbf{Note:} Descriptive statistics for main Prolific data used in Section \ref{sec:scale_use_evidence}.
\end{minipage}

\end{table}  % Descriptives for Prolific data

\begin{landscape}
\begin{table}[!h]
\caption{Descriptive statistics for September 2024 Prolific and \citet{jump} data}
\label{tab:descriptives_prolific}
\centering
\begin{tabular}{l*{1}{cccccccccccccccccccccc}}
\toprule
& \multicolumn{5}{c}{\textbf{September 24 Prolific}} & \multicolumn{5}{c}{\textbf{\citet{jump}}} \\
\cmidrule(lr){2-6} \cmidrule(lr){7-11} 
            &       N &        Mean &          SD &         Min &         Max       &       N &        Mean &          SD &         Min &         Max \\
\midrule
\textbf{Satisfaction measure} & & & & & \\
Life satisfaction (discrete)&        1238&        6.28&        2.07&        0.00&       10.00 & 508 & 6.73 & 1.93 & 0.00 & 10.00 \\
LS (discrete unprompted)&         621&        6.38&        1.97&        0.00&       10.00\\
LS (discrete linear prompt)&         617&        6.18&        2.16&        0.00&       10.00\\
Life satisfaction (continuous)&        1216&        6.42&        2.07&        0.00&       10.00\\
LS (continuous unprompted)&         613&        6.49&        2.05&        0.00&       10.00\\
LS (continuous linear prompt)&         603&        6.35&        2.08&        0.20&       10.00\\
\midrule
\textbf{Height \& weight} & & & & & \\
Height(cm)  &        1185&      171.13&       10.37&      129.69&      198.12\\
Weight(kg)  &        1186&       81.60&       24.83&       40.82&      192.32\\
\midrule
\textbf{Slider values} & & & & & \\
Slider 1    &         606&        1.07&        0.84&        0.00&        8.60 &         508&        0.72&        0.49&        0.35&        6.90\\
Slider 2    &         606&        1.95&        1.07&        0.00&        8.60  &         508&        1.47&        0.77&        0.69&        7.24 \\
Slider 3    &         606&        2.85&        1.20&        0.40&        8.60 &         508&        2.26&        0.96&        1.03&        7.59\\
Slider 4    &         606&        3.86&        1.16&        0.70&        8.70 &         508&        3.10&        1.11&        1.03&        7.93\\
Slider 5    &         606&        4.94&        1.10&        1.20&        8.90 &         508&        4.03&        1.20&        1.72&        8.28\\
Slider 6    &         606&        6.03&        1.18&        1.30&        9.30 &         508&        5.04&        1.25&        2.07&        8.62\\
Slider 7    &         606&        7.05&        1.24&        1.30&       10.00  &         508&        6.14&        1.20&        2.41&        8.97\\
Slider 8    &         606&        7.98&        1.05&        4.30&       10.00 &         508&        7.30&        1.02&        2.76&        9.31\\
Slider 9    &         606&        8.94&        0.69&        6.60&       10.00 &         508&        8.56&        0.71&        3.10&        9.66\\
% \midrule
% \textbf{Demographics} & & & & & \\
% Ln(Income)  &        1127&        7.21&        1.81&        0.00&        9.39\\
% Not Unemployed  &        1127&        0.93&        0.25&        0.00&        1.00\\
% Age         &        1122&       45.54&       15.56&       18.00&       81.00\\
% Age Squared &        1122&       23.16&       14.42&        3.24&       65.61\\
% Has partner &        1127&        0.61&        0.49&        0.00&        1.00\\
% Higher education&        1127&        0.60&        0.49&        0.00&        1.00\\
% Non-White   &        1127&        0.25&        0.43&        0.00&        1.00\\
% Female      &        1111&        0.51&        0.50&        0.00&        1.00\\
% Household Size&        1127&        2.59&        1.28&        1.00&        7.00\\
% Has Children&        1127&        0.27&        0.45&        0.00&        1.00\\
% Homeowner   &        1127&        0.62&        0.49&        0.00&        1.00\\
\bottomrule
\end{tabular}

\vspace{5pt}

\begin{minipage}{0.95\textwidth}
\small
\noindent \textbf{Note:} Descriptive statistics for the data used in Section \ref{sec:complement}.
\end{minipage}

\end{table}  % Descriptives for Prolific data
\end{landscape}

\end{document}